\documentclass[twocolumn]{aastex631}
\usepackage{amsmath}
\usepackage{amssymb}
\usepackage{euscript}
\usepackage{algorithm2e,algorithmic}
\usepackage{bbold}
\usepackage{paralist}
\usepackage{hyperref}
\usepackage{booktabs}
\usepackage{rotating}
\hypersetup{         
    unicode=true,          colorlinks=true,       linkcolor=red,          citecolor=blue,        filecolor=magenta,      urlcolor=red           }

\def\gtorder{\mathrel{\raise.3ex\hbox{$>$}\mkern-14mu
             \lower0.6ex\hbox{$\sim$}}}
\def\ltorder{\mathrel{\raise.3ex\hbox{$<$}\mkern-14mu
             \lower0.6ex\hbox{$\sim$}}}

\def\gtorder{\mathrel{\raise.3ex\hbox{$>$}\mkern-14mu
             \lower0.6ex\hbox{$\sim$}}}
\def\ltorder{\mathrel{\raise.3ex\hbox{$<$}\mkern-14mu
             \lower0.6ex\hbox{$\sim$}}}

\newcommand{\package}[1]{\texttt{#1}}

\newcommand{\swift}{\textit{Swift}}

\newcommand{\ultrasat}{\textit{ULTRASAT}}

\newcommand{\kms}{km$\,$s$^{-1}$}

\newcommand{\msun}{$M_{\odot}$}

\shortauthors{Irani et al.}

\newcommand{\referee}[1]{#1}
\newcommand{\refereetwo}[1]{#1}

\begin{document}

\title{The Early Ultraviolet Light-Curves of Type II Supernovae and the Radii of Their Progenitor Stars}

\author[0000-0002-7996-8780]{Ido ~Irani}
\affiliation{Department of Particle Physics and Astrophysics,
             Weizmann Institute of Science,
             234 Herzl St, 7610001 Rehovot, Israel}
\email{idoirani@gmail.com}

\author{Jonathan Morag}
\affiliation{Department of Particle Physics and Astrophysics,
             Weizmann Institute of Science,
             234 Herzl St, 7610001 Rehovot, Israel}

\author[0000-0002-3653-5598]{Avishay ~Gal-Yam}
\affiliation{Department of Particle Physics and Astrophysics,
             Weizmann Institute of Science,
             234 Herzl St, 7610001 Rehovot, Israel}

\author{Eli ~Waxman}
\affiliation{Department of Particle Physics and Astrophysics,
             Weizmann Institute of Science,
             234 Herzl St, 7610001 Rehovot, Israel}

\author[0000-0001-6797-1889]{Steve ~Schulze}
\affiliation{Department of Physics, 
             The Oskar Klein Center, Stockholm University, 
             AlbaNova, SE-10691 Stockholm, Sweden}

\author[0000-0003-1546-6615]{Jesper~Sollerman}
\affiliation{Department of Astronomy, 
             The Oskar Klein Center, Stockholm University, 
             AlbaNova, SE-10691 Stockholm, Sweden}

\author{K-Ryan Hinds}
\affiliation{Astrophysics Research Institute, Liverpool John Moores University, IC2 Liverpool Science Park, 146 Brownlow Hill, Liverpool L3 5RF, UK}

\author{Daniel A. Perley}
\affiliation{Astrophysics Research Institute, Liverpool John Moores University, IC2 Liverpool Science Park, 146 Brownlow Hill, Liverpool L3 5RF, UK}

\author[0000-0003-0853-6427]{Ping Chen}
\affiliation{Department of Particle Physics and Astrophysics,
             Weizmann Institute of Science,
             234 Herzl St, 7610001 Rehovot, Israel}

\author[0000-0002-4667-6730]{Nora L. Strotjohann}
\affiliation{Department of Particle Physics and Astrophysics,
             Weizmann Institute of Science,
             234 Herzl St, 7610001 Rehovot, Israel}

\author[0000-0002-0301-8017]{Ofer Yaron}
\affiliation{Department of Particle Physics and Astrophysics,
             Weizmann Institute of Science,
             234 Herzl St, 7610001 Rehovot, Israel}

\author[0000-0001-8985-2493]{Erez A. Zimmerman}
\affiliation{Department of Particle Physics and Astrophysics,
             Weizmann Institute of Science,
             234 Herzl St, 7610001 Rehovot, Israel}

\author{Rachel Bruch}
\affiliation{Department of Particle Physics and Astrophysics,
             Weizmann Institute of Science,
             234 Herzl St, 7610001 Rehovot, Israel}

\author[0000-0002-6786-8774]{Eran O. Ofek}
\affiliation{Department of Particle Physics and Astrophysics,
             Weizmann Institute of Science,
             234 Herzl St, 7610001 Rehovot, Israel}
             
\author[0000-0001-6753-1488]{Maayane T. Soumagnac}
\affiliation{Department of Physics, Bar Ilan University, Ramat-Gan, 52900, Israel}
\affiliation{ Lawrence Berkeley National Laboratory, 1 Cyclotron Road, Berkeley, CA 94720, USA}                
          
\author[0000-0002-6535-8500]{Yi Yang}
\affiliation{Department of Astronomy, University of California, Berkeley, CA 94720-3411, USA}
\affiliation{Bengier-Winslow-Robertson Postdoctoral Fellow}

\author[0000-0001-5668-3507]{Steven L. Groom}
\affiliation{IPAC, California Institute of Technology, 1200 E. California
             Blvd, Pasadena, CA 91125, USA}

\author[0000-0002-8532-9395]{Frank J. Masci}
\affiliation{IPAC, California Institute of Technology, 1200 E. California
             Blvd, Pasadena, CA 91125, USA}
\author{Marie Aubert}
\affiliation{Universit\'e Clermont Auvergne, CNRS/IN2P3, LPC, F-63000 Clermont-Ferrand, France}
    
\author{Reed Riddle}
\affiliation{Caltech Optical Observatories, California Institute of Technology, Pasadena, CA 91125, USA}

\author[0000-0001-8018-5348]{Eric C. Bellm}
\affiliation{DIRAC Institute, Department of Astronomy, University of Washington, 3910 15th Avenue NE, Seattle, WA 98195, USA}
\author{David Hale}
\affiliation{Caltech Optical Observatories, California Institute of Technology, Pasadena, CA  91125}

\begin{abstract}
We present a sample of 34 normal SNe II detected with the Zwicky Transient Facility, with multi-band UV light-curves starting at $t\leq4\,$days after explosion, and X-ray observations. We characterize the early UV-optical color, provide empirical host-extinction corrections and show that the $t>2\,$days UV-optical colors and the blackbody evolution of the sample are consistent with shock-cooling (SC) regardless of the presence of `flash ionization" features. We present a framework for fitting SC models which can reproduce the parameters of a set of multi-group simulations up to 20\% in radius and velocity.  Observations of 15 SNe II are well-fit by models with breakout radii $<10^{14}\,$cm. 18 SNe are typically more luminous, with observations at $t\geq1\,$day that are better fit by a model with a large $>10^{14}\,$cm breakout radius. However, these fits predict an early rise during the first day that is too slow. We suggest these large-breakout events are explosions of stars with an inflated envelope or with confined circumstellar material (CSM). Using the X-ray data, we derive constraints on the extended ($\sim10^{15}\,$cm) CSM density independent of spectral modeling, and find most SNe II progenitors lose $\dot{M}<10^{-4}M_{\odot}\,\rm yr^{-1}$ up to a few years before explosion. We show that the overall observed breakout radius distribution is skewed to higher radii due to a luminosity bias. We argue that the $66^{+11}_{-22}\%$ of red supergiants (RSG) explode as SNe II with breakout radii consistent with the observed distribution of RSG, with a tail extending to large radii, likely due to the presence of CSM. 
\newpage
\end{abstract}

\section{Introduction}
\label{sec:introduction}
The progenitor stars of the majority of spectroscopically regular \citep{galyam2017} supernovae (SNe) II are red super-giants (RSG), as confirmed by pre-SN detections \citep[see][and references therein]{smartt2009,smartt2015a,vandyk2017}. While this is the case, we do not yet know if all RSG stars explode as SNe, and the details of the latest stages of stellar evolution are not accurately known. As we cannot know which star will explode as a SN ahead of time, the only way of systematically observing the short-lasting final stages of stellar evolution are through their terminal explosions as SNe. Using this approach, the properties of a progenitor star immediately prior to explosion can be connected to its observed supernova.
Connecting the progenitors to the SN explosions they create has been a long-lasting goal of supernova studies \citep{Galyam2007,smartt2015a,Modjaz2019}. In the last decade, large statistical studies of SNe have become commonplace. While these can place some constraints on the progenitor properties, the progenitor radius, ejected mass and explosion energy have degenerate effects on the SN light curves \citep[][]{Goldberg2019,Dessart2019}. Acquiring independent estimates of these properties through their peak and plateau properties remains a difficult and unsolved problem.

Measuring the progenitor radius is possible by observing the earliest phase of the SN explosion.  The first photons emitted from the SN explosion will be the result of shock breakout of the radiation-mediated shock from the stellar surface - the breakout pulse. The photons that were captured in the shock transition region escape on a timescale of \referee{minutes to hours if breakout will occur at the edge of the stellar envelope, or longer if it occurs in the surrounding circumstellar material (CSM)}. Typically, this allows us to constrain the progenitor radius directly from the duration of the breakout pulse \citep[for a review on the subject, see][ and references therein]{Waxman2017}.  The shocked material, which has been compressed and heated, is then ejected and quickly reaches a state of homologous expansion \citep{Matzner_1999}. From the moment of shock-breakout and in the absence of interaction with pre-existing material above the photosphere, the dominant emission mechanism is the cooling of this heated envelope, which evolves according to simple analytic solutions until hydrogen recombination becomes significant. 

This stage, called the shock-cooling phase, typically lasts a few days for normal SNe II, and less than a day for stripped-envelope supernovae and 1987A-like SNe II. During this time, the temperature and luminosity evolution are highly sensitive to the progenitor radius and to the shock velocity - allowing to constrain these parameters \citep{Chevalier1992,Nakar2010,Rabinak2011}. Since the first generation of models, theoretical advancements have extended the applications of shock-cooling models to low-mass envelopes \citep{Piro2015,Piro2021} and later times \citep{sapir2016}. Recently, \citet[][hereafter \refereetwo{M23}]{Morag2023} interpolated between the planar and spherical phases, extending the validity of the model of \cite{sapir2016} to earlier times, and treated the suppression of flux in UV due to line absorption \referee{and emission} \citep[][\refereetwo{M24}]{Morag2024}. \referee{This model, as well as its predecessors, are valid prior to hydrogen recombination at $0.7$ eV.}

In the past decade, high-cadence and wide-field surveys have enabled the early time detection and multi-band followup of SNe.  
The Palomar Transient Factory (PTF; \citealt{law2009,kulkarni2013}), the Astroid-Terrestrial impact Last Alert System (ATLAS; \citealt{Tonry2018}), the Zwicky Transient Facility (ZTF; \citealt{Bellm2019b,Graham2019}), the Distance Less than 40\,Mpc Survey \citep[DLT40;][]{Tartaglia2018}, and most recently the Young Supernovae Experiment (YSE;  \citealt{Jones2021}) have been conducting 1--3\,day cadence wide-field surveys and regularly detect early phase SNe \citep[e.g.][]{Hachinger2009, galyam2011,arcavi2011,nugent2011,galyam2014, Benami2014, Khazov2016, Yaron2017,Hosseinzadeh_2018,Ho2019,Soumagnac2020,Bruch2021, galyam2022,Perley2022,Terreran2022,JacobsonGalan2022b,Tinyanont2022,Hosseinzadeh2022, Irani2024}. 

Previous attempts to model the early-phase emission of SNe II yield mixed results. Many studies fit the analytical shock cooling models of \cite{Nakar2010} or \cite{Rabinak2011}. These models require multiband photometry extending to the early time and the UV, as the model parameters are highly sensitive to the temperature $\sim$1 day after explosion. Many works find radii that are small compared to the observed RSG distribution from the Small and Large Magellanic Clouds (SMC,LMC). For example, \cite{Gonzalez2015} and \cite{Gall2015} compile large optical light curve samples, fitting $ugriz$ and $r$ band photometry respectively, and \referee{assume} a \referee{fixed} validity \referee{time} for the models \referee{(i.e., not dependent on the model parameters)}. While \cite{Rubin_2016,Rubin_2017} demonstrated that adopting a fixed validity introduces a bias in the parameter inference, a fixed validity remains commonplace \citep[e.g.,][]{Hosseinzadeh_2018}. Recent attempts by \cite{Soumagnac2020,Ganot_2022} and 
\cite{Hosseinzadeh2023} find large RSG radii $\sim1000\, R_{\odot}$ by fitting early UV-optical light-curves, in tension with previous results, while \cite{Vallely2021} fit single band high-cadence \textit{Transiting Exoplanet Survey Satellite} \citep[\textit{TESS};][]{Ricker2014} light curves and find unrealistically small RSG progenitor radii, which they calibrate to numerical simulations.

While some large samples by \cite{valenti2016,Faran2018} fit the luminosities and temperatures of SNe II using multi-band UV-optical datasets, these did not extend to the very early times. However, these studies demonstrate that the blackbody evolution is in agreement with the expectations of the shock-cooling framework of a cooling blackbody with $T\sim t^{-0.5}$ \citep{Faran2018}.

A different approach to analytic cooling models is the use of numerical hydrodynamical simulations. Motivated by the fact that narrow features from CSM interaction are commonly observed in SNe II \citep{galyam2014,Khazov2016,Yaron2017,Bruch2021,bruch2022}, these models include a dense shell of CSM, ejected from the progenitor before explosion. This results in an extended non-polytropic density profile extending to few $10^{14}\, \rm cm$ from the progenitor star prior to explosion.  \cite{Morozova2018} shows the early time multi-band evolution of a sample of SNe II is better explained by models with dense CSM compared to models which do not include CSM. The breakout radii in this case are typically at the edge of the CSM, at large radii ($\lesssim3000 R_{\odot}$).  \cite{Dessart2017,Dessart2019} fit the early ($>$ few days) spectroscopic and photometric sequence of SNe with a grid of non-LTE simulations, and find a small amount of CSM improves the match of the models with the early time photometry. \cite{forster2018} fit a sample of 26 (\refereetwo{photometrically classified)} SNe II to a grid of hydrodynamical models and argue that they observe a delayed rise in the majority of SNe II explained by the presence of CSM.

In this paper we present a sample of spectroscopically regular SNe II with well-sampled UV-optical light curves. We present our sample selection strategy in $\S$~\ref{sec:sample}, and the details of our photometric and X-ray follow-up in  $\S$~\ref{sec:observations}. In $\S$~\ref{sec:results} we analyze the color evolution ($\S$~\ref{subsec:color}), and blackbody evolution ($\S$~\ref{subsec:blackbody}) of the SNe. In $\S$~\ref{subsec:SC_fitting} we model the light curves during the shock cooling phase. We discuss our results and their implications to the SN progenitors in $\S$~\ref{sec:discussion}. 

Throughout the paper we use a flat $\Lambda$CDM cosmological model with H$_{0} = 67.4$\,\kms\,Mpc$^{-1}$, $\Omega_{M} = 0.315$, and $\Omega_{\Lambda} = 0.685$ \citep{Planck2018a}. 
\pagebreak

\section{Sample}
\label{sec:sample}
\subsection{Observing strategy}
In \cite{Bruch2021}, we described the selection process of infant SNe from the ZTF alert stream. Using a custom filter, we select transient in extragalactic fields ($|b|>14$ deg), with a non detection-limit $<2.5$ days from the first detection, and from a non-stellar origin. These candidates are routinely manually inspected by a team of duty astronomers in Europe and Israel during California night-time in order to reject false positives (such as stellar flares, galactic transients, and active galactic nuclei).  Management of follow-up resources and candidates was performed through the GROWTH marshal \citep{Kasliwal2019} and  Fritz/\package{SkyPortal} platforms \citep{vanderWalt2019,Coughlin_2023}.  Promising candidates rising by at least $0.5$ mag from the previous non-detection are followed-up with optical spectroscopy, optical photometry (various instruments) and UV photometry using the UV-Optical Telescope (UVOT) onboard the \textit{Neil Gehrels Swift Observatory} \citep{Gehrels2004,Roming2005}. We also followed-up publicly announced infant SNe II which pass our criteria, with ZTF data during the first week. \referee{While initially we required a blue color $g-r<0$ mag for triggering UVOT, This assumption was later relaxed.} For this paper, we consider all ZTF infant SNe with UV photometry in the first 4 days after estimated explosion and which are classified as spectroscopically regular SNe II at peak light. We consider SNe which are detected until Dec 31st, 2021. Classification references are listed in Table \ref{tab:sn_list}. 

\begin{deluxetable*}{llcccccccl}
\centering
\label{tab:sn_list}
\tablecaption{List of 34 SNe included in this study}
\tablewidth{34pt} 
\tablehead{\colhead{SN} &\colhead{ZTF ID} & \colhead{$\alpha$ (J2000)} & \colhead{$\delta$ (J2000)} & \colhead{z} & \colhead{$d^{a}$ [Mpc]} & \colhead{$t_{ND}$ [JD]} & \colhead{$t_{exp}$ [JD]} & \colhead{$\tau_{flash}^{b}$ [days]} & \colhead{Reference$^{c}$}} 
\tabletypesize{\scriptsize} 
\startdata
SN\,2018cxn & ZTF18abckutn & 237.026897 & 55.714855 & 0.0401 & 186.6 & 2458289.7490 & $2458289.76\pm0.01$ & $<0.0$ & [1] \\
SN\,2018dfc & ZTF18abeajml & 252.032360 & 24.304095 & 0.0365 & 170.0 & 2458302.7103 & $2458303.8\pm0.009$ & $6.2\pm2.8$ & [1] \\
SN\,2018fif & ZTF18abokyfk & 2.360629 & 47.354083 & 0.0172 & 76.5 & 2458349.8973 & $2458350.874\pm0.002$ & $1.6\pm1.0$ & [1.2]\\
SN\,2019eoh & ZTF19aatqzim & 195.955635 & 38.289155 & 0.0501 & 229.6 & 2458601.7817 & $2458606.683\pm0.031$ & $<0.0$ & [3] \\
SN\,2019gmh & ZTF19aawgxdn & 247.763189 & 41.153961 & 0.0307 & 141.3 & 2458633.8250 & $2458634.324\pm0.444$ & $<1.4$ & [3] \\
SN\,2019nvm & ZTF19abqhobb & 261.411100 & 59.446730 & 0.0181 & 86.4 & 2458713.7416 & $2458714.69\pm0.007$ & $1.9\pm1.0$ & [3,4] \\
SN\,2019omp & ZTF19abrlvij & 260.142987 & 51.632780 & 0.0450 & 206.9 & 2458717.7910 & $2458718.713\pm0.0$ & $<0.0$ & [3] \\
SN\,2019oxn & ZTF19abueupg & 267.803290 & 51.382550 & 0.0200 & 90.3 & 2458723.7895 & $2458724.342\pm0.129$ & $<0.4$ & [3] \\
SN\,2019ozf & ZTF19abulrfa & 279.817010 & 54.287872 & 0.0480 & 221.2 & 2458723.7900 & $2458724.728\pm0.005$ & $<0.0$ & [3] \\
SN\,2019ust & ZTF19acryurj & 13.593396 & 31.670182 & 0.0220 & 96.0 & 2458799.8053 & $2458800.04\pm0.177$ & $5.0\pm0.5$ & [3] \\
SN\,2019wzx & ZTF19aczlldp & 37.782326 & 4.311291 & 0.0275 & 124.9 & 2458833.7282 & $2458835.506\pm0.092$ & $<1.1$ & [11] \\
SN\,2020cxd & ZTF20aapchqy & 261.621953 & 71.094063 & 0.0039 & 23.7 & 2458896.0296 & $2458896.671\pm0.671$ & $<2.4$ & [5,6] \\
SN\,2020dyu & ZTF20aasfhia & 184.913047 & 33.040393 & 0.0500 & 230.7 & 2458911.9254 & $2458912.814\pm0.021$ & $<0.0$ & [3] \\
SN\,2020fqv & ZTF20aatzhhl & 189.138576 & 11.231654 & 0.0075 & 15.0 & 2458936.9007 & $2458939.43\pm0.16$ & $<0.9$ & [7] \\
SN\,2020jfo & ZTF20aaynrrh & 185.460355 & 4.481697 & 0.0052 & 14.7 & 2458971.7751 & $2458975.231\pm0.424$ & $<0.5$ & [8] \\
SN\,2020lfn & ZTF20abccixp & 246.737033 & 20.245906 & 0.0440 & 202.2 & 2458995.8154 & $2458996.701\pm0.018$ & $4.2\pm1.5$ & [3] \\
SN\,2020mst & ZTF20abfcdkj & 281.793965 & 60.496802 & 0.0590 & 274.0 & 2459012.8161 & $2459013.689\pm0.067$ & $<0.1$ & [3] \\
SN\,2020nif & ZTF20abhjwvh & 196.057282 & -10.351002 & 0.0104 & 50.5 & 2459021.7334 & $2459023.783\pm0.765$ & $<0.9$ & [11] \\
SN\,2020nyb & ZTF20abjonjs & 29.783900 & 86.676205 & 0.0155 & 72.1 & 2459026.9709 & $2459033.849\pm0.014$ & $<0.0$ & [11] \\
SN\,2020pni & ZTF20ablygyy & 225.958184 & 42.114032 & 0.0169 & 83.2 & 2459045.7542 & $2459046.638\pm0.004$ & $5.0\pm1.0$ & [9] \\
SN\,2020pqv & ZTF20abmoakx & 220.498180 & 8.462724 & 0.0338 & 160.2 & 2459046.7104 & $2459048.646\pm0.023$ & $5.2\pm2.5$ & [3] \\
SN\,2020qvw & ZTF20abqkaoc & 250.983335 & 77.879897 & 0.0500 & 230.7 & 2459065.8438 & $2459066.222\pm0.417$ & $<0.6$ & [11] \\
SN\,2020afdi & ZTF20abqwkxs & 224.868111 & 73.898678 & 0.0239 & 110.9 & 2459069.7995 & $2459070.277\pm0.341$ & $1.3\pm0.5$ & [3] \\
SN\,2020ufx & ZTF20acedqis & 322.652706 & 24.673752 & 0.0500 & 230.7 & 2459116.8338 & $2459117.752\pm0.015$ & $4.9\pm1.0$ & [3] \\
SN\,2020uim & ZTF20acfdmex & 28.188740 & 36.623160 & 0.0185 & 80.5 & 2459117.8602 & $2459118.823\pm0.0$ & $<0.1$ & [3] \\
SN\,2020xhs & ZTF20acknpig & 30.742868 & 45.020286 & 0.0244 & 106.6 & 2459138.8669 & $2459138.936\pm0.301$ & $<1.8$ & [3] \\
SN\,2020xva & ZTF20aclvtnk & 263.035128 & 53.653989 & 0.0240 & 108.7 & 2459141.7258 & $2459142.69\pm0.69$ & $<1.0$ & [3] \\
SN\,2020aavm & ZTF20acrinvz & 116.681975 & 18.113551 & 0.0450 & 227.2 & 2459168.9788 & $2459169.935\pm0.746$ & $<1.0$ & [11] \\
SN\,2020abue & ZTF20acvjlev & 121.084598 & 56.302082 & 0.0280 & 126.9 & 2459188.0078 & $2459189.663\pm0.118$ & $<0.2$ & [11] \\
SN\,2020acbm & ZTF20acwgxhk & 40.074159 & 2.427067 & 0.0217 & 93.1 & 2459192.7093 & $2459193.654\pm0.022$ & $<0.1$ & [11] \\
SN\,2021apg & ZTF21aafkwtk & 205.330192 & 24.495531 & 0.0269 & 128.4 & 2459228.0064 & $2459230.722\pm0.186$ & $<1.2$ & [11] \\
SN\,2021ibn & ZTF21aasfseg & 132.558710 & 37.026990 & 0.0442 & 197.2 & 2459306.7862 & $2459307.252\pm0.295$ & $<0.4$ & [11] \\
SN\,2021skn & ZTF21abjcjmc & 246.204167 & 39.734653 & 0.0297 & 139.5 & 2459397.8203 & $2459398.735\pm0.743$ & $<1.1$ & [11] \\
SN\,2021yja & ZTF21acaqdee & 51.088215 & -21.565626 & 0.0053 & 22.6 & 2459459.4000 & $2459464.4\pm0.06$ & $<2.3$ & [10] \\
\hline
\enddata
\tablenotetext{a}{Corrected for Virgo, Great Attractor, and Shapley supercluster infall.}
\tablenotetext{b}{In rest-frame days, calculated using the last spectrum showing flash features or by taking the first spectrum as an upper limit.}
\tablenotetext{c}{[1] \cite{Bruch2021}, [2] \cite{Soumagnac2020}, [3] \cite{bruch2022}, [4] \cite{Vallely2021}, [5] \cite{Yang2021}, [6] \cite{Valerin2022}, [7] \cite{Tinyanont2022}, [8] \cite{Sollerman2021}, [9] \cite{Terreran2022}}, [10] \cite{Hosseinzadeh2022}, [11] TNS classification reports: \cite{Perley_tns_19,Hiramatsu2020,Perley_tns_2020,Dahiwale2020c,Weil2020_tns,Dahiwale2020d,Pessi2020_tns,Delgado2021_tns,Deckers2021_tns,Siebert_2021_tns}.
\tablenotetext{d}{This table is available in machine-readable format.}

\end{deluxetable*} 

\subsection{Distance}
\begin{figure}[t]
\centering
\includegraphics[width=\columnwidth]{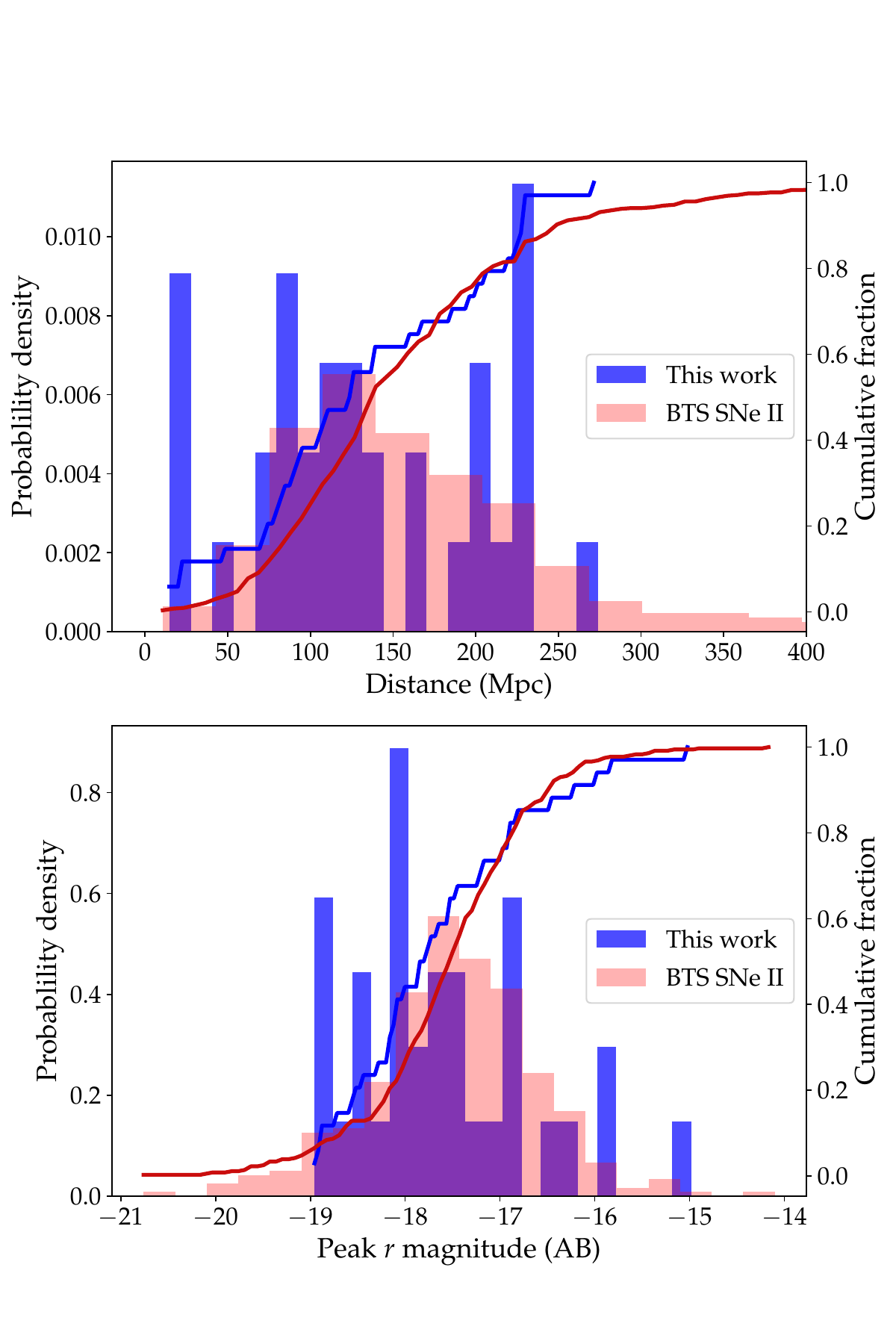} \\
\caption{In the top panel, we show the distribution of distances to the SNe in our sample, compared to the distribution of BTS SNe II. We truncate the plot at 400 Mpc for clarity. In the bottom panel, we show the distribution of peak $r$-band magnitude compared to BTS SNe II. In both panels, we show histograms and the cumulative distributions.}
\label{fig:distance_peak}
\end{figure}

We adopt Hubble-flow distances using the NASA Extragalactic Database (NED)\footnote{\hyperlink{https://ned.ipac.caltech.edu/}{https://ned.ipac.caltech.edu/}} and using their online calculator to correct the redshift-distance for Virgo, Great Attractor, and Shapley supercluster infall \citep[based on the work of ][]{Mould2000}. The top panel of Fig.~\ref{fig:distance_peak} shows the distribution of distances in our sample compared to that of a magnitude-limited and spectroscopically complete sample from ZTF Bright Transient Survey \citep{fremling2020,perley2020}.\footnote{\href{http://sites.astro.caltech.edu/ztf/rcf/explorer.php}{http://sites.astro.caltech.edu/ztf/rcf/explorer.php}}

\subsection{Extinction}
We correct for foreground Galactic reddening using the \cite{schlafly2011} recalibration of the \cite{schlegel1998} extinction maps, and assuming a \cite{Cardelli1989} Milky Way extinction law with $R_{V}=3.1$. These corrections are applied to all photometry data appearing in this paper. We do not correct the photometry for host-galaxy extinction, and treat this effect separately in $\S$~\ref{subsec:SC_fitting}. 

\subsection{Time of zero flux}
\begin{figure}[t]
\centering
\includegraphics[width=\columnwidth]{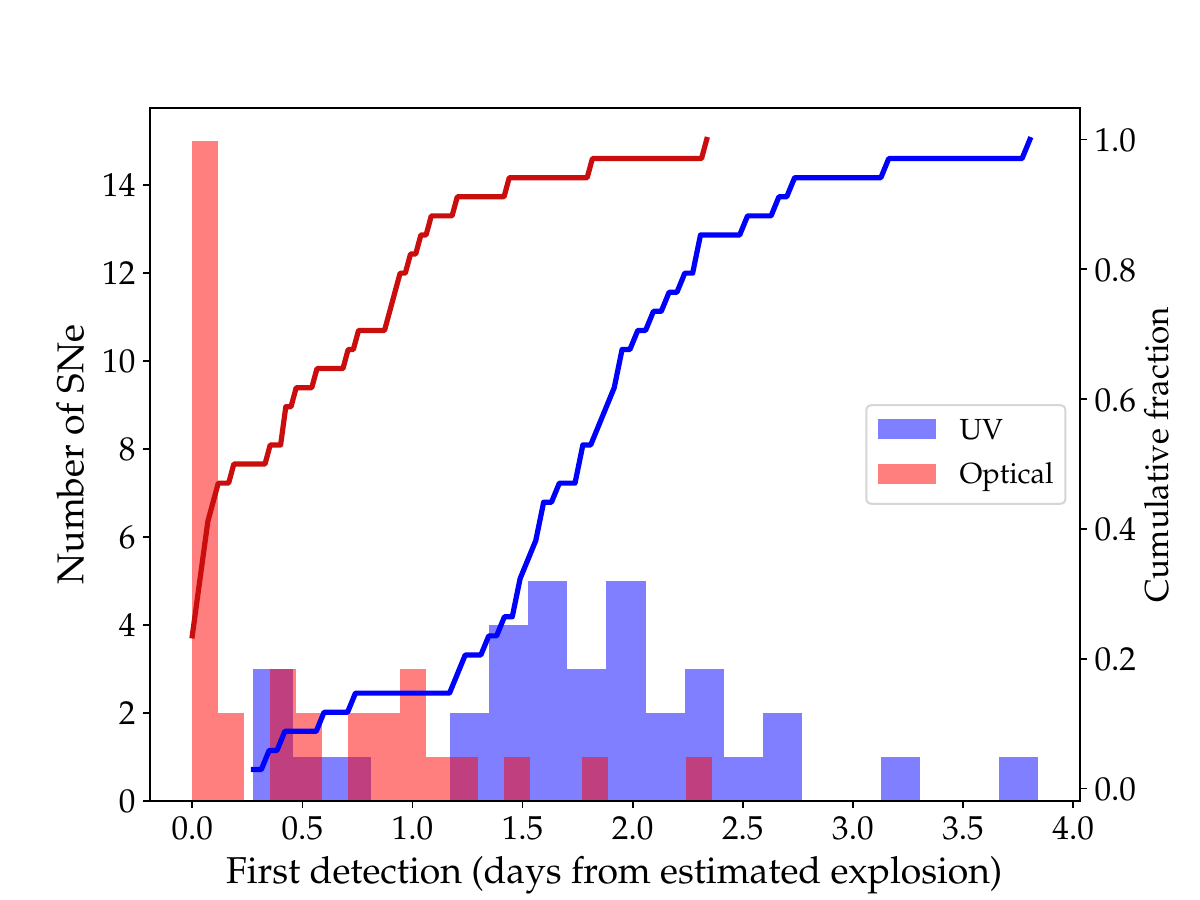} \\
\caption{The times of first detection relative to the estimated time of zero flux, in UV and in optical bands. Both a histogram and a cumulative distribution is shown.}
\label{fig:first_time}
\end{figure}

We acquire an initial estimate of the time of zero flux $t_{0}$ using a power-law extrapolation of the forced-photometry flux to 0. Using both $g$-band and $r$-band data, we fit a function $f_{\lambda}=f_{0}(t-t_{0})^{n}$ with a slope of $0<n<5$, and allow values of $t_{0}$ between the first detection of the SN and the last non-detection. We then estimate the error on $t_{0}$ as the scatter in $t_{0,best}$ over all allowed values of $n$, and choose to use the band with the best constraint on $t_{0}$. In Fig.~\ref{fig:first_time}, we show the distribution of detection times \referee{in both UV and optical bands} relative to the estimated time of zero flux \referee{computed from optical data}. We find a large fraction of the SNe have $t_{0}$ close to their first detections. Most of these are SNe where first detection in the forced photometry light-curve are recovered from a non-detection in the \referee{automated ZTF} alert photometry - resulting in a sharp rise \referee{and $t_0$ estimates that are very close to the time of first detection}. As the SN time of zero flux should not correlate with the time of first detection, we expect a uniform distribution in $t_{0}$ \referee{and in $t_{first}$}. $t_{first} - t_{0}$ should then be a rising and falling distribution.  The fact that our results deviate from such a distribution indicates a systematic deviation from a power-law rise in flux - a model which is not physically motivated. \cite{Hosseinzadeh2023} fit the early light curve of the recently discovered Type II SN\,2023ixf \citep{Itagaki2023}, and show that the rise is comprised from 2 phases - a slower phase followed by a sharply rising phase. For such a light curve, extrapolating based on the sharply rising phase would result in a time of first light too late by several hours, and the first point on the rise would be close to the fit $t_{0}$. Our fit provides preliminary evidence this is the case for the majority of SNe II. 

\subsection{Flash feature timescale}
\cite{Bruch2021} define flash-features based on the presence of the $\lambda4686$ \ion{He}{2} feature before broad H recombination features appear. The flash feature duration $\tau_{flash}$ is defined through the half-time between the last spectrum showing  $\lambda4686$ \ion{He}{2} emission and the subsequent epoch \citep{bruch2022}. We adopt these definitions and the measurements of \cite{bruch2022} throughout our paper. We extend the estimation to the SNe not included in \cite{bruch2022} using all available spectroscopy, which will be released in a future publication. \\

In Table~\ref{tab:sn_list} we list the 34 SNe in our sample, as well as their median alert coordinates, redshifts, distance estimates, non-detection limits, estimated time of zero flux and their flash feature timescales, if applicable. 

\subsection{RSG radiation-hydrodynamic simulations}

\label{subsec:MG_sims}
When comparing data to semi-analytic models, which are calibrated to numerical simulations, it is unclear how the calibration scatter and theoretical uncertainties will propagate to observed fluxes. These could potentially manifest as correlated residuals when the model is compared to the data, and subsequently create biases in the fit parameters. In order to demonstrate and account for such effects in our analysis, we repeat some of the analysis we perform throughout the paper to a set of 28 multi-group radiation-hydrodynamical simulations of RSG described in detail in \refereetwo{M24}. These simulations are generated by relaxing the assumption of local thermal equilibrium (LTE) and instead solving the radiation transfer using multiple photon groups and a realistic opacity table with free-free, bound-free and bound-bound opacities at different densities, temperatures, and compositions. \referee{Thus, these simulations account for the effects of line blanketing and line emission.} The simulations allow us to generate synthetic data sets with arbitrary sampling in time with any set of filters. Unless mentioned otherwise, we use the sampling, filters, and error-bars of the light curves of SN\,2020uim, arbitrarily chosen from our sample as a representative SN. We do not add simulated noise, and all points are assumed to be detected regardless of luminosity unless otherwise mentioned.

\section{Observations}
\label{sec:observations}
\subsection{Optical photometry}
\label{subsec:opt-photometry}
ZTF photometry in the \textit{gri} bands were acquired using the ZTF camera \citep{Dekany2020} mounted on the 48\,inch (1.2\,m) Samuel Oschin Telescope at Palomar Observatory (P48). These data were processed using the ZTF Science Data System \citep[ZSDS;][]{Masci2019}. 

\referee{While scanning was preformed using the automated alert photometry pipeline, the} light curves \referee{reported in this work} were obtained using the ZTF forced-photometry service.\footnote{See ztf\_forced\_photometry.pdf under \url{https://irsa.ipac.caltech.edu/data/ZTF/docs}} The forced photometry is performed on difference images produced using the optimal image subtraction algorithm of Zackay, Ofek and Gal-Yam \citep[ZOGY;][]{Zackay2016} at the position of the SN, calculated from the median ZTF alert locations which are listed in Table \ref{tab:sn_list}. We removed images that have flagged difference images (with problem in the subtraction process), bad pixels close to the SN position, a large standard deviation in the background region, or a seeing of more than 4\arcsec. We performed a baseline correction to ensure the mean of the pre-SN flux is zero. We report detections above a $3\sigma$ threshold, and use a $5\sigma$ threshold for upper limits. 

In addition to the ZTF photometry, we also used the following instruments to collect early multi-band light-curves:
\begin{itemize} 

\item 

The Optical Imager (IO:O) at the 2.0\,m robotic Liverpool Telescope  \citep[LT; ][]{Steele2004} at the Observatorio del Roque de los Muchachos. We used the Sloan Digital Sky Survey \citep[SDSS;][]{york2000} $u$, $g$, $r$, $i$ and $z$ filters. Reduced images were downloaded from the LT archive and processed with custom image-subtraction and analysis software (K. Hinds and K. Taggart et al., in prep.)  Image stacking and alignment is performed using \package{SWarp} \citep{Bertin2010} where required.  Image subtraction is performed using a pre-explosion reference image in the appropriate filter from the Panoramic Survey Telescope and Rapid Response System 1 \citep[Pan-STARRS1;][]{Chambers2016} or SDSS.  The photometry are measured using PSF fitting methodology relative to Pan-STARRS1 or SDSS standards and is based on techniques in \cite{fremling2016}. For SDSS fields without u-band coverage, we returned to these fields after the SN had faded on photometric nights to create deep stacked u-band reference imaging. We then calibrated these field using IO:O standards taken on the same night at varying airmasses and used these observations to calibrate the photometry \citep{Smith2002}.

\item The Rainbow Camera \citep{Blagorodnova2018a} on the Palomar 60\,inch (1.5\,m) telescope (P60; \citealt{Cenko2006}). Reductions were performed using the automatic pipeline described by \cite{fremling2016}. 
\end{itemize}

In addition to the above, we use early optical light curves from the literature. These include the multi-band light curves covering the rise of SN\,2021yja \citep{Hosseinzadeh2022} and light curves from the TESS  for SN\,2020fqv \citep{Tinyanont2022} and SN\,2020nvm \cite{Vallely2021}. 

\subsection{UV photometry}
\begin{figure*}[t]
\centering
\includegraphics[width=\textwidth]{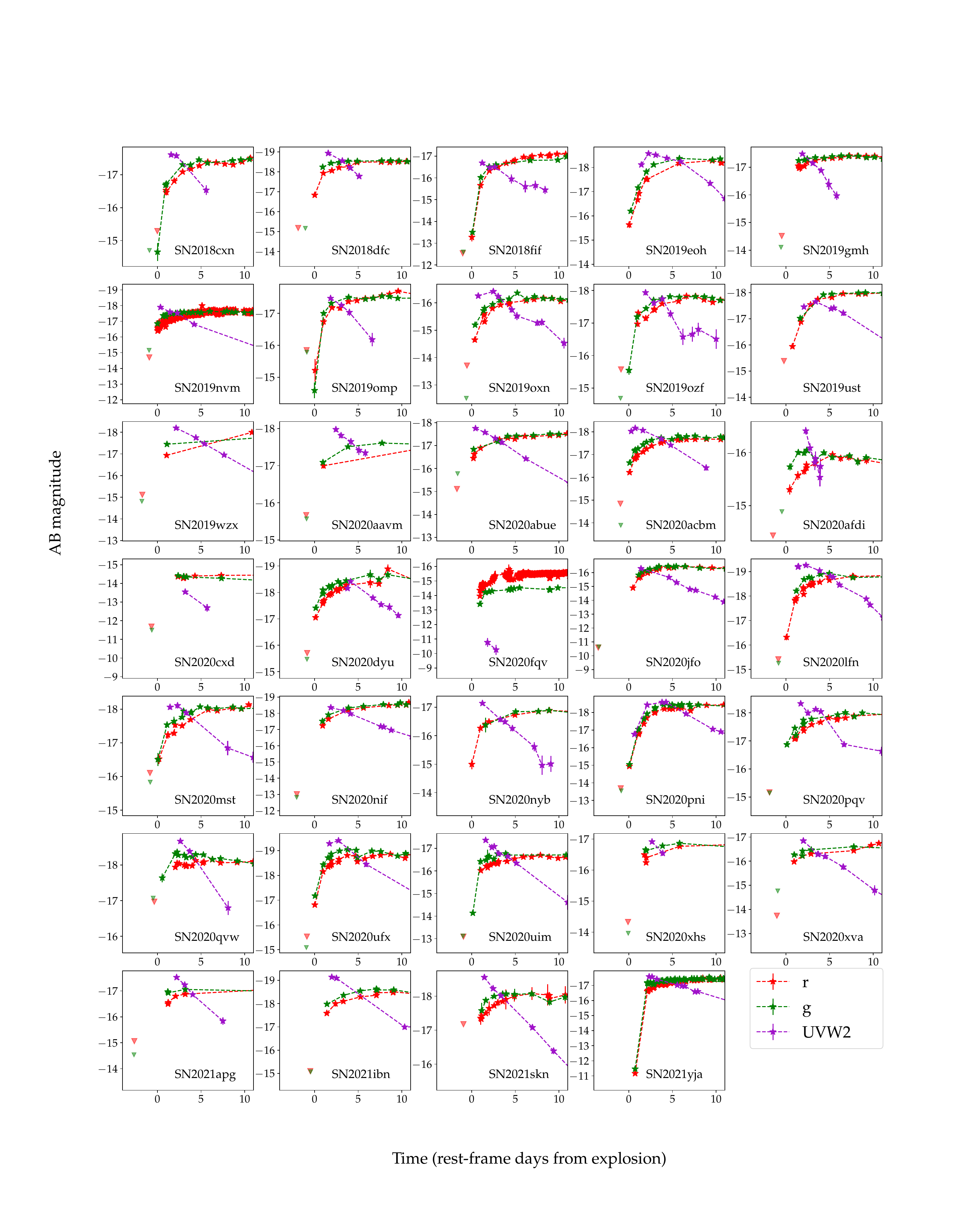} \\
\caption{UVW2 (magenta stars), $g$ and $r$ (green and red stars) light curves for all of the objects in our samples. The latest upper limits before discovery are marked with a downward facing triangle. We note that 
some points which are marked as limits in the alert photometry became detections using forced photometry.}
\label{fig:lc_grid}
\end{figure*}

\begin{figure}[t]
\centering
\includegraphics[width=\columnwidth]{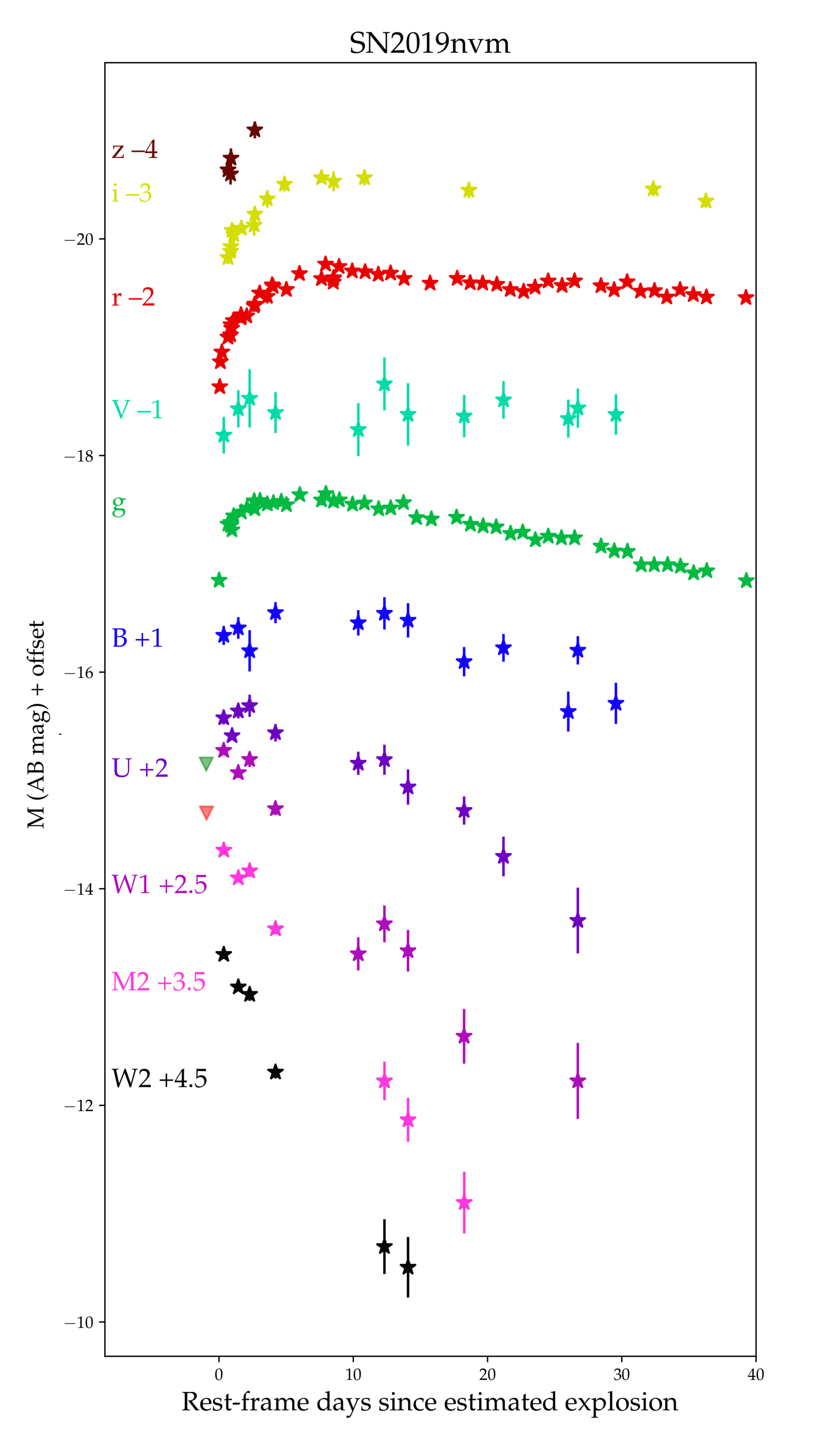} \\
\caption{A representative example of the multi-band light curves of SN\,2019nvm in the first 40 days. The complete figure set (34 images) is available in the online journal. The photometry is also available as the data behind the figure.}
\label{fig:example_lc}
\end{figure}

UV photometry were acquired for all SNe using UVOT onboard the \textit{Neil Gehrels Swift Observatory} \citep{Gehrels2004,Roming2005}. We reduced the images using the \swift\ \package{HEAsoft}\footnote{\url{https://heasarc.gsfc.nasa.gov/docs/software/heasoft/}
v. 6.26.1.} toolset. Individual exposures comprising a single \referee{visit} were summed using  \package{uvotimsum}. Source counts were then extracted using \package{uvotsource} from the summed images using
a circular aperture with a radius of 5{\arcsec}. The background was estimated from several larger regions surrounding the host galaxy. These counts were then converted to fluxes using the photometric zero points of \cite{Breeveld2011} with the latest calibration files from September 2020, and including a small scale sensitivity correction with the latest map of reduced sensitivity regions on the sensor from March 2022. A UV template image was acquired for all SNe and for all bands after the SN had faded, with an exposure time twice as long as used for the deepest image of the SN. These images were then summed with any archival images of the site and used to estimate the host flux at the SN site. We remove the local host-galaxy contribution by subtracting the SN site flux from the fluxes of the individual epochs. In Fig.~\ref{fig:lc_grid} we show the early $g,r$ and $UVW2$ light curves of the SNe in our sample. In Fig.~\ref{fig:example_lc} we show a representative example of the multi-band light curves in our sample. We make the multi-band light curve figures of individual SNe available through the journal website and WISeREP. Finally, we show the full ZTF forced photometry light curves in Fig.~\ref{fig:full_lc}.

\figsetstart
\figsetnum{4}
\figsettitle{Multi-band light curves}

\figsetend

\subsection{X-ray observations}
\label{subsec:xray_obs}
While the SNe were monitored with UVOT, \swift\ also observed the field between 0.3 and 10\,keV with its onboard X-ray telescope (XRT) in photon-counting mode \citep{Burrows2005a}. We analyzed these data with the online tools provided by the UK \swift\ team.\footnote{\href{https://www.swift.ac.uk/user_objects/}{https://www.swift.ac.uk/user\_objects}} These online tools use the methods of \citet{Evans2007a, Evans2009a} and the software package \package{HEASoft} v. 6.29 to generate XRT light curves and upper limits, perform PSF fitting, and provide stacked images.

In most cases, the SNe evaded detection at all epochs. We derive upper limits by calculating the median $3\sigma$ count-rate limit of each observing block in the 0.3--10\,keV band, determined from the local background. We stack all data \referee{(acquired during UV observations of the SNe at early times and when creating the UV templates)}, and converting the count-rates to unabsorbed flux by assuming a power-law spectrum with a photon index of 2, and taking into account the Galactic neutral hydrogen column density at the location of the SN \citep{HI4PI2016a}. 

In several cases (SN\,2020jfo, SN\,2020nif and SN\,2020fqv) we find spurious detections which are likely associated with a nearby constant source, identified by inspecting co-added X-ray images over all epochs, and by comparing to archival survey data through the HILIGT server \citep{Saxton2022}. We treat the measured flux as upper limits on the SN flux. 

\begin{figure}[t]
\centering
\includegraphics[width=\columnwidth]{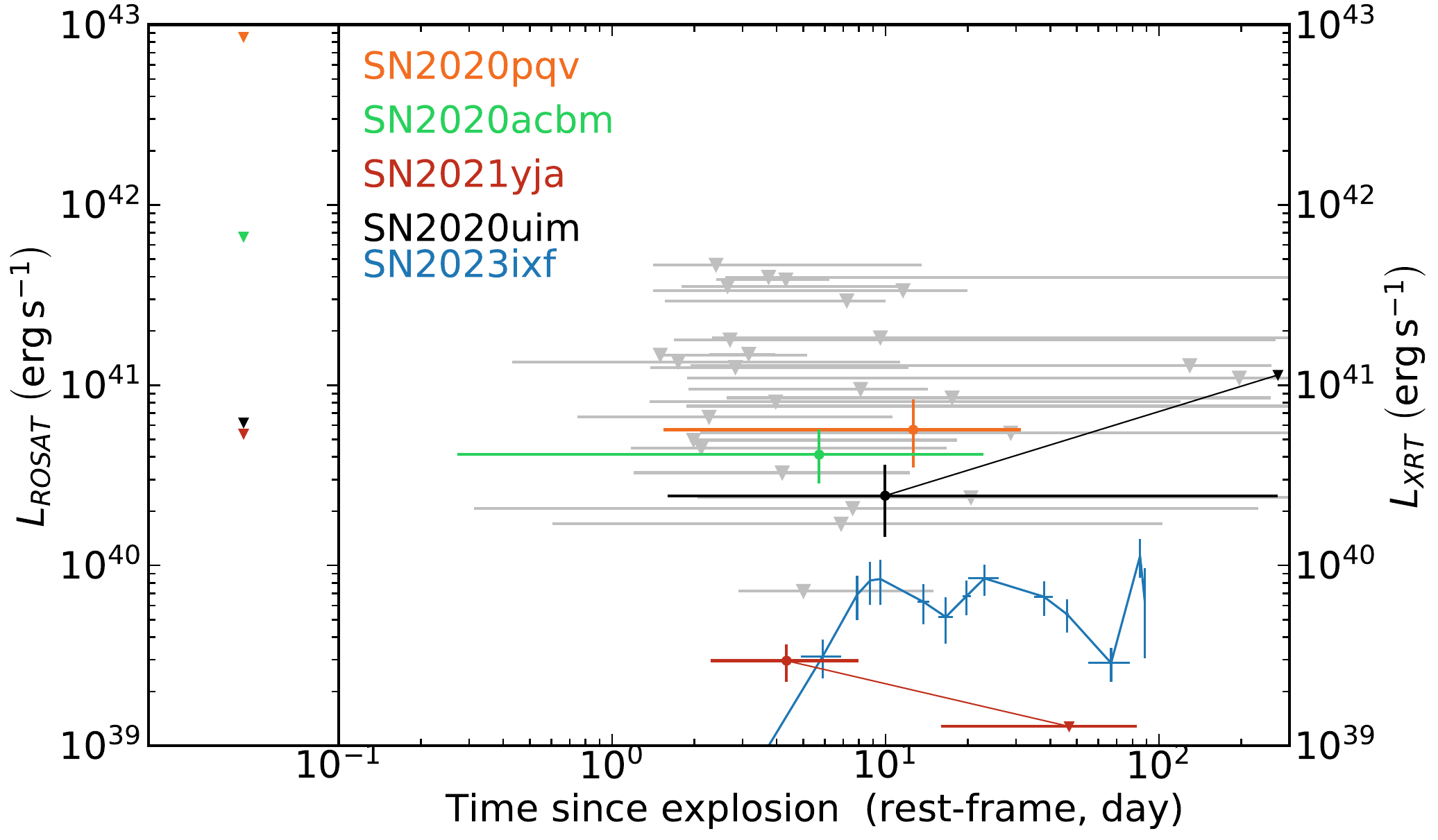} \\
\caption{The right panel shows the XRT binned detections and upper limits for the SNe in our sample. Measurements were binned over the duration of the \swift\ observations, and the time of detections and upper limits is set to the mean photon arrival time. The left panel shows upper limits on the emission for the SN location for the 4 XRT detections from archival ROAST survey data. We also show the XRT light curve of the nearby Type II SN\,2023ixf \citep{Zimmerman2023}. }
\label{fig:XRT_lc}
\end{figure}

For SN\,2020acbm and SN\,2020uim, we report $>3\sigma$ X-ray detections from the binned exposures.\footnote{We note that while the detection significance $S/\sqrt(B)>3$, where $S$ is the source flux and $B$ is the background level, taking into account the source flux in the error calculation results in a $< 3 \sigma$ measurement error, since the measurement signal-to-noise is $S/\sqrt(B+S)$. These approximations for the signal to noise hold in the Gaussian limit, which is approximately correct in our case.} For both SNe, the SN location is within 90\% error region of the source PSF. In the case of SN\,2020pqv we report a detection 11" from the SN where the source 90\% localization region is $8\farcs5$. We lack constraining limits on the quiescent flux at the location of all three SNe when comparing to archival ROSAT data or compared to the late-time XRT exposures. For SN\,2021yja, we report a source 2\farcs6 from the SN site from observations in the first 10 days, brighter by a factor $4.2\pm1.8$ than the derived $3\sigma$ upper limit from observation in subsequent epochs - robustly indicating the emission is related to the SN. We report our measurements in Table \ref{tab:XRT}, and show our results in Fig. \ref{fig:XRT_lc}. 

\begin{deluxetable*}{ccccccc}
\centering
\label{tab:XRT}
\tablecaption{XRT photometry for SNe included in this study}
\tablewidth{34pt} 
\tablehead{\colhead{SN} &\colhead{$t$ [day]} & \colhead{ $t_{\rm max}$ [day]} & \colhead{$t_{\rm min}$ [day]} & \colhead{XRT count rate [s$^{-1}$]} & \colhead{Flux [$10^{-14}$ erg s$^{-1}$ cm$^{-2}$]} & \colhead{Luminosity [$10^{40}$ erg s$^{-1}$]}} 
\tabletypesize{\scriptsize} 
\startdata
SN\,2018cxn & 7.2 & 10.0 & 1.6 & $<0.002$ & $<7.1$ & $<28.3$ \\
SN\,2018dfc & 1.5 & 5.2 & 1.5 & $<0.0011$ & $<4.3$ & $<14.08$ \\
SN\,2018fif & 2.1 & 16.8 & 1.2 & $<0.0015$ & $<6.4$ & $<4.56$ \\
SN\,2019eoh & 11.6 & 20.0 & 1.4 & $<0.0015$ & $<5.3$ & $<33.68$ \\
SN\,2019gmh & 479.9 & 480.4 & 1.9 & $<0.0009$ & $<3.2$ & $<7.41$ \\
SN\,2019nvm & 7.6 & 230.2 & 0.3 & $<0.0006$ & $<2.3$ & $<1.85$ \\
SN\,2019omp & 2.6 & 11.2 & 1.8 & $<0.0019$ & $<6.9$ & $<35.07$ \\
SN\,2019oxn & 2.3 & 10.6 & 0.7 & $<0.0018$ & $<6.8$ & $<6.6$ \\
SN\,2019ozf & 196.6 & 391.3 & 1.9 & $<0.0005$ & $<1.9$ & $<10.9$ \\
SN\,2019ust & 20.5 & 325.5 & 2.1 & $<0.0005$ & $<2.2$ & $<2.55$ \\
SN\,2019wzx & 28.7 & 692.5 & 2.1 & $<0.0008$ & $<2.9$ & $<5.39$ \\
SN\,2020aavm & 4.3 & 6.2 & 2.4 & $<0.0016$ & $<6.2$ & $<31.65$ \\
SN\,2020abue & 1.7 & 11.3 & 0.4 & $<0.0018$ & $<7.0$ & $<13.45$ \\
SN\,2020acbm & 5.7 & 22.8 & 0.3 & $0.0011 \pm 0.0004$ & $4.0 \pm 1.5$ & $4.56 \pm 1.71$ \\
SN\,2020afdi & 3.2 & 4.0 & 2.3 & $<0.0027$ & $<10.1$ & $<14.05$ \\
SN\,2020cxd & 5.0 & 14.9 & 2.9 & $<0.0028$ & $<10.7$ & $<0.38$ \\
SN\,2020dyu & 9.6 & 476.7 & 2.3 & $<0.0008$ & $<2.9$ & $<18.22$ \\
SN\,2020fqv & 1.7 & 59.0 & 0.0 & $<0.0072$ & $<43.1$ & $<5.79$ \\
SN\,2020jfo & 1.4 & 84.5 & 0.0 & $<0.0017$ & $<6.3$ & $<0.41$ \\
SN\,2020lfn & 4.0 & 119.5 & 1.4 & $<0.0004$ & $<1.7$ & $<8.03$ \\
SN\,2020mst & 2.4 & 13.5 & 1.4 & $<0.0013$ & $<5.2$ & $<46.13$ \\
SN\,2020nif & 3.3 & 16.8 & 0.0 & $<0.006$ & $<22.9$ & $<5.87$ \\
SN\,2020nyb & 4.2 & 12.3 & 1.2 & $<0.0012$ & $<5.3$ & $<3.05$ \\
SN\,2020pni & 6.9 & 103.1 & 0.6 & $<0.0006$ & $<2.1$ & $<1.41$ \\
SN\,2020pqv & 12.6 & 31.3 & 1.5 & $0.0005 \pm 0.0002$ & $1.8 \pm 0.9$ & $5.2 \pm 2.42$ \\
SN\,2020qvw & 3.7 & 484.5 & 2.6 & $<0.0016$ & $<6.2$ & $<39.45$ \\
SN\,2020ufx & 2.7 & 267.2 & 1.7 & $<0.0007$ & $<2.8$ & $<17.74$ \\
SN\,2020uim & 272.2 & 272.5 & 272.0 & $<0.0036$ & $<14.7$ & $<12.14$ \\
SN\,2020uim & 10.0 & 271.8 & 1.6 & $0.0008 \pm 0.0004$ & $3.1 \pm 1.5$ & $2.6 \pm 1.26$ \\
SN\,2020xhs & 17.5 & 256.7 & 2.6 & $<0.0014$ & $<6.3$ & $<9.06$ \\
SN\,2020xva & 2.0 & 18.3 & 1.9 & $<0.0009$ & $<3.5$ & $<4.92$ \\
SN\,2021apg & 8.1 & 14.3 & 1.9 & $<0.0014$ & $<4.8$ & $<8.53$ \\
SN\,2021ibn & 129.6 & 257.3 & 1.9 & $<0.0008$ & $<2.8$ & $<13.54$ \\
SN\,2021skn & 2.8 & 12.1 & 1.4 & $<0.0015$ & $<5.4$ & $<11.68$ \\
SN\,2021yja & 4.3 & 8.0 & 2.3 & $0.0013 \pm 0.0003$ & $4.8 \pm 1.2$ & $0.32 \pm 0.08$ \\
SN\,2021yja & 46.9 & 83.2 & 15.9 & $<0.0006$ & $<2.1$ & $<0.14$ \\
\hline
\enddata
\tablenotetext{a}{All times are reported in rest-frame days}
\tablenotetext{b}{We report 3$\sigma$ upper limits, or measurements with a significance of 3$\sigma$ above the background level. }
\tablenotetext{c}{Fluxes are corrected for galactic neutral hydrogen column density, and converted from count-rates assuming a power-law spectrum with a photon index of 2.}
\tablenotetext{d}{For SN\,2020jfo, SN\,2020fqv and SN\,2020nif we report quiescent host-galaxy detections as upper limits on the SN flux.}
\end{deluxetable*} 

\section{Results}
\label{sec:results}

\subsection{Color evolution}
\label{subsec:color}
Before recombination begins, and although the external layers of the SN ejecta are not in LTE, the spectrum of a SN II is expected to be well approximated by a blackbody \citep{Baron_2000,Blinnikov_2000,Nakar2010,Rabinak2011,Morag2024}. However, several reasons exists to expect deviations of the spectrum from a perfect blackbody:
\begin{itemize}
    \item Extinction can contribute significantly to deviations from blackbody. While the exact applicable extinction law has a modest effect on the optical colors, it can create major differences in the UV and UV-optical colors. Large $R_{V}$ values will cause bluer UV-optical colors compared to an $R_{V}=3.1$ MW extinction law. Many star-forming galaxies lack the characteristic ``bump" at 220 nm, which will mostly affect the UVM2-band photometry \citep{Calzetti2000a, Salim2020}. For both SNe Ia and stripped-envelope SNe, sample color-curves have been used to derive a ``blue edge" where the amount of extinction is assumed to be zero 
    \citep{Phillips1999,stritzinger2018b}. This in turn \referee{has been} used to estimate the host-galaxy extinction in the line of sight to the SN, typically performed at phases for which the intrinsic scatter in color is minimal. 
    \item While a frequency-independent opacity is expected to yield a blackbody continuum, a frequency-dependent opacity will create deviations. These will manifest as emission and absorption features - particularly line blanketing in the UV, as well as broad deviations from blackbody in the continuum. \referee{These effects strongly depend on the temperature of the ejecta}. \refereetwo{M24} characterize these deviations using multi-group radiation hydrodynamical simulations \referee{(including temperature and density dependent line-opacity)} and these are included in their latest analytical model. \textcolor{blue}{...simulations including line opacity, and confirmed against a separate high frequency resolution ($\Delta\lambda/\lambda\sim 10^{-5}$) calculation that incorporates Doppler expansion opacity. These effects are included in their latest analytical model.}
    Line blanketing in the UV is observed in the few early time UV spectra of SNe II \citep[][]{Brown2007,Vasylyev2022,Vasylyev2023a,Bostroem2023a, Zimmerman2023}. \referee{Recently, \cite{Zimmerman2023} confirmed the presence of emission lines from highly ionized species in the UV, as well as photospheric absorption features that appear in the UV while the optical spectrum is still a smooth continuum around $T\sim15,000$K}.  
    \item  CSM interaction is suggested to create bluer UV-optical colors, to be associated with a higher luminosity, and with spectral signatures indicating the presence of CSM  \citep{Ofek2010,Katz2011,Chevalier2011, Hillier2019}. CSM interaction is typically accompanied by strong line emission \citep{Yaron2017}, possibly in the UV, which can create deviations from blackbody. 
\end{itemize}

\begin{figure*}[t]
\centering
\includegraphics[width=\textwidth]{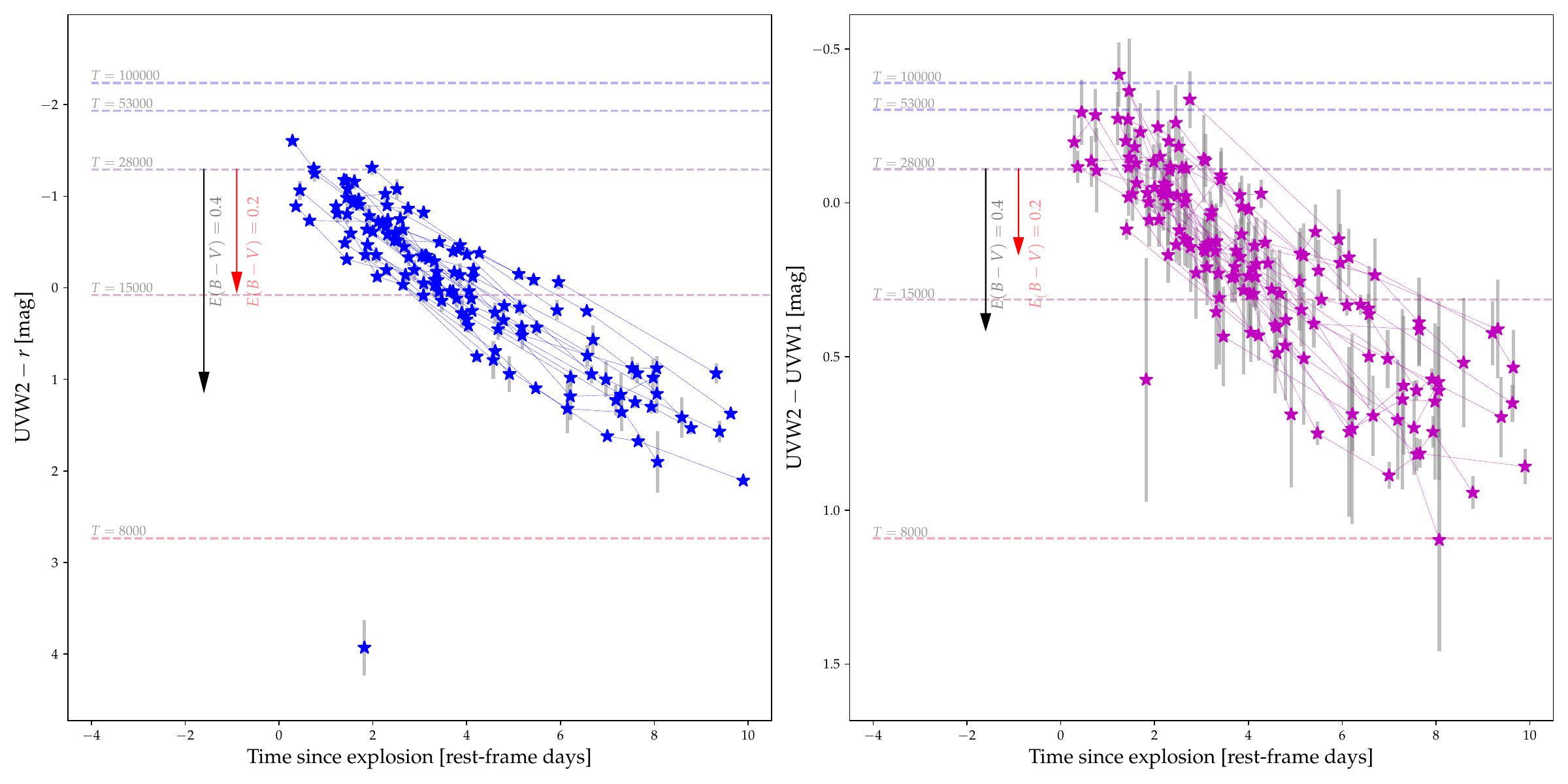} \\
\caption{The color evolution of SNe II in our sample in the UVW2 -- $r$ bands (left panel) and UVW2 -- UVW1 (right panel) bands. Each curve represents a single SN. The dashed lines are the colors of blackbodies at various temperatures. The arrows show the color difference due to extinction with $\rm{E(B-V)}=0.2\ \rm{mag}$ (red arrow) and with $\rm{E(B-V)}=0.4\ \rm{mag}$ (black arrow), assuming a Milky Way extinction curve with $R_{\rm V} = 3.1$. The outlier in the left plot is the highly extinguished SN\,2020fqv.}
\label{fig:color_evo1}
\end{figure*}

\begin{figure*}[t]
\centering
\includegraphics[width=\textwidth]{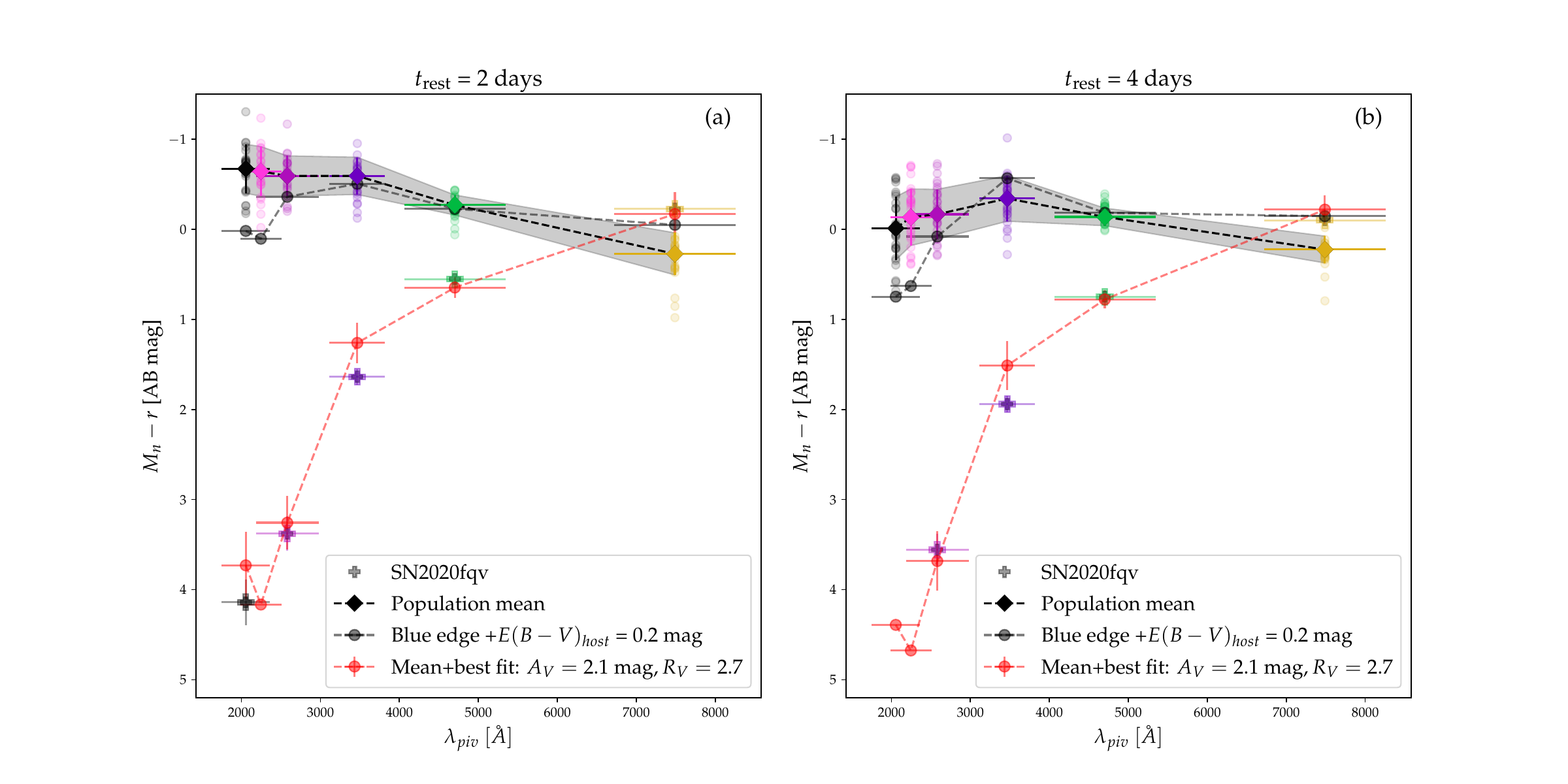} \\
\caption{The mean colors of a sample of SNe II at (a) $t=2$ d and (b) $t=4$ d. The solid points and gray shaded regions show the mean color and the scatter of each color. The transparent points are individual SN colors. Both the mean and individual SN colors are color-coded by wavelength. The gray points demonstrate the effect of applying $E(B-V)=0.2$ mag with an $R_{V}=3.1$ \cite{Cardelli1989} extinction law to the bluest colors, demonstrating the extinction in our sample is smaller than this value. The colored plus are the colors of the highly reddened SN\,2020fqv. The red curve shows the effect of reddening the mean colors using the best fit extinction curve, which reproduces the colors of SN\,2020fqv to within the errorbars for all wavelengths. }
\label{fig:sed_II}
\end{figure*}

Using our well sampled light curves, we constrain the deviations from a blackbody spectral energy distribution (SED) in our sample, as well as attempt to isolate their main source (i.e., physical or extinction).

First, we consider the effect of extinction. In Fig.~\ref{fig:color_evo1}, we show the $UVW2-r$ and $UVW2-UVW1$ color curves for our sample. On both plots, we illustrate the effect of applying galactic extinction with $E(B-V)$ of $0.2,0.4$ mag with red and black arrows respectively. We show dashed lines showing the expected colors of a blackbody with various temperatures in the background. The scatter in the color curves represents the variance in temperature and in extinction. A significant variance in temperature (and thus in color) is expected if these SNe are powered by shock-cooling, as the temperature evolution is sensitive to the shock-breakout radius. Despite this, all SNe in our sample besides the highly extinguished SN\,2020fqv \citep{Tinyanont2022} fall within $E(B-V)=0.2$ mag of the bluest SN in the sample. We consider this value an upper limit on the reddening affecting these SNe.\footnote{Our sample does not include other extinguished SNe since we require a blue color to trigger UVOT. In the case of SN\,2020fqv, UVOT was triggered by another group, and thus had early UV and is included in this study.}

In Fig.~\ref{fig:sed_II}, we show the $M_{n}-M_{r}$ color distributions in our sample at $t=2$ and $t=4$ days (panels (a) and (b), respectively), where $n\in \{UVW2,UVM2,UVW1,U,g,i\}$. For each band, the transparent data points show the interpolated color, the solid diamonds and black dashed lines show the average color, and the error bars and gray shaded regions show the standard deviation of the color. A extinction corresponding to a galactic extinction curve with $E(B-V)=0.2$ mag applied to the bluest \referee{SN in the $UVW2-r$} color (transparent points with highest $M_{n}-M_{r}$) is indicated by the gray transparent data points. For the \referee{of the $UVW2-r$ and $UVM2-r$} colors, which are most sensitive to extinction, this mild amount of extinction is sufficient to account for the full scatter in all SNe besides SN\,2020fqv. \referee{However, since $E(B-V)=0.2$ mag is not enough to account for the optical scatter (as indicated by the trend of the gray line), it is likely that this scatter is explained by differences in temperature and that the typical extinction of the sample is lower.} Assuming that SN\,2020fqv is well represented by our sample in its intrinsic SED, we use the average colors to calculate its extinction curve. In each curve, we determine $E(n-r)$ from the color difference at $t=2$ days, and fit a \cite{Cardelli1989} extinction curve with free $R_{V}$ and $A_{V}$. In Fig. \ref{fig:sed_II} we show both the colors of SN\,2020fqv (solid plus) and the best fit  extinction curve applied to the average SED (red points), which match well at both times. Here and in the rest of the paper, we assign wavelengths to filters using the pivot wavelength for a flat spectrum $\lambda_{piv} = \sqrt{\frac{\int T\left(\lambda\right)\lambda d\lambda}{\int T\left(\lambda\right)\frac{d\lambda}{\lambda}}}$, where $T$ is the filter transmission curve, downloaded from the Spanish Virtual Observatory \citep[SVO;][]{Rodrigo2012,Rodrigo2020}.\footnote{\referee{We estimate that $\lambda_{piv}$ of a blackbody with $10,000>T>30,000$ (relevant to this study) will be within 5\% of the $\lambda_{piv}$ assuming a flat spectrum for all filters used in our study. While the flux conversion factors depend on the spectral shape \citep[e.g.][]{brown2016}, we estimate this effect to be less than 10\% for a blackbody within this temperature range. }} 

\begin{figure}[t]
\centering
\includegraphics[width=\columnwidth]{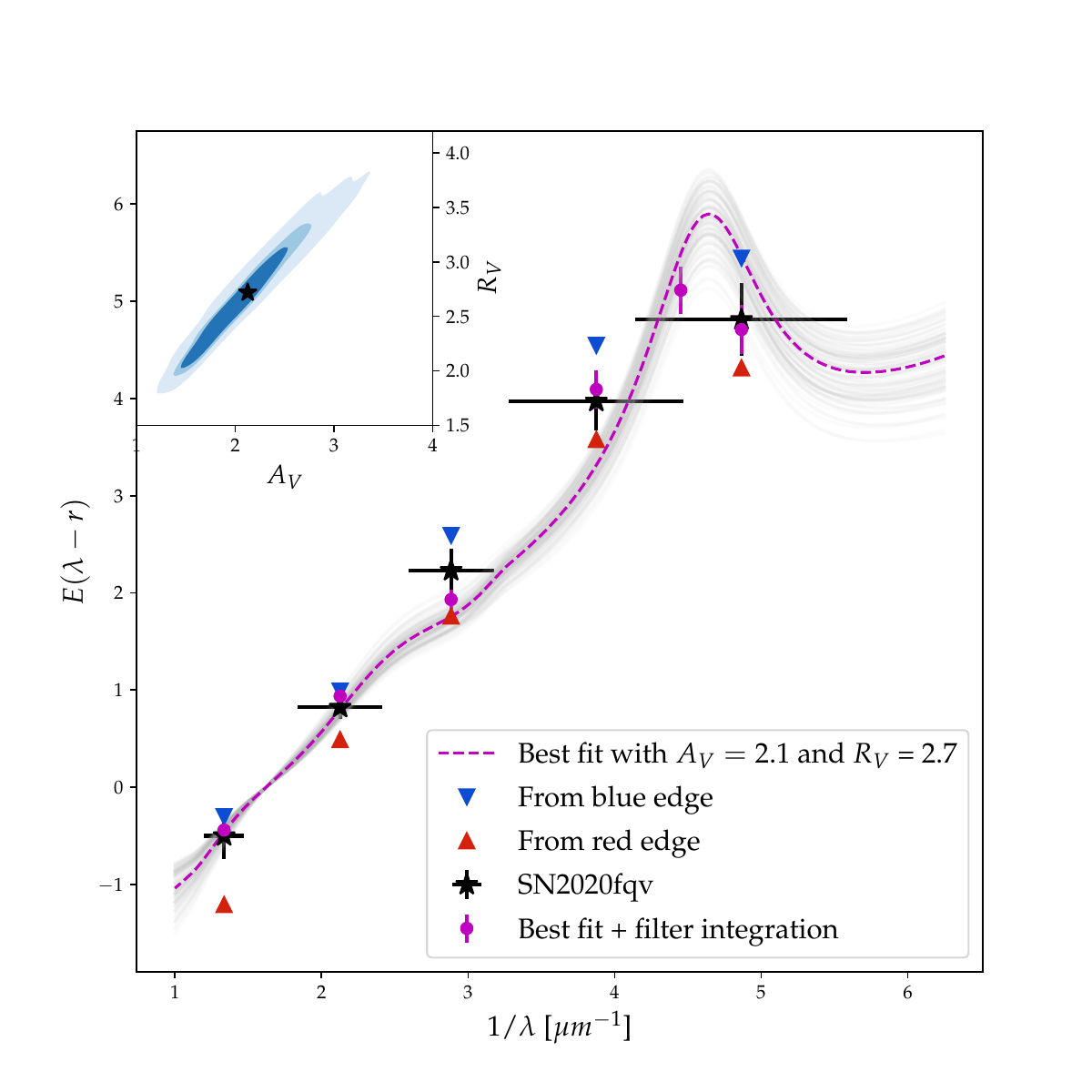} \\
\caption{The best fit extinction curve we find for SN\,2020fqv by correcting it to the mean colors of SNe II. In each band, the downward (upward) pointing blue (red) triangle shows the limits of the value of $A_{\lambda}$ from the bluest (reddest) objects in the sample. The black points show the color difference from the sample. The purple points are the best fit extinction curve, applied on a spectrum of \referee{a blackbody with $T=20,000$K}, and integrated over the filter bandpass. The purple curve is the best fit extinction laws, and the gray transparent curves are 50 randomly drawn curves from the posterior distribution. In the inset, we show the posterior distribution of our fit, with colors indicating the \referee{50\%, 68\% and 95\% confidence} regions.  }
\label{fig:Alam_fqv}
\end{figure}

In Fig.~\ref{fig:Alam_fqv}, we show the calculated $E(n-r)$ for SN\,2020fqv along with the best fitting extinction curves. The computed posterior probability distribution in the $A_{V}-R_{V}$ plane is shown in the inset. Using our results, we can determine extinction to $E(B-V)=0.1$ mag on average, and with a maximum systematic uncertainty of $E(B-V)=0.2$ mag. The case of SN\,2020fqv demonstrates that for highly extinguished SNe, a tight constraint can be acquired on $R_{V}$. As $UVM2$ measurement for SN\,2020fqv were not acquired, we cannot discriminate between extinction curves with and without the 220\,nm feature. However, these can likely be distinguished if such measurements were available. For mildly extinguished SNe, one may limit the extinction using these data. In Table \ref{tab:colors}, we report the color for $t=1$ to $t=5$ days. When using this method to measure the extinction, we caution against using a single epoch to estimate the extinction, as it can be degenerate with a temperature difference from the SN II population.

\begin{figure*}[t]
\centering
\includegraphics[width=\textwidth]{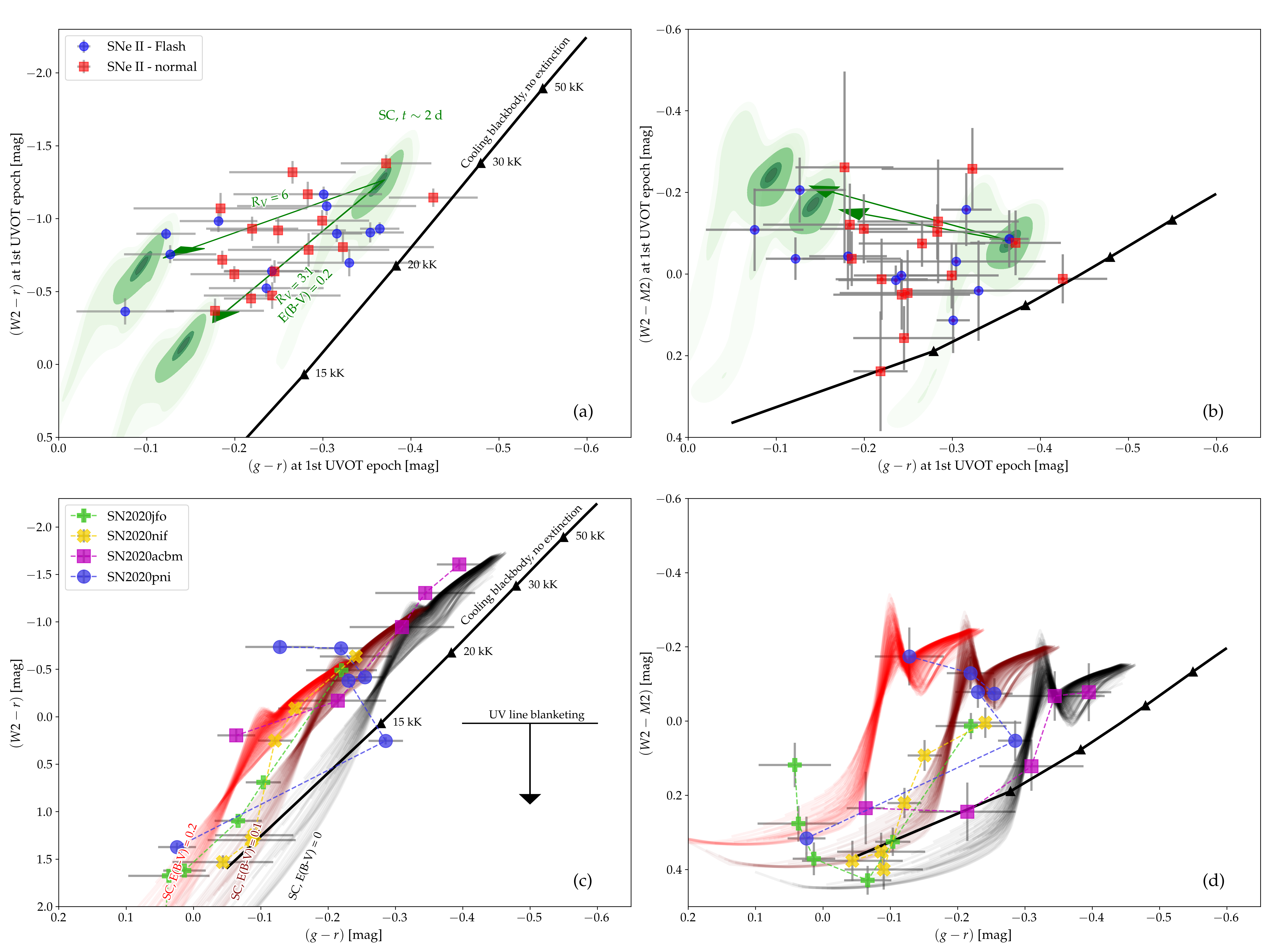} \\
\caption{(a) Color-color diagram of the UVW2-$r$ and the optical $g-r$ color. The data points represent different SNe at their 1st UV epoch with (blue circles) and without (red squares) flash features. The solid black curve represent the colors of a blackbody with temperatures between 100\,kK and 5\,kK. \referee{The green contours show the expected color-color values of the models of \refereetwo{M24} at $t=1.5-2.5$ d for a range of models parameters. The effect of adding extinction with $E(B-V) = 0.2$ mag different $R_{V}$ values is illustrated using green arrows}. (b) is similar to (a), but for $UVW2-UVM2$. (c) and (d) are similar to (a) and (b), but showing the color evolution of 4 SNe before $t<7$ days. We also show model Shock-cooling curves using the models of \refereetwo{M24} with increasing $E(B-V)$ (\refereetwo{0, 0.1 and 0.2 mag}) using a MW extinction law with $R_{V} = 3.1$. The distance from the black line corresponds to the deviation from blackbody, which is present in all SNe studied in this work. \referee{For clarity and due to its red color, SN\,2020fqv is not shown in this plot.} }
\label{fig:1st_UVOT}
\end{figure*}

We next consider intrinsic deviations from blackbody. In Fig. \ref{fig:1st_UVOT}, we show color-color plots of the SNe in our sample at the first UV epoch. In panel (a) we plot the  $W2-r$ and $g-r$ colors, and in panel (b) the $UVW2-UVM2$ and $g-r$ colors. Data points indicate the colors of the SNe II at their first UVOT visit, where blue and red colors represent SNe with and without flash features in their early spectra, respectively. The solid black line corresponds to a blackbody with 0 extinction between $10,000\,$K and $100,000\,$K. \referee{The green contours show the expected color-color values of the models of \refereetwo{M24} at $t=1.5-2.5$ d for a range of models parameters. The effect of adding extinction with $E(B-V) = 0.2$ mag different $R_{V}$ values is illustrated using green arrows}. 

The positions that various SNe occupy in the Fig. \ref{fig:1st_UVOT} demonstrate a clear deviation from a non-extinguished blackbody (black curve). SNe with and without flash features occupy the same area in the parameter space, indicating that this deviation from blackbody is not related to the presence of optically thin CSM. Pure reddening can explain some of the deviation, but requires $R_{V}>3.1$, a high temperature close to $100,000\,$K, and $E(B-V)$ of up to $0.4$ mag for some of the objects - more than the 0.2 mag that we infer based on the scatter in color curves. \referee{Relative to the expected color-color values predicted from the envelope cooling models of \refereetwo{M24}, an extinction law with $R_{V} \geq3.1$ is required to explain the position of all points. While other effects could mimic the bluer UV-optical colors of some of the points, }
A difference in $R_{V}$ seems a better explanation. It is consistent with the colors of the various SNe in both the $W2-r$ color, where the value of $R_{V}$ has a large effect on the color and the $W2-M2$ color, which is relatively unaffected by the value of $R_{V}$. \referee{On the other hand, a deviation caused by a line, e.g., in the W2 band, would have a more significant effect on the $W2-M2$ color. }

In panels (c) and (d) we show the expected color-color values from the analytic shock cooling models of \refereetwo{M24}, at $E(B-V) = 0-0.4$ mag, including time-dependent deviations from blackbody. Colored points represent a subset of SNe from our sample and their evolution in their first week. The time-dependent nature of the color curves (evolving from blue to red) conclusively indicates some of the deviation is intrinsic \referee{(i.e., due to evolving line blanketing and line emission)}. For many of the objects, the color evolution is similar to the expected color evolution in the shock cooling models, and a combination of mild $E(B-V)<0.2$ mag, intrinsic deviations from blackbody, and in some cases $R_{V}>3.1$, can fully explain all SN colors. \referee{We note that line blanketing alone cannot explain the observed deviations, since the UV-optical colors are bluer than the blackbody that fits the optical colors alone.} The color evolution of SN\,2020pni (blue stars) stands out in our sample. Its $g-r$ color becomes bluer in the first few days of its evolution. \cite{Terreran2022} argue the early light curve of this SN is powered by a shock breakout in an extended wind, rather than cooling of a shocked envelope. This non-monotonic color evolution was also observed for \referee{SN\,2018zd \citep{Hiramatsu2021}} and the nearby SN\,2023ixf \citep[][]{Zimmerman2023,JacobsonGalan2023,Hiramatsu2023, Gaici2023}, also suspected as a wind breakout.

To conclude, 33 of 34 SNe in our sample show $UVW2-r$ colors that become redder with time, consistent with a cooling behaviour. Using the mean colors, the extinction of any SNe can be constrained to better than $E(B-V) = 0.2$ mag. The early UV-optical colors of SNe II indicate deviations from blackbody that are consistent with the expected deviations due to extinction and the expected intrinsic deviations from blackbody in a cooling envelope, with no additional CSM interaction required.

\begin{deluxetable*}{lcccccccc}
\label{tab:bb_trunc}
\centering
\tablecaption{Early-time blackbody fits of SNe included in this work (truncated)}
\tablewidth{34pt} 
\tablehead{\colhead{SN} & \colhead{t [rest-frame days]} & \colhead{$\rm T_{bb}\ [K]$} & \colhead{$\rm R_{BB}\ [10^{14} \rm cm]$ } & \colhead{$\rm L_{pseudo}\ [10^{42} \rm erg\ \rm s^{-1}]$} & \colhead{$\rm L_{pseudo,extrap}\  [10^{42} \rm erg\ \rm s^{-1}]$}& \colhead{$\chi^{2}/dof$}  } 
\tabletypesize{\scriptsize} 
\startdata
SN2018cxn & 1.6 & $23500\pm1500$ & $2.15\pm0.16$ & $3.75\pm3.75$ & $10.01\pm0.7$ & 1.36 \\
SN2018cxn & 2.19 & $20800\pm1400$ & $2.58\pm0.22$ & $3.77\pm3.77$ & $8.69\pm0.54$ & 2.22 \\
SN2018cxn & 5.6 & $12700\pm600$ & $5.3\pm0.42$ & $3.5\pm3.5$ & $5.08\pm0.13$ & 2.0 \\
SN2018cxn & 9.59 & $10500\pm300$ & $6.82\pm0.34$ & $2.83\pm2.83$ & $3.88\pm0.04$ & 0.57 \\
SN2018dfc & 1.58 & $22000\pm800$ & $4.31\pm0.2$ & $13.18\pm13.18$ & $31.38\pm1.18$ & 1.0 \\
SN2018dfc & 3.19 & $17000\pm500$ & $5.99\pm0.25$ & $12.46\pm12.46$ & $21.67\pm0.37$ & 1.01 \\
SN2018dfc & 4.1 & $15000\pm300$ & $6.92\pm0.21$ & $10.69\pm10.69$ & $17.05\pm0.16$ & 0.46 \\
SN2018dfc & 5.03 & $13300\pm200$ & $8.19\pm0.29$ & $9.88\pm9.88$ & $14.7\pm0.12$ & 0.55 \\
SN2018fif & 1.22 & $20600\pm1500$ & $1.68\pm0.14$ & $1.67\pm1.67$ & $3.69\pm0.27$ & 2.66 \\
SN2018fif & 1.25 & $20100\pm1100$ & $1.73\pm0.12$ & $1.65\pm1.65$ & $3.54\pm0.18$ & 2.13 \\
SN2018fif & 2.1 & $15600\pm500$ & $2.7\pm0.15$ & $1.85\pm1.85$ & $3.08\pm0.06$ & 1.65 \\
SN2018fif & 2.65 & $15100\pm700$ & $2.87\pm0.21$ & $1.81\pm1.81$ & $2.96\pm0.07$ & 2.87 \\
SN2018fif & 4.57 & $12000\pm600$ & $4.22\pm0.3$ & $1.98\pm1.98$ & $2.65\pm0.05$ & 3.6 \\
SN2018fif & 6.15 & $11200\pm600$ & $4.86\pm0.42$ & $2.04\pm2.04$ & $2.68\pm0.05$ & 4.41 \\
SN2018fif & 6.17 & $11100\pm600$ & $4.87\pm0.42$ & $2.04\pm2.04$ & $2.68\pm0.05$ & 4.41 \\
SN2018fif & 7.31 & $10600\pm600$ & $5.33\pm0.46$ & $2.04\pm2.04$ & $2.66\pm0.04$ & 4.21 \\
SN2018fif & 8.33 & $10000\pm500$ & $5.86\pm0.49$ & $2.0\pm2.0$ & $2.6\pm0.04$ & 3.5 \\
\hline
\hline
\enddata
\tablenotetext{a}{A full version of this table is made available through the journal website.}
\tablenotetext{b}{A 0.1 mag systematic error was adopted for the fitting.}
\end{deluxetable*}

\subsection{Blackbody evolution}
\label{subsec:blackbody}
We linearly interpolate the UV-optical light curves of the sample SNe to the times of UV observations and construct an SED. Using the \package{Scipy} \package{curve\_fit} package \citep{Virtanen2020}, we fit this SED to a Planck function and recover the evolution of the blackbody temperature, radius, and luminosity parameters $T_{\rm eff}$, $R_{\rm BB}$, and $L_{\rm BB}$, respectively. We assume a $0.1$\,mag systematic error in addition to the statistical errors to account for imperfect cross-instrument calibration. In addition to the best-fit blackbody luminosity, we calculate a pseudobolometric luminosity by performing a trapezoidal integration of the interpolated SED and extrapolating it to the UV and infrared (IR) using the blackbody parameters. The fit results are reported in Table~\ref{tab:bb_trunc}. 

\begin{figure}[t]
\centering
\includegraphics[width=\columnwidth]{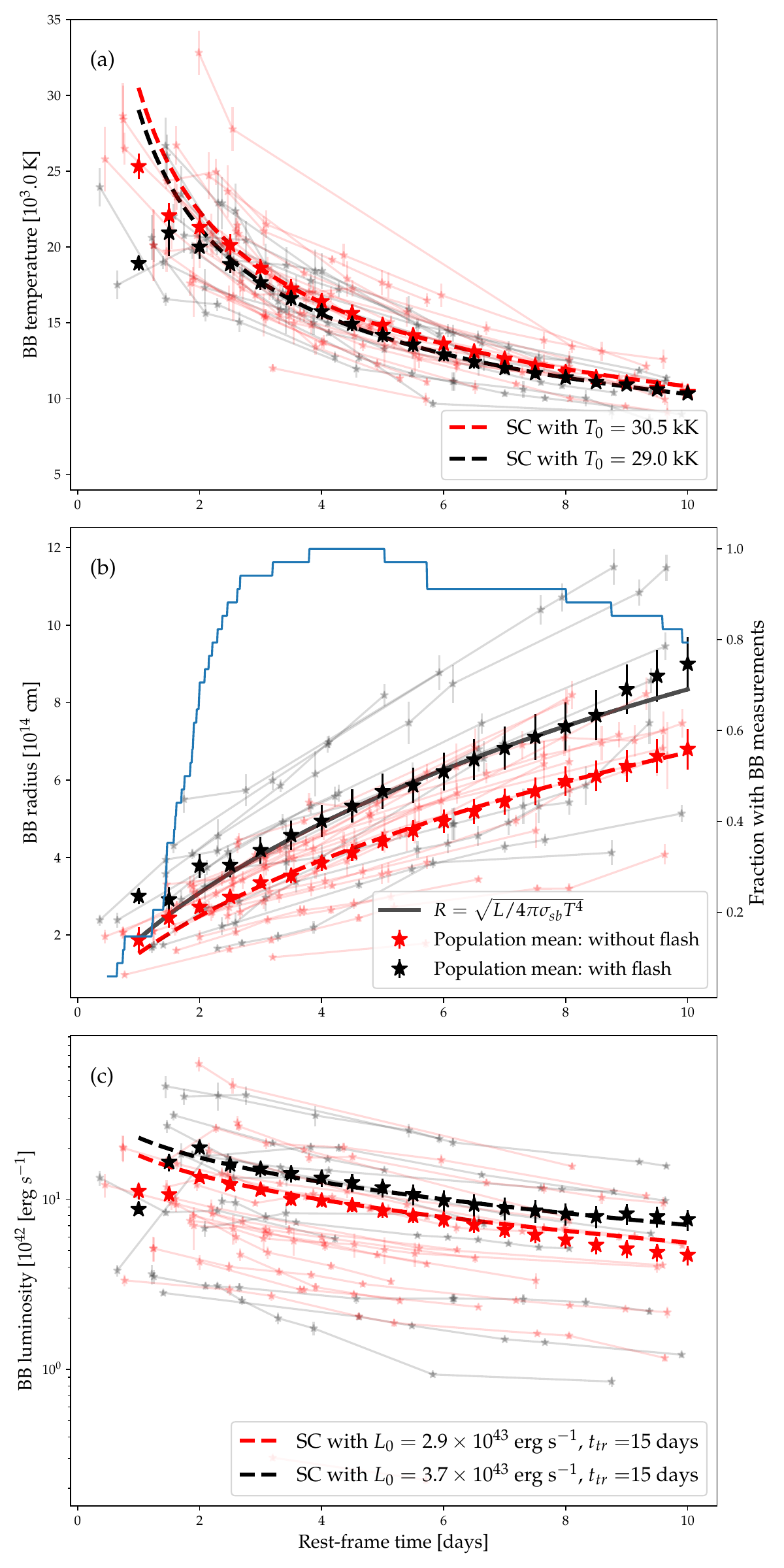} \\
\caption{The blackbody evolution of a sample of SNe II during the first 10 days. The transparent points represent individual SNe, color coded according to the presence of flash features (black) or lack thereof (red). The blue curve indicates the fraction of the sample with blackbody fits at each time step. The solid points show the population mean, and the dashed curves show the predicted evolution according to spherical phase shock cooling. Panels (a)-(c) shows the blackbody temperature, radius and luminosity, respectively. The match between the predictions of spherical phase shock-cooling models and the population blackbody evolution motivates the use of these models to fit individual SN light curves.}
\label{fig:bb_samp}
\end{figure}

In Fig.~\ref{fig:bb_samp} we show the blackbody evolution for our sample SNe, as well as the mean blackbody evolution of the population. To do so, we interpolate the temperatures, radii and luminosities with 0.5 day intervals, and take the population mean separately for SNe with and without flash ionization features as determined by \cite{Bruch2021} and \cite{bruch2022}. We estimate the error on the population mean through a bootstrap analysis \citep{EfroTibs93}. We draw 34 SNe, allowing for repetitions. We then draw samples from the blackbody parameters of each SN assuming a Gaussian distribution for every fit point. We then interpolate to the same time grid and calculate the population mean at every time step. The blue histogram shows the fraction of SNe in our sample with blackbody fit as a function of time. 

\begin{figure}[t]
\centering
\includegraphics[width=\columnwidth]{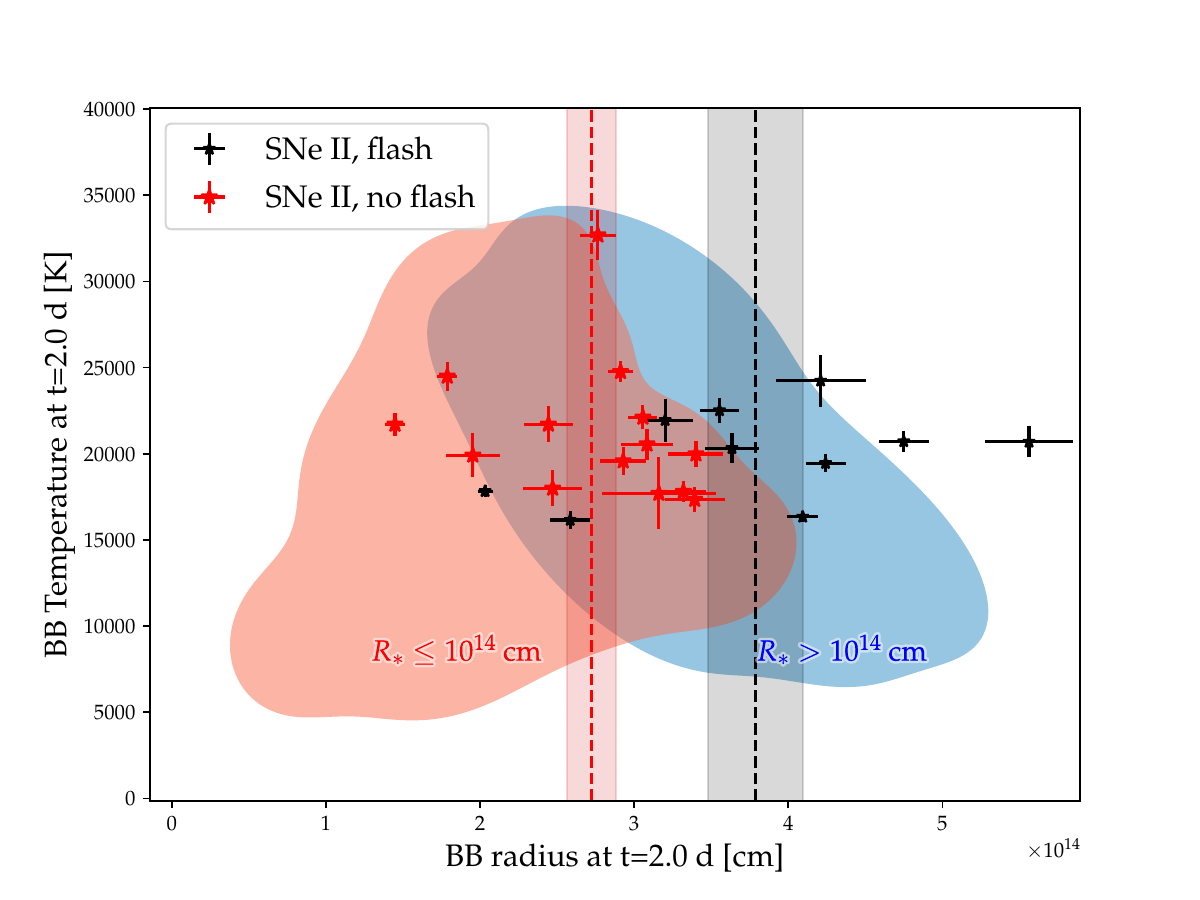} \\
\caption{The distribution of blackbody temperature and radius, interpolated to $t=2$ days. Black and red points are SNe II with and without flash features, respectively. The dashed lines and shaded regions show the population mean and standard error. The red and blue colored regions show the area occupied by simulated progenitors with $<10^{14}$ cm and $>10^{14}$ cm respectively. These are generated by fitting a blackbody to synthetic datasets constructed from the MG simulations of \refereetwo{M24}. Of the 23 SNe with a measurement at this time, 7 are only consistent with simulations that have a breakout radii $>10^{14}$ cm, or with a shock velocity parameter $v_{s*}\gtrsim 6000 {\rm km\, s^{-1}}$}
\label{fig:BB_data_rad}
\end{figure}

We find that SNe with flash features have a blackbody temperature  $6.3\%\pm{4.1\%}$ cooler and a radius (or photospheric velocity) $28\%\pm{11\%}$ larger than SNe without flash features. This difference is highlighted in Fig. \ref{fig:BB_data_rad} where we show the radius and temperature distribution of SNe with and without flash features, interpolated to $t=2$ days after explosion. At all times where a significant $>50\%$ fraction of the sample have measurements, the mean blackbody properties are well described by the predictions of spherical phase shock cooling, \referee{(fit to the population mean evolution)}. Our results indicate that the population of SNe II is well described by a cooling blackbody following shock-breakout at the edge of a shell of material with a steep density profile.


\subsection{Shock-cooling fitting}
\label{subsec:SC_fitting}
\subsubsection{Method and validation}
\label{subsec:method}
As the population blackbody evolution is well described by shock cooling, we fit individual SN light-curves to shock-cooling models. We do this using the model presented in \refereetwo{M23} and \refereetwo{M24}, which interpolated between the planar phase (i.e., when $r\approx R_{bo}$) and spherical phase (i.e., when $vt \gtrsim R_{bo} $) of shock cooling, and predicts the deviations of the SED from blackbody as a function of model parameters. The full model is described in \citet{Morag2023,Morag2024} and is briefly summarized in $\S$ \ref{ap:SC_mod}. 

The model has four independent physical parameters: The progenitor radius $R = R_{13}\, 10^{13}\ \rm cm$, the shock velocity parameter $v_{s*} = v_{s*,8.5}\, 10^{8.5}\, \rm{cm\, s}^{-1}$, the product of density numeric scale factor $f_\rho$ and the progenitor mass $M_{*} = M\, M_{\odot}$ (treated as a single parameter) and the envelope mass $M_{env} = M_{env,\odot}\, M_{\odot}$.  In addition to these parameters, we also fit for the extinction curve, parameterized as a \cite{Cardelli1989} law with free $R_{V}$ and $E(B-V)$, and the breakout time $t_{0}$. 

As demonstrated in \citet{Rubin_2016}, adopting a fixed validity domain will create a bias against some large radius models.  For every model realization, we calculate the validity domain, omitting the points outside this validity range from consideration. In order to properly compare between models with a different number of valid points, we we adopt a likelihood function based on the $\chi^{2}$ probability density function (PDF), as described in detail in \citet{Soumagnac2020}. 

Shock cooling models are expected to have residuals in temperature on order 5\% -- 10\% from model predictions \citep{Rabinak2011,sapir2016} when an average opacity is assumed, and additional systematics due to the presence of lines. \refereetwo{M24} expect the residuals on the flux to be of order 20\% -- 40\%, which will be correlated in time and wavelength. These residuals determine the appropriate covariance matrix to use in the $\chi^{2}$ statistic. They will also provide a criterion through which we can reject fits to a given data set. Indeed, when comparing the light curves of our sample of hydrodynamical simulations to the analytical model predictions, we find that in 50\% of the data points have residuals extending to $0.17$ mag and 95\% have residuals extending to $0.45$ mag. To incorporate the correlation between residuals into our analysis, we construct a likelihood function using the following steps:
\begin{itemize}
    \item Given a set of light curves, we construct a set of synthetic measurements from the set of hydrodynamical simulations of \refereetwo{M24} at the same times and photometric bands, by integrating the simulated SED with the appropriate transmission filters. 
    \item From each simulation, we construct a set of residuals from the analytic model predicted by the physical parameters of each simulation. 
    \item For each light-curve point, we calculate the covariance term as the mean over all simulations, taking into account only simulations which are valid at that time.
    \item Since the covariance matrix has too many parameters to be accurately estimated in full, we take the singular value decomposition (SVD) of the mean covariance and keep the top 3 eigenvalues.\footnote{This choice accounts for $>80\%$ of the variance, while preventing negative eigenvalues for any sampling used in our work.}  We then add this covariance matrix with a diagonal covariance matrix constructed from the observational errors in each data point, and add a 0.1 mag systematic error for cross-instrument calibration. 
    \item The likelihood of a model given the data is taken to be $\mathcal{L}=PDF(\chi^2,\nu)$ where $\chi^2=(d_i-m_i)cov^{-1}_{ij}(d_j-m_j)$, where $\vec{d},\vec{m}$ are the data and the model respectively, $\nu$ is the number of points where the model is valid, and PDF is the $\chi^{2}$ distribution PDF.
\end{itemize}

\begin{figure}[t]
\centering
\includegraphics[width=\columnwidth]{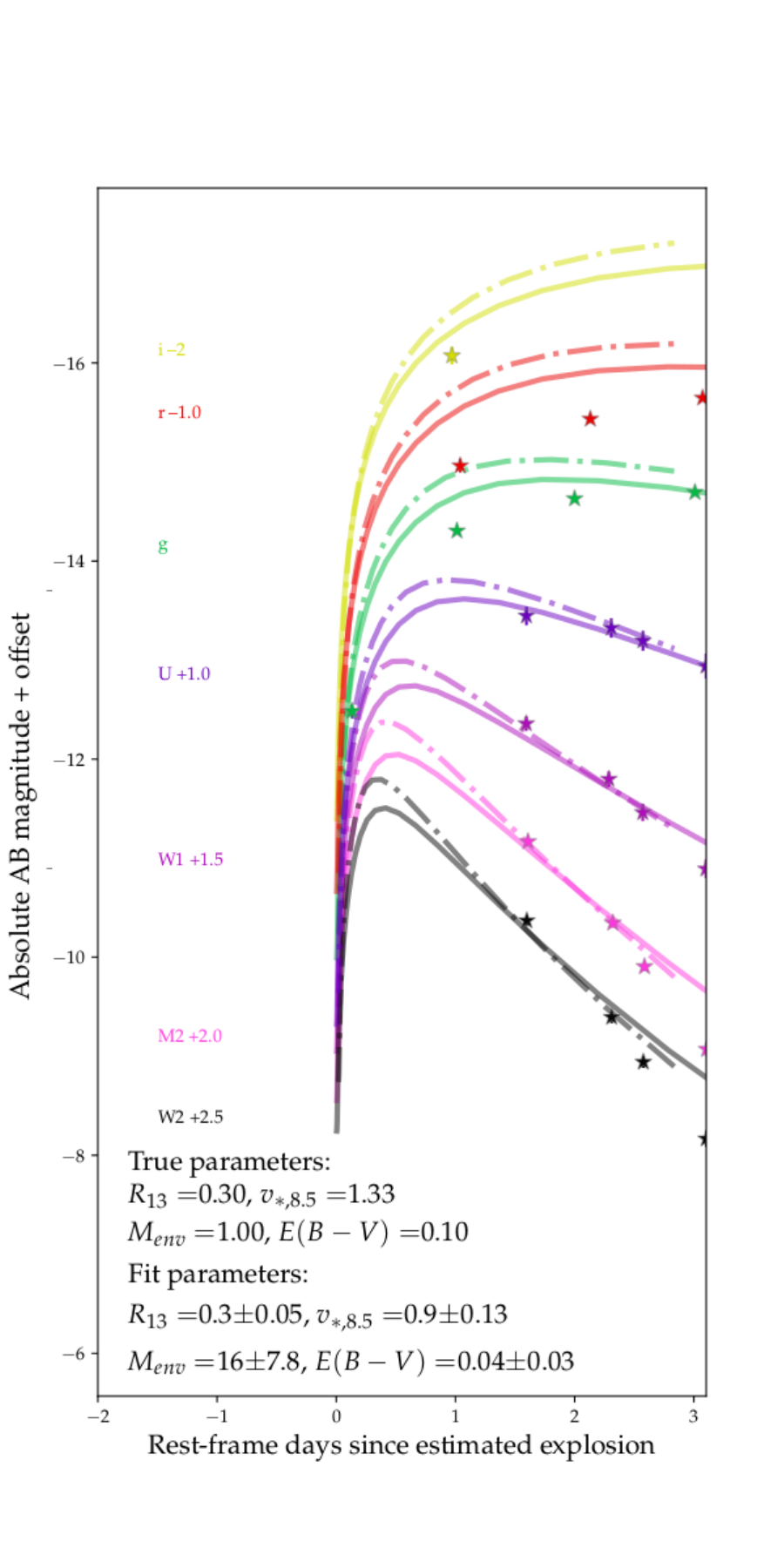} \\
\caption{Example of shock-cooling fits to a multi-band synthetic dataset, compared to the models generated from the physical simulation parameters. The solid lines are the average fits from the posterior, and the dot-dashed lines are model generated from the physical simulation parameters. The model light curve typically deviate by up to 20\% (calibration uncertainty) from the simulations, and are expected to deviate by up to 40\% in band specific flux due to theoretical uncertainty. We show the model until to its upper validity time. The best fit model accurately reproduces the breakout radius, velocity and finds a similar $E(B-V)$, but cannot reproduce the envelope mass or other model parameters. }

\label{fig:example_syn}
\end{figure}

Using this likelihood, we fit the model to the photometry using the nested-sampling \citep{Skilling2006} package
\package{dynesty} \citep{Higson2019,Speagle2020a}. We validate our method by testing that even in the presence of such residuals, we can still recover the true model parameters from simulated data sets. We fit all simulated data sets using this method, and compare the fit parameters with the physical parameters used in the simulations. In Fig. \ref{fig:example_syn}, we show an example of such a fit for a simulation generated with $R_{13} = 0.3$, $v_{s*,8.5} = 1.33, M_{env} = 1\, M_{\odot}$ with $E(B-V) = 0.1$ mag extinction added. We recover  $R_{13} = 0.3 \pm 0.05$, $v_{s*,8.5} = 0.9\pm0.13, M_{env} = 16\pm 7.8\, M_{\odot}$ and $E(B-V) = 0.04\pm0.03$ mag.

\begin{figure}[t]
\centering
\includegraphics[width=\columnwidth]{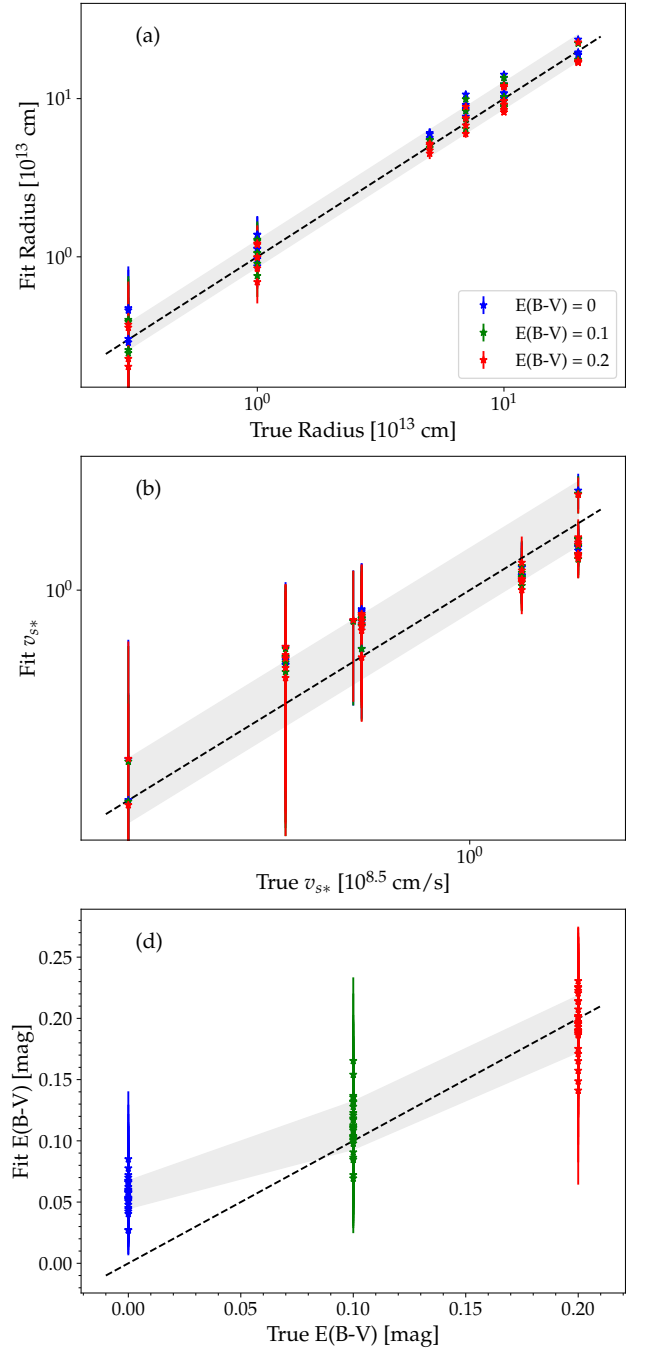} \\
\caption{Parameter recovery when fitting a sample of synthetic light curves with analytic shock-cooling models. In panels (a) and (b), we show the fit and true parameters for $R_{13}$ and $v_{s*,8.5}$, respectively. In panel (c), we show the recovery accuracy of $E(B-V)$. The dashed line represents a perfect recovery, and the shaded regions represent the $68\%$ interval over the full parameter space. }
\label{fig:accu}
\end{figure}

In Fig. \ref{fig:accu} we show the fit and true radii $R_{13}$ and shock velocity parameter $v_{s*,8.5}$, compared to the parameters used in the simulations. The 90\% confidence intervals for parameter recovery are $30\%$ for $R_{13}$, $26\%$ for $v_{s*,8.5}$ and better than $0.05$ mag in $E(B-V)$ over the entire parameters space of our simulations. However, we cannot recover $M_{env}$ or $f_{\rho}M_{tot}$ to better than an order of magnitude, and our fit results are highly sensitive to our choice of prior in those parameters, indicating they cannot be effectively constrained from shock-cooling modelling. 

Our results demonstrate that even given significant residuals, one may still fit these analytic models and recover the shock velocity, progenitor radius and the amount of dust reddening with no significant biases. Our results also demonstrate that rejecting shock-cooling as the main powering mechanism of the early light curves requires residuals larger than $\sim0.5$ mag.

\figsetstart
\figsetnum{14}
\figsettitle{Multi-band light curve fits}

\figsetend

\begin{figure*}[t]
\centering
\includegraphics[width=\textwidth]{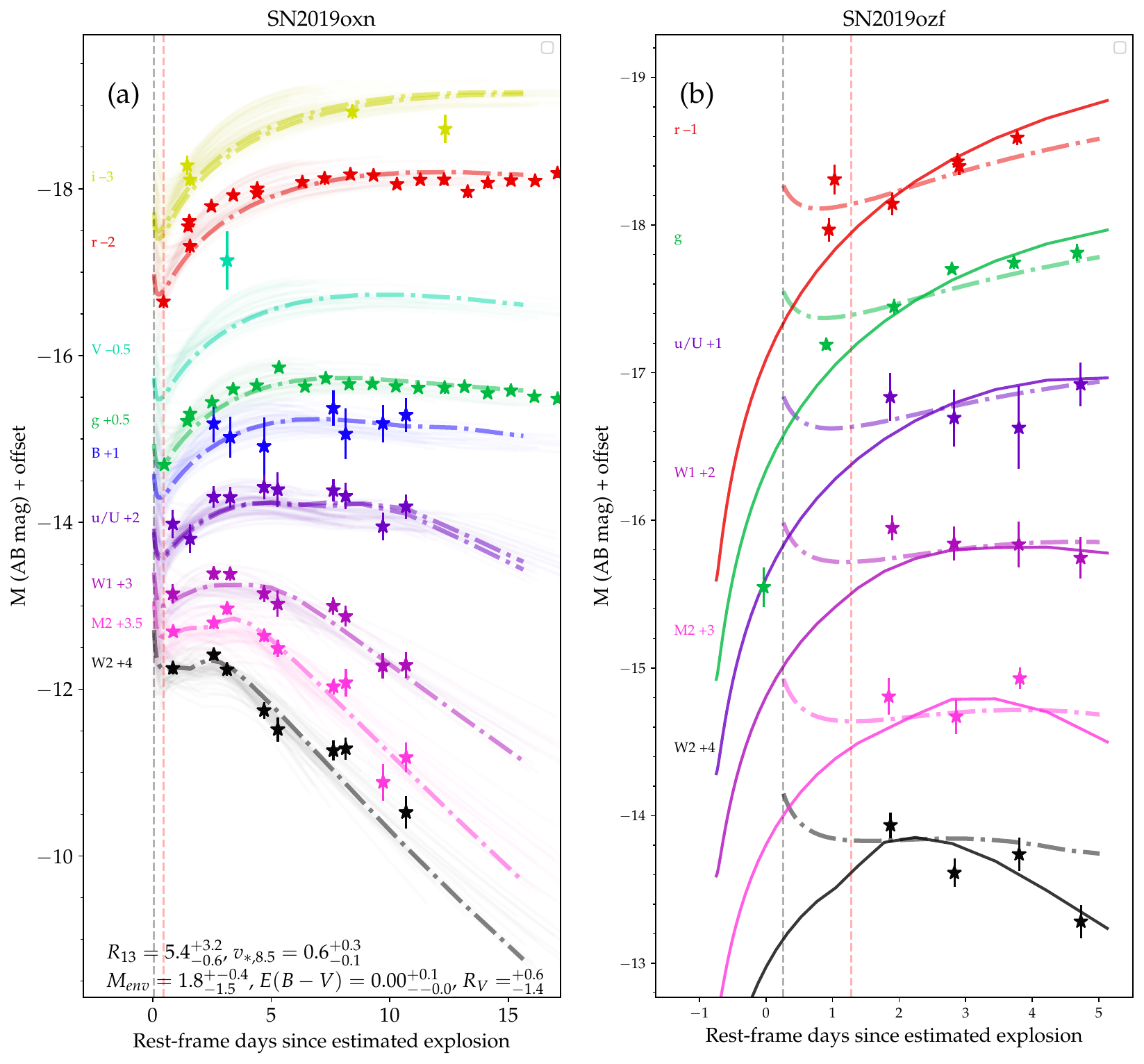} \\
\caption{(a) An example of a fit to a SN dataset from our sample. The dot-dashed curves are the best fits in each band. The transparent curves are 50 random samples from the posterior distribution. The vertical dashed lines indicate the best fit lower validity domain (gray) and the transition from planar to spherical phase (orange). (b) An example of a fit which misses the rise (the first $g$-band point) for the best fit model ($R_{13} = 22.2, v_{s*,8.5} = 1.7$, dot-dashed lines), but to which a reasonable lower radius fit exists ($R_{13} = 4.0, v_{s*,8.5} = 3.3$, solid lines). The complete figure set (33 images) is available in the online journal.} \label{fig:example_fit_SN}
\end{figure*}

\begin{figure}[t]
\centering
\includegraphics[width=\columnwidth]{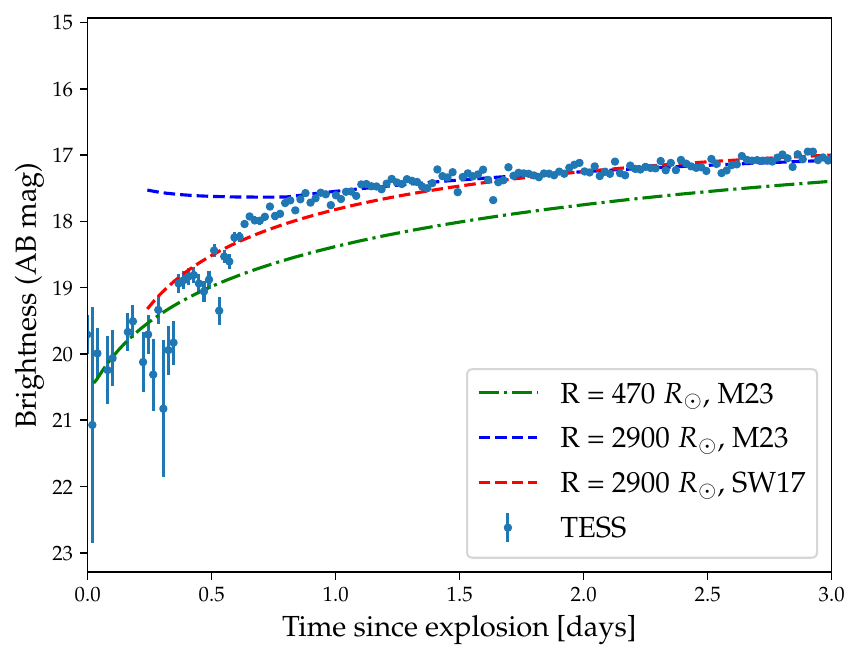} \\
\caption{Best fit shock-cooling models to the early time \textit{TESS}\ light curve of SN 2020nvm. The blue curve shows the best fit \refereetwo{M24} model to the multi-band light curve, which misses the rise during the planar phase. The green curve shows the best fit for a narrow radius prior, and the red curve shows the same model as the blue curve but accounting only for the spherical phase with the model of \cite{sapir2016}. While the spherical phase only can reproduce the full light curve, taking the planar phase into account results in different early time light curve. When including the planar phase, no good fit is found which can describe the entire light curve.  }
\label{fig:TESS_nvm}
\end{figure}

\subsubsection{Light-curve fits}
\label{subsec:SNfits}
We ran our fitting routine on all sample SNe. We used log-uniform priors for $R_{13}\in[0.1,30]$, $v_{s*,8.5}\in[0.1,6]$, $f_{\rho}M\in[0.1,200]$, $M_{env,\odot}\in[0.3,30]$. We also fit $t_{exp}\in[t_{ND}-1,t_{first}]$ with a uniform prior, where $t_{ND}$ is the last non-detection and $t_{first}$ is the first detection, respectively (relaxing the prior on $t_{ND}$ does not significantly impact our fit). Motivated by our analysis in $\S$~\ref{subsec:color} we also fit for host-galaxy extinction by assuming a \cite{Cardelli1989} reddening law with uniform priors on $E(B-V)$ and $R_{V}$ in the range $E(B-V)\in[0,0.25]$ mag and $R_{V}\in[2,5]$. For SN\,2020fqv, we fit with a wide prior of $E(B-V)\in[0.25,1]$ mag, given the high host-extinction we inferred from its color evolution. 

In addition to the flat priors on the parameters, we include non-rectangular priors through the model validity domain. This is done to prevent fits that exclude most data points from the validity range for parameter combinations with high  $v_{s*,8.5}$ and low $M_{env}$. We assign 0 probability to models that have no photometry data within their validity domain. While this does not impact our results in this work, fitting models without good non-detection limits shortly before explosion, or that are expected to have short validity times (e.g., due to small radii, or high velocity to envelopes mass ratios), might be affected by this demand. In \citet{Soumagnac2020}, we assigned priors on the recombination time at 
$0.7~\rm{eV} = 8120$ K ($t_{0.7\ \rm eV}\sim R_{13}^{0.56}v_{s*,8.5}^{0.16}$) of the SN through it spectral sequence. However, in some of the simulations of \refereetwo{M24}, we start seeing signs of hydrogen emission already at $20,000$ K. Instead, we use priors derived from the blackbody sequence of the SN. Since there are residuals in color between the simulations and models, and since the effect of host galaxy extinction is known to better than $0.2$ mag, the fit temperature assuming $E(B-V) = 0$  mag might not always be accurately used to determine the true photospheric temperature. We quantify the maximal effect of these systematics on the photospheric temperature near $0.7~\rm{eV} = 8120$ K. We fit all synthetic datasets (with an extinction of up to $E(B-V) = 0.2$ mag) with blackbody SEDs assuming no host extinction, and find that demanding that $T>10,700$ K is enough to determine that $t>t_{0.7 eV}$, and $T<5500$ is enough to determine that $t<t_{0.7 eV}$ for any combination of parameters, as long as $E(B-V)\leq0.2$ mag. These physically motivated priors on the recombination time have a significant effect on our fitting process.

Due to the peculiar temperature and luminosity evolution of SN\,2020pni, which does not fit the general predictions of spherical phase shock cooling, we omit this SN from the fitting process. We will treat the modelling of this SN in detail in Zimmerman et al. (in perp.). 

In Table \ref{tab:SC_res} we report the parameters of our posterior sampling at the 10th, 50th and 90th percentiles. In all cases, we find good fits for the light curves at $t>1$ days after explosion. Our fits divide into 2 cases: (1) For 15 SNe, we find good fits to the UV-optical SN light curves throughout the evolution. These models are characterized by a radius under $10^{14}$ cm, and residuals better than 0.42 mag (95\%) throughout the first week. (2) For the remaining 18 SNe, the early optical light curve points do not match the rise of the models - either pushing it out of the model validity domain or missing it completely by more than 1 mag. These models are exclusively characterized by a large radius ($>10^{14}$ cm) required to account for a high luminosity, but do not show the shallow rise or double peaked feature expected for planar phase shock cooling of such a star.\footnote{In the Rayleigh-Jeans limit, this can be intuitively understood as $f_{\lambda} \propto T_{BB}R_{BB}^2$, resulting in $f_{\lambda}\propto t^{-1/3}$ in the planar phase, and $f_{\lambda}\propto t^{1.15}$ early in the spherical phase.} After the first day from estimated explosion, these fits have comparable residuals to group (1). If forced to fit a radius of $<10^{14}$ cm - a reasonable fits achieved in about half of the cases. For the rest of the objects in this group, forcing a small radius results in a bad overall fit. 

Since the spherical phase luminosity $L_{{\rm RW}}\sim R_{13}v_{{\rm s*,8.5}}^{1.91}$, these fits are characterized by a higher $v_{\rm s*,8.5}$ and more host-galaxy extinction to decrease the temperature as $T_{{\rm ph,t=1\ \rm d}}\sim R_{13}^{1/4}v_{{\rm s*,8.5}}^{0.07}$.  We show examples of fits of both cases in Fig. \ref{fig:example_fit_SN}, and make all figures of all light curve fits available as online figures through the journal website upon publication. In Fig  \ref{fig:TESS_nvm}, we show the illuminating example of SN\,2020nvm, which was observed by \textit{TESS} throughout its rise. We show that a model accounting only for the spherical phase will artificially create a much sharper rise compared to a model which fits the peak. In this case, our best small-radius fit did not match the observed light curve well, and the large radius model (one of the largest values in our sample) misses the rise. The clear first peak expected in planar phase cooling is not observed even at early times.\footnote{We note some features are present in the very early light curve. These are also present in some of the simulations of \refereetwo{M24}, and could be the result of lines. This is likely not the shock breakout signal, which is expected to be very faint in this band \citep{sapir2013,katz2013,sapir2014}} The \cite{sapir2016} model fits the rise much better, although it is not physical at early times. 

\begin{figure}[t]
\centering
\includegraphics[width=\columnwidth]{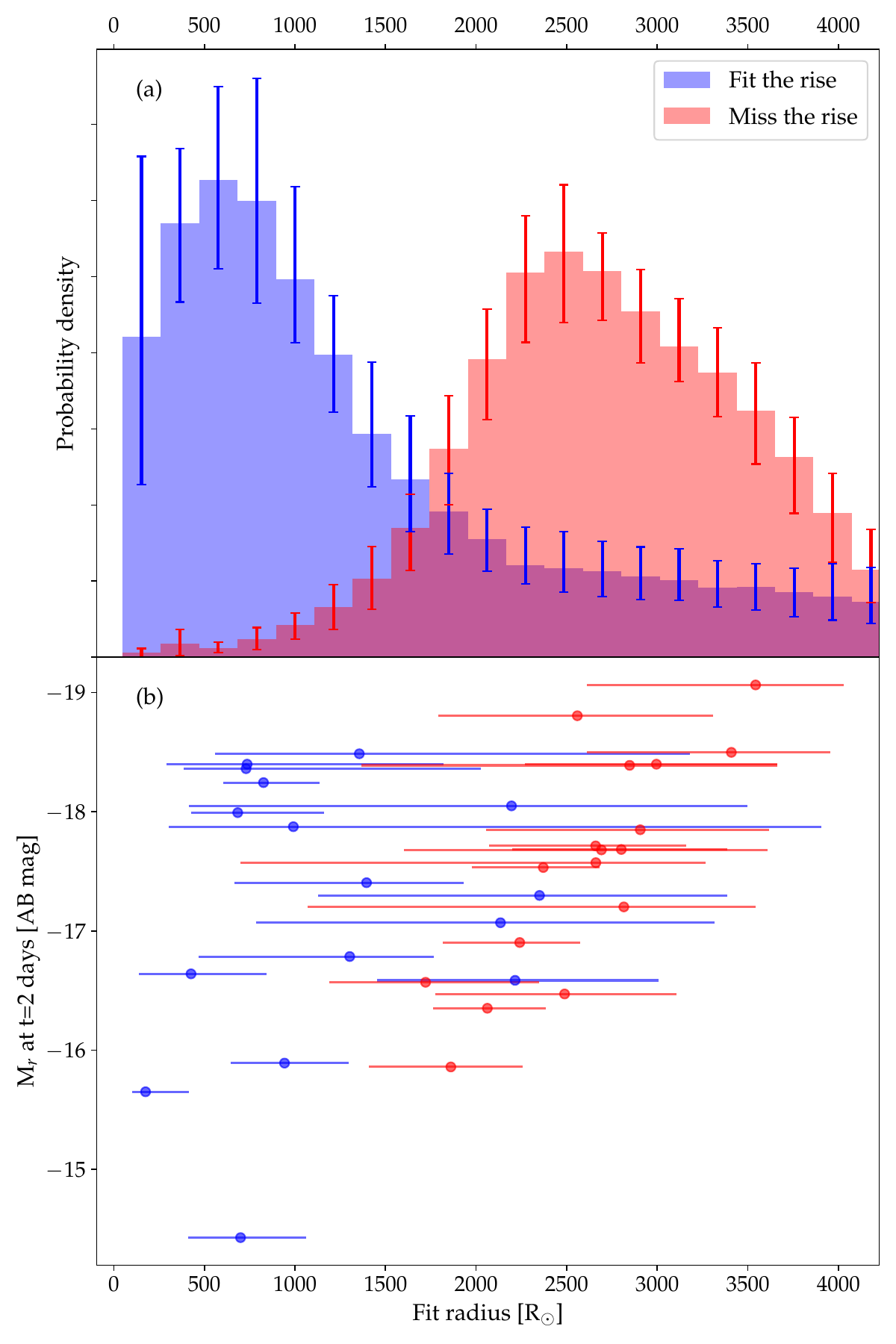} \\
\caption{(a) Posterior probability distribution of the breakout radius, for SNe whose fit misses the rise and SNe whose best fit does not miss the rise. (b) A scatter plot showing the correlation between the best fit radius and the $r$-band magnitude at $t=2$ days.  A large fit radius  is strongly associated with with missing the rise during the first day, and is associated with a brighter $r$-band light curve.}
\label{fig:miss_good_Rhist}
\end{figure}

In Fig. \ref{fig:miss_good_Rhist}, we present the posterior probability for the radius of best-fit models that miss the rise, and those that match the rise. We find no statistically significant difference between SNe with and without flash features (which could perhaps be detected given a larger sample). 

\begin{figure*}[t]
\centering
\includegraphics[width=\textwidth]{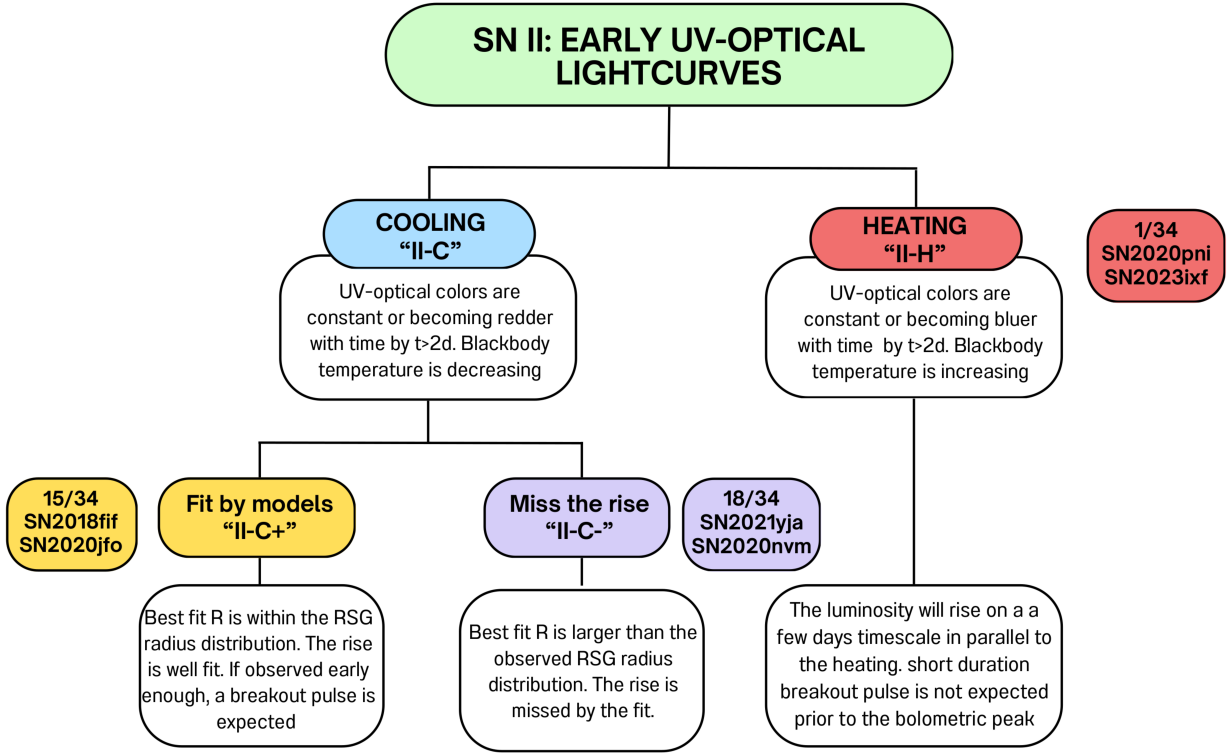} \\
\caption{Schematic classification of the early light curves of SNe II. They roughly divide into 2 groups: (1) SNe with increasing temperatures at early times, which call ``II-H", and (2) with decreasing temperatures or ``II-C". We further divide the later into 2 groups: (a) which are well fit by shock cooling models at early times, have a good early fit, and a small fit radius. We call these ``II-C+". (b) which are not well fit at early times, are more luminous as a population, and have larger fit radii. We call these ``II-C-". Next to each group we denote the number of SNe in the sample which belong to it, as well as example SNe. }
\label{fig:flowchart}
\end{figure*}

We summarize the different categories our objects fall into in Fig.~\ref{fig:flowchart}. Most SNe II are cooling at early times, showing constant or reddening UV-optical colors. We refer to these as ``II-C". SNe II which are heating and showing a bluer UV-optical color with time are referred to as ``II-H". We further subdivide the II-C group into SNe with small fit radius (``II-C+"), which are well fit at early times, and those with large fit radius (``II-C-"), which are not well fit by shock cooling models at early times.

\begin{deluxetable*}{lcccccccc}
\centering
\label{tab:SC_res}
\tablecaption{Best fit parameters for shock cooling fitting}
\tablewidth{38pt} 
\hspace{3pt}
\tablehead{\colhead{SN} & \colhead{$R_{13}$} & \colhead{$v_{s*,8.5}$ } & \colhead{$f_{M}$} & \colhead{$E(B-V)$ [mag]} & \colhead{$R_{V}$} & \colhead{$t_{0}$ [JD]} & \colhead{$t_{0.7}$ [days]} & \colhead{$t_{tr}$ [days]}} 
\tabletypesize{\scriptsize} 
\startdata
SN2019eoh & $7.7^{+0.9}_{-0.9}$ & $2.0^{+0.3}_{-0.3}$ & $0.5^{+0.5}_{-0.4}$ & $0.0^{+0.0}_{-0.0}$ & $4.0^{+0.9}_{-1.3}$ & $-0.02^{+0.05}_{-0.06}$ & 26.7 & 16.3 \\
SN2020aavm & $13.4^{+4.1}_{-4.3}$ & $0.9^{+0.3}_{-0.3}$ & $83.7^{+85.7}_{-70.8}$ & $0.1^{+0.1}_{-0.1}$ & $3.9^{+1.0}_{-1.2}$ & $0.19^{+0.45}_{-0.50}$ & 22.2 & 47.1 \\
SN2020fqv & $3.4^{+1.6}_{-1.7}$ & $1.0^{+0.4}_{-0.3}$ & $99.4^{+76.4}_{-75.7}$ & $0.8^{+0.0}_{-0.0}$ & $2.8^{+0.5}_{-0.5}$ & $0.40^{+0.26}_{-0.28}$ & 13.8 & 47.4 \\
SN2019oxn & $5.4^{+0.8}_{-0.8}$ & $0.7^{+0.1}_{-0.1}$ & $1.4^{+1.7}_{-1.2}$ & $0.0^{+0.0}_{-0.0}$ & $3.9^{+0.9}_{-1.2}$ & $0.01^{+0.18}_{-0.21}$ & 20.2 & 46.8 \\
SN2020ufx & $15.2^{+3.8}_{-4.2}$ & $2.2^{+0.5}_{-0.4}$ & $83.0^{+81.7}_{-64.5}$ & $0.0^{+0.0}_{-0.0}$ & $3.6^{+1.1}_{-1.2}$ & $-0.33^{+0.27}_{-0.33}$ & 27.7 & 4.6 \\
SN2019ozf & $16.5^{+2.8}_{-3.4}$ & $0.7^{+0.2}_{-0.2}$ & $94.5^{+77.6}_{-71.0}$ & $0.0^{+0.0}_{-0.0}$ & $3.7^{+1.0}_{-1.2}$ & $-0.44^{+0.33}_{-0.33}$ & 20.6 & 30.9 \\
SN2018cxn & $13.7^{+3.2}_{-3.1}$ & $0.6^{+0.1}_{-0.1}$ & $109.7^{+68.1}_{-69.8}$ & $0.1^{+0.1}_{-0.1}$ & $3.8^{+1.0}_{-1.2}$ & $-0.00^{+0.01}_{-0.01}$ & 16.7 & 38.9 \\
SN2020cxd & $3.5^{+1.3}_{-1.0}$ & $0.3^{+0.1}_{-0.0}$ & $53.0^{+77.3}_{-47.4}$ & $0.1^{+0.1}_{-0.1}$ & $3.7^{+1.0}_{-1.2}$ & $0.48^{+0.53}_{-0.60}$ & 12.4 & 32.9 \\
SN2019nvm & $17.2^{+2.1}_{-2.2}$ & $0.7^{+0.1}_{-0.1}$ & $111.2^{+68.4}_{-70.3}$ & $0.1^{+0.0}_{-0.0}$ & $3.6^{+1.1}_{-1.1}$ & $-0.51^{+0.38}_{-0.34}$ & 25.6 & 38.2 \\
SN2020lfn & $17.5^{+2.1}_{-2.6}$ & $1.5^{+0.3}_{-0.2}$ & $98.5^{+72.3}_{-67.1}$ & $0.0^{+0.0}_{-0.0}$ & $3.8^{+1.0}_{-1.2}$ & $-0.24^{+0.22}_{-0.24}$ & 27.0 & 8.8 \\
SN2020jfo & $6.3^{+1.3}_{-1.3}$ & $0.6^{+0.1}_{-0.1}$ & $72.3^{+76.8}_{-56.8}$ & $0.0^{+0.0}_{-0.0}$ & $3.6^{+1.1}_{-1.2}$ & $0.11^{+0.23}_{-0.27}$ & 14.3 & 28.5 \\
SN2020nyb & $14.8^{+3.1}_{-3.0}$ & $0.4^{+0.1}_{-0.1}$ & $60.0^{+81.4}_{-53.1}$ & $0.1^{+0.1}_{-0.1}$ & $3.9^{+0.9}_{-1.1}$ & $-0.55^{+0.47}_{-0.60}$ & 21.9 & 60.6 \\
SN2019wzx & $14.0^{+4.9}_{-6.2}$ & $1.0^{+0.4}_{-0.3}$ & $92.6^{+79.7}_{-73.3}$ & $0.0^{+0.0}_{-0.0}$ & $3.6^{+1.1}_{-1.2}$ & $-0.26^{+0.51}_{-0.55}$ & 20.2 & 31.2 \\
SN2019gmh & $14.9^{+4.6}_{-6.6}$ & $0.8^{+0.3}_{-0.2}$ & $132.4^{+55.0}_{-68.5}$ & $0.1^{+0.0}_{-0.0}$ & $4.1^{+0.7}_{-0.9}$ & $-0.30^{+0.20}_{-0.16}$ & 24.5 & 10.1 \\
SN2020afdi & $8.0^{+1.7}_{-1.6}$ & $0.4^{+0.1}_{-0.1}$ & $49.1^{+77.2}_{-45.2}$ & $0.0^{+0.0}_{-0.0}$ & $3.8^{+1.0}_{-1.2}$ & $-0.07^{+0.27}_{-0.28}$ & 21.8 & 48.5 \\
SN2020pqv & $17.3^{+2.2}_{-2.7}$ & $0.8^{+0.2}_{-0.2}$ & $80.0^{+83.6}_{-64.6}$ & $0.0^{+0.0}_{-0.0}$ & $3.7^{+1.0}_{-1.2}$ & $-0.61^{+0.46}_{-0.50}$ & 29.5 & 13.5 \\
SN2020mst & $14.7^{+4.2}_{-4.5}$ & $0.9^{+0.2}_{-0.2}$ & $90.1^{+80.7}_{-72.2}$ & $0.1^{+0.1}_{-0.1}$ & $3.6^{+1.1}_{-1.2}$ & $-0.35^{+0.31}_{-0.33}$ & 24.6 & 23.6 \\
SN2020dyu & $16.0^{+3.2}_{-3.4}$ & $1.2^{+0.3}_{-0.2}$ & $105.0^{+71.6}_{-70.4}$ & $0.0^{+0.0}_{-0.0}$ & $3.6^{+1.1}_{-1.1}$ & $-0.54^{+0.28}_{-0.24}$ & 26.0 & 14.0 \\
SN2021apg & $14.9^{+3.9}_{-4.7}$ & $0.6^{+0.2}_{-0.2}$ & $87.9^{+80.8}_{-70.7}$ & $0.1^{+0.1}_{-0.1}$ & $3.7^{+1.0}_{-1.2}$ & $-0.31^{+0.76}_{-0.83}$ & 23.0 & 12.9 \\
SN2020xva & $10.6^{+3.0}_{-2.7}$ & $0.5^{+0.1}_{-0.1}$ & $66.8^{+84.8}_{-58.6}$ & $0.1^{+0.1}_{-0.1}$ & $4.0^{+0.8}_{-1.2}$ & $0.22^{+0.40}_{-0.46}$ & 18.6 & 44.5 \\
SN2020nif & $13.0^{+5.1}_{-5.0}$ & $1.6^{+0.4}_{-0.3}$ & $87.7^{+81.4}_{-70.5}$ & $0.1^{+0.0}_{-0.0}$ & $3.6^{+1.1}_{-1.2}$ & $-0.07^{+0.41}_{-0.46}$ & 23.1 & 9.0 \\
SN2020acbm & $15.5^{+2.3}_{-2.4}$ & $0.7^{+0.1}_{-0.1}$ & $92.6^{+77.5}_{-70.7}$ & $0.0^{+0.0}_{-0.0}$ & $3.7^{+1.1}_{-1.2}$ & $-0.46^{+0.33}_{-0.33}$ & 26.7 & 10.8 \\
SN2021yja & $15.2^{+3.8}_{-4.6}$ & $0.6^{+0.2}_{-0.1}$ & $90.2^{+80.7}_{-72.7}$ & $0.0^{+0.0}_{-0.0}$ & $3.6^{+1.1}_{-1.2}$ & $0.39^{+1.08}_{-1.08}$ & 21.7 & 54.2 \\
SN2019omp & $13.5^{+4.9}_{-5.0}$ & $0.8^{+0.2}_{-0.2}$ & $105.5^{+72.9}_{-74.3}$ & $0.1^{+0.1}_{-0.1}$ & $3.7^{+1.0}_{-1.2}$ & $-0.48^{+0.33}_{-0.31}$ & 24.4 & 33.1 \\
SN2021ibn & $8.0^{+2.8}_{-2.6}$ & $1.9^{+0.5}_{-0.5}$ & $89.1^{+80.3}_{-69.5}$ & $0.0^{+0.0}_{-0.0}$ & $3.7^{+1.0}_{-1.2}$ & $0.19^{+0.15}_{-0.18}$ & 17.6 & 6.8 \\
SN2020qvw & $15.2^{+3.7}_{-4.4}$ & $1.4^{+0.3}_{-0.3}$ & $90.1^{+80.7}_{-70.3}$ & $0.0^{+0.0}_{-0.0}$ & $3.6^{+1.1}_{-1.2}$ & $-0.10^{+0.25}_{-0.21}$ & 24.0 & 6.5 \\
SN2018fif & $12.2^{+3.5}_{-3.3}$ & $0.7^{+0.1}_{-0.1}$ & $63.8^{+79.4}_{-54.2}$ & $0.2^{+0.1}_{-0.1}$ & $4.2^{+0.6}_{-0.9}$ & $-0.25^{+0.22}_{-0.29}$ & 20.6 & 47.2 \\
SN2020pni & $14.8^{+4.2}_{-5.1}$ & $1.4^{+0.5}_{-0.4}$ & $92.3^{+79.0}_{-71.1}$ & $0.0^{+0.0}_{-0.0}$ & $3.7^{+1.0}_{-1.2}$ & $0.14^{+0.73}_{-0.68}$ & 25.5 & 10.8 \\
SN2021skn & $12.8^{+5.4}_{-5.0}$ & $1.1^{+0.4}_{-0.3}$ & $105.7^{+71.5}_{-73.3}$ & $0.0^{+0.0}_{-0.0}$ & $3.6^{+1.1}_{-1.2}$ & $-0.48^{+0.38}_{-0.32}$ & 23.3 & 10.8 \\
SN2020uim & $13.4^{+1.8}_{-1.7}$ & $0.4^{+0.1}_{-0.1}$ & $52.7^{+79.2}_{-47.6}$ & $0.0^{+0.0}_{-0.0}$ & $3.7^{+1.1}_{-1.2}$ & $-0.24^{+0.30}_{-0.37}$ & 20.7 & 55.0 \\
SN2018dfc & $15.1^{+4.0}_{-4.7}$ & $1.4^{+0.5}_{-0.4}$ & $105.5^{+71.7}_{-72.1}$ & $0.0^{+0.0}_{-0.0}$ & $3.6^{+1.1}_{-1.2}$ & $-0.76^{+0.29}_{-0.23}$ & 27.8 & 13.7 \\
SN2020xhs & $14.1^{+3.3}_{-3.7}$ & $0.5^{+0.2}_{-0.1}$ & $96.4^{+78.4}_{-75.3}$ & $0.1^{+0.1}_{-0.0}$ & $3.6^{+1.1}_{-1.2}$ & $0.59^{+0.61}_{-0.51}$ & 19.5 & 49.9 \\
SN2019ust & $10.0^{+2.2}_{-2.1}$ & $2.1^{+0.4}_{-0.4}$ & $17.1^{+24.9}_{-15.7}$ & $0.2^{+0.0}_{-0.0}$ & $4.1^{+0.7}_{-0.9}$ & $0.57^{+0.12}_{-0.14}$ & 23.1 & 4.8 \\
SN2020abue & $13.9^{+3.0}_{-3.0}$ & $0.6^{+0.1}_{-0.1}$ & $97.0^{+77.3}_{-73.6}$ & $0.0^{+0.0}_{-0.0}$ & $3.7^{+1.1}_{-1.2}$ & $-0.86^{+0.50}_{-0.48}$ & 22.5 & 26.8 \\
\enddata
\tablenotetext{a}{all times are in rest-frame days}
\tablenotetext{b}{Uncertainties reflect the 10th-90th percentiles of the posterior probability distribution}
\tablenotetext{c}{This table will be made available in machine-readable format through the journal website upon publication.}
\end{deluxetable*}

\begin{figure*}[t]
\centering
\includegraphics[width=\textwidth]{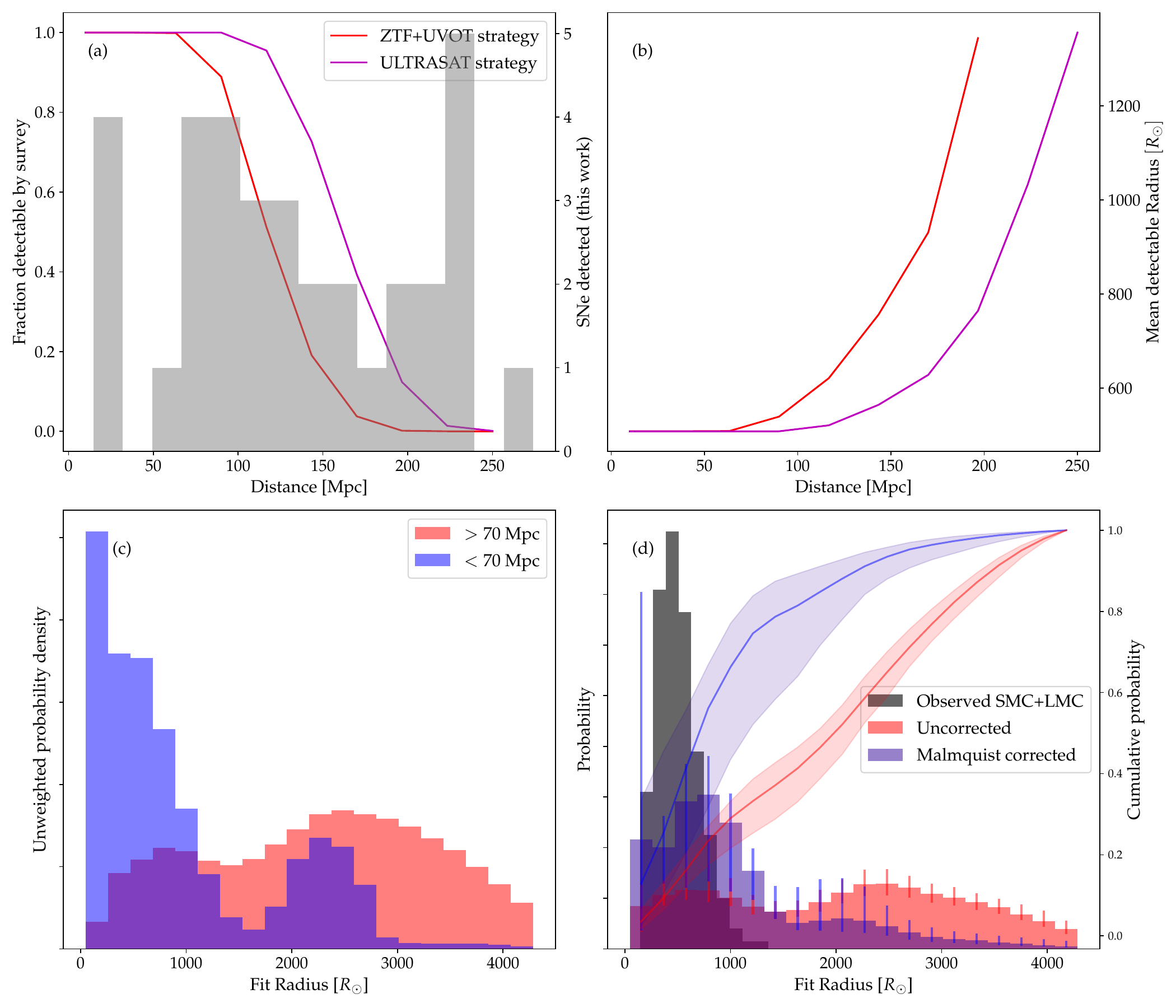} \\
\caption{(a) A histogram of the distances of SNe in this work, and the fraction of simulated SNe light curves which would be followed up with our observations study, and in the \textit{Ultraviolet Transient Astronomy Satellite} (\ultrasat) survey. We assume the radius distribution of \cite{Davies2013} for the models. (b) The mean radius of the detected SNe, demonstrating a luminosity bias at $d>70 $ Mpc. (c) The unweighted posterior probability distribution of the breakout radius, and $d$ above or below 70 Mpc. (d) The posterior distribution of the full sample, corrected and uncorrected for the luminosity bias. The gray histogram is a distribution of RSG radii from \cite{Davies2013}. We also shock the cumulative distribution of the observed and corrected posterior distribution, with $68\%$ confidence intervals. While the observed fraction of SNe with large $>1000\,R_{\odot}$ radius is $71^{+7}_{-4}\%$, they only account for $34^{+23}_{-11}\%$ of exploding RSGs.  }
\label{fig:Rfit_hist}
\end{figure*}

\section{Discussion}
\label{sec:discussion}
\subsection{RSG radius distribution}
\label{subsec:rad_dist}
\subsubsection{What can the early-time fits teach us?}

In $\S$~\ref{subsec:method}, we demonstrated that with a typical set of UV-optical light curves, we can recover the breakout radius and shock velocity parameter from the simulations of \refereetwo{M24} for a wide range of parameters. When applying our method to the SNe of our sample, we found good fits to roughly half of the SNe, with radii consistent with the observed RSG radius distribution (II-C+). The remaining SNe systematically miss the rise and are characterized either by a high $R_{13}$ or a high $v_{s*,8.5}$ due to the higher luminosity of this group compared to other SNe (II-C-). Since there are acceptable fits for roughly half of such SNe, and as the blackbody radius and temperatures of the majority of the sample evolve according to the predictions of spherical phase shock cooling, we cannot rule out that it is the primary powering mechanism of these SNe. Our lack of early-time UV-optical colors and of high quality sampling in the first hours of the SN explosions prevents us from testing whether the blackbody evolution in the very early times evolves according to the predictions of planar phase shock cooling. However, we note that when optical colors are available during these first phases, the colors are consistent with that of a hot $>15,000$ K blackbody. With this in mind, there are several possibilities to explain the large radius fits:
\begin{enumerate}
    \item These SNe are powered by shock cooling only, and have a small radius. The failure to fit the rise is due to correlated residuals not present in the simulations, and thus is not modeled in the covariance matrix we used - creating a bias to larger radii in some cases, or they did not cover this particular combination of shock velocity and radius. This possibility is likely what happens in half of the cases, where a good fit is acquired if the fit is forced to a small radius. In other cases, the small radius fit still misses the rise or a unrealistically high $v_{s*}$ is required.  
    \item These SNe have a large progenitor radius, and their early time evolution does not fit the predictions of planar phase shock cooling from a spherical RSG envelope. Recent work by \cite{Goldberg_2022a,Goldberg_2022b} shows that the turbulent 3D structure of the outer regions of the envelope, or a non-spherical breakout surface could possibly extend the duration of shock breakout and affect the early stages of shock cooling up to a timescale of $R/v \lesssim 1$ day. If this is the case for the majority of similar fits, the large radius of the progenitor star would be consistent with a shell of dense CSM or an inflated envelope at $<3\times10^{14}\, \rm cm$, with the breakout occurring at the edge of the shell. This interpretation is also supported by spectropolarimetric observations of SN\,2021yja \citep{Vasylyev2023b}, showing a high degree of continuum polarization during the early photospheric phase ($t>25$ days). SN\,2021yja is well fit by a large radius model during its full evolution, but misses the rise by several magnitudes. The large radius fit is also noted by \cite{Hosseinzadeh2022}, that fit the spherical phase model of \cite{sapir2016} and acquire very similar parameters, but their fit matches the rise at early times due to lacking an accurate description of the planar phase. A similar case is demonstrated in Fig. \ref{fig:TESS_nvm}.  
    \item These SNe are the result of a breakout from the edge of a shell of dense CSM on a several hours timescale, and the early (few days) light curve is characterized by the subsequent cooling. The intrinsic timescale (i.e., ignoring light travel time) for shock breakout from any spherical density profile is $\frac{\Delta R}{v} = \frac{c}{v_{bo}^2\, \kappa\, \rho_{bo}}$ $\Delta R$ is the width of the breakout shell, and $v_{bo},\rho_{bo}$ are the velocity and density at breakout \citep[][and references therein]{Waxman2017}. A shock breakout in a slowly declining and extended density profile will be characterized by a density of $\lesssim10^{-12}\, \rm{g\,cm}^{-3}$ and occur on a few days timescale. This is likely what occurred during the explosions of SN\,2020pni,  \referee{SN\,2018zd \citep{Hiramatsu2021}},  and more recently for SN\,2023ixf \citep{Zimmerman2023}, where a rise in temperature was observed during the first few days. In both cases, breakout occurred from a shell of dense CSM confined to $<2\times10^{14}$ cm. 
    If the mass of this shell is higher, breakout will occur at the edge of the shell at densities of $\rho_{bo}\sim 10^{-11}\, \rm{g\,cm}^{-3}$, resulting in an hours long breakout which will power the optical rise.  Since we do not include breakout in our modelling (assumed to occur before observations began) the early time light curve will be missed by the fit. 
    After breakout, the cooling should still evolve according to the predictions of spherical or planar phase shock cooling, which are insensitive to the exact shape of the density profile \citep{Sapir2011, Rabinak2011,sapir2016}. The parameter inference will likely be wrong in this case, since cooling is measured relatively to the peak of breakout. A delay of $\delta t_{\rm d} = 0.12 \frac{\Delta R}{10^{13}\, \rm cm}\frac{v}{10^{9}\, {\rm cm\, s^{-1}}}^{-1} \rm day$ will result in an increase of $(1+\delta t_{\rm d})^{1.8}$ in the fit progenitor radius, but will not change the general conclusion that the radius is large enough to reach such low $\rho_{bo}$.     
    This scenario is seemingly challenged by the lack of strong association between the presence of flash ionization features and a large fit radius. However flash features trace the CSM density profile at $\sim 10^{15}\, \rm cm$ \citep{Yaron2017} rather than $R\sim 10^{14}$ cm required for this effect to become significant. This scenario is consistent with the conclusions of \cite{Morozova2018}, who fit a grid of hydrodynamical models of progenitors surrounded by dense CSM at $<10^{14}$ cm, and found that they are consistent with the light curves of observed Type II SNe, with breakout occurring at the edge of the dense CSM. 
    \end{enumerate}

Similarly to the heating defining the extended breakout of the II-H category, an optical rise while the temperature is heating is the unambiguous marker of an increase in the bolometric luminosity, expected only during breakout itself. Observing or ruling out such heating during the first day of the explosion through high-cadence UV-optical observations thus has the potential to resolve any remaining ambiguity regarding SNe in the II-C- group, since all three options presented above have different predictions for the breakout pulse itself:
\begin{enumerate}
    \item The breakout pulse occurs at densities of $\sim 10^{-9}\, \rm{g\,cm}^{-3}$. The breakout duration is likely dominated by the light travel time, lasting minutes to an hour. Breakout will likely peak at tens of eV. 
    \item The breakout pulse occurs at densities of $\sim 10^{-9}\, \rm{g\,cm}^{-3}$. The asymmetric nature of the breakout shell caused a smearing of the breakout to a timescale of a few hours. Locally, the width of the shock transition is still similar, so that breakout would still likely peak at tens of eV. 
    \item The breakout pulse occurs at densities of $\lesssim 10^{-11}\, \rm{g\,cm}^{-3}$. The low density causes the intrinsic breakout timescale to last a few hours, dominating over the light travel time. Locally, the width of the shock transition is large, so that breakout might be peaking at $\sim 10$ eV, and could contribute significantly to the optical during the early rise. No additional short duration pulse can be observed. 
\end{enumerate}
 \subsubsection{The intrinsic 
 progenitor radius distribution}
To connect the observed parameter distribution to the intrinsic progenitor radius distribution, we account for the selection effects and biases introduced by our observation strategy and the dependence of the luminosity on the breakout radius. We calculate model light curves for the sample of RSG of \cite{Davies2018}. We calculated the radii from the observed effective temperatures and luminosities, and generate a set of light curves with a velocity parameter $v_{s*,8.5}$ in the range $0.5-1.5$, with the rest of the model parameters set to unity and assuming no host or galactic extinction along the line of sight. We test what fraction of the models is recovered by our observation strategy as a function of distance, demanding a blue color ($g-r<0\,$mag) at $t=1$ day, and the object brighter than $19.5\,$mag at the same time, which is the typical brightness limiting our ability to classify the object as an SN II, a criterion for followup in our program. We repeat this analysis for an \ultrasat\ strategy - demanding an optical brightness above 19.5 mag at peak for spectroscopic classification, and that the light curve is higher than the limiting magnitude of $22.5\,$mag at 1d \citep{Shvartzvald2023}. 

We find that as the distance increases above $70$ Mpc, we are increasingly biased towards higher progenitor radii. In panel (a) of Fig. \ref{fig:Rfit_hist}, we show the fraction of RSG explosions recovered as a function of distance with each strategy, and histogram of the distances of our sample. In panel (b), we show the mean radius of the recovered sample, as a function of distance. In panel (c), we show the posterior distribution of the SNe radius above and below a distance of 70 Mpc. The radius posterior distribution of closer SNe is highly skewed towards radii below $1000 R_{\odot}$, while the distribution of SNe at larger distances is skewed to values above $1000 R_{\odot}$. 

We correct the Malmquist bias following the treatment of \cite{Rubin_2016}. For each point in the posterior sample, we calculate a weight factor $w_{i}=\frac{D_{i}^{3}}{\sum_{j}D_{j}^{-3}}$ where $M_{i}+17 = 5\log(\frac{D_{i}^{*}}{/10\, \rm pc})$. We show the resulting corrected posterior distribution in Fig. \ref{fig:Rfit_hist} panel (d), along with the unweighted distribution and the distribution of RSG radii of \cite{Davies2018}. The error bars are calculated by bootstrapping the posterior distribution: for every realization, we recalculate the posterior for 33 SNe randomly sampled from the list of SNe with viable fits, while allowing for repetition. We repeat this process 500 times and plot the mean and standard deviation on each bin of the histogram.

Our analysis shows that even if most ($67^{+9}_{-5}\%$) of the observed SNe have large ($R>1200\, R_{\odot}$) breakout radii, the breakout radius distribution would be consistent with the observed RSGs radius distribution ($R<1200\, R_{\odot}$) in $69^{+13}_{-26}\%$ of SNe II explosions. Hinds et al. (in prep.) will analyze the optical light curves of SNe II in the magnitude-limited BTS survey, and reaches a similar conclusion. We further note that for SNe with a CSM breakout such as SN\,2020pni,  \referee{SN\,2018zd} or SN\,2023ixf, a breakout radius of $\sim 1500-3000\,R_{\odot}$ is needed to explain the breakout timescale and would be consistent with the distribution we report here \citep{Zimmerman2023}. In the case of SN\,2023ixf, constraints on the SN progenitor from pre-explosion data confirms a dusty shell at a similar radius \citep[e.g.,][]{Qin2023}. This supports the the idea that SNe II-C- have large radii due to a shell of CSM from which shock breakout occurs.

\subsection{X-ray emission and constraints on extended CSM density}
\label{subsec:xray_csm}
Following SN shock breakout, the accelerated ejecta will expand into the surrounding optically thin CSM, acting as a piston and creating a shock in the CSM. For typical CSM densities, this shock is expected to be collisionless, heat the gas to $\sim 100$ keV temperatures and produce X-ray emission \citep{Fransson1996,Katz2011,Chevalier_2012,Svirski2012,Ofek2014a}. In $\S$~\ref{subsec:xray_obs}, we reported the XRT detections and upper limits at the SN location, binned over the duration of the \swift\ observations (typically $\sim 10,000$ ks). The limits we acquire are several orders of magnitude deeper than the optical emission, reaching as deep as few SNe II previously detected by XRT. SN\,2005cs \citep{Brown2007}, SN\,2006bp \citep{Brown2007}, SN\,2012aw \citep{Immler2012}, SN\,2013ej \citep{Margutti2013_b}, and recently SN\,2023ixf \citep{Grefenstette2023}. 

\begin{figure}[t]
\centering
\includegraphics[width=\columnwidth]{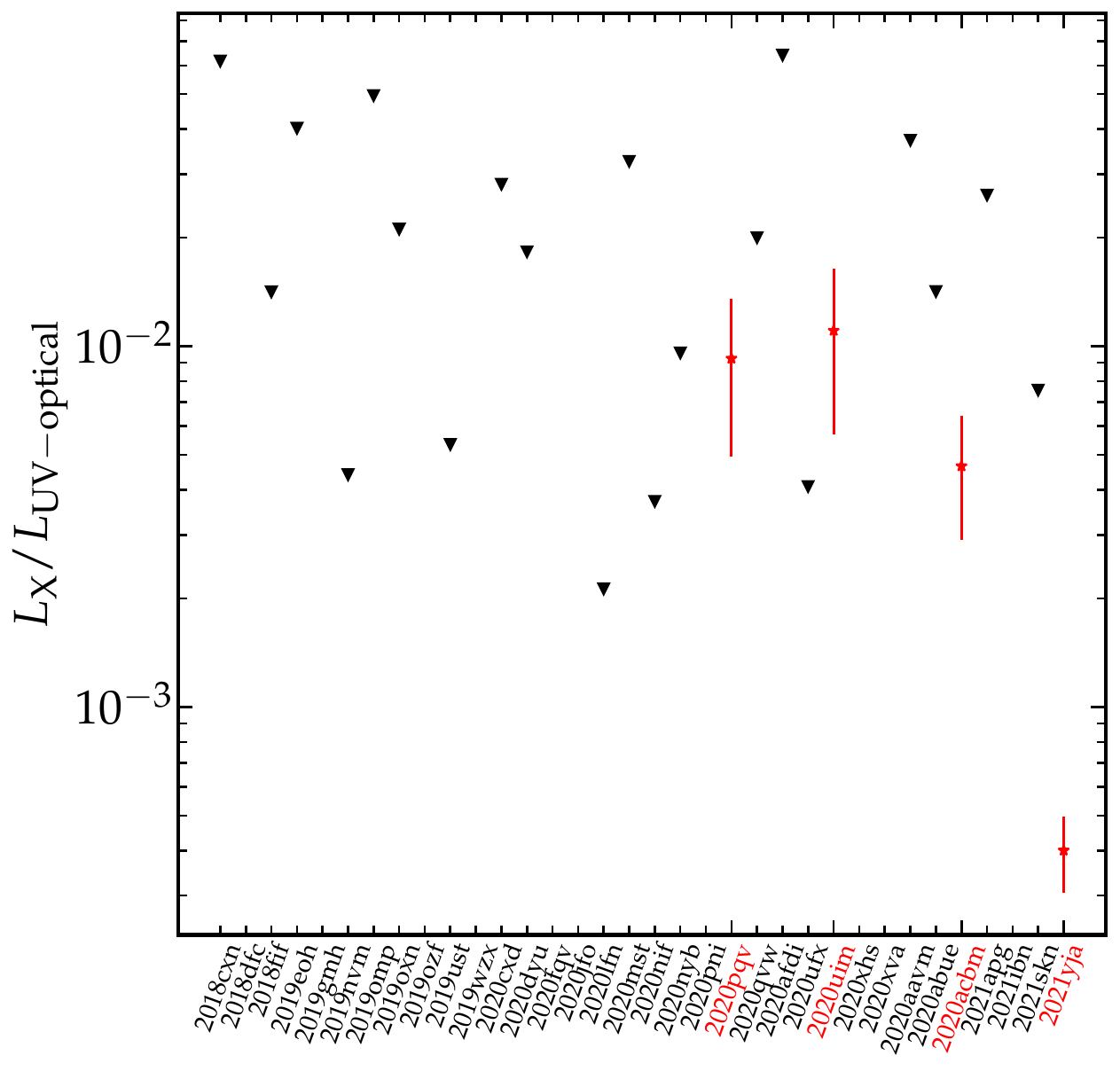} \\
\caption{The ratio of X-ray to UV-optical emission, measured at the same times and averaged over the duration of the \swift\ observations for the different SNe in our sample. Upper limits are shown \referee{as black triangles}, and the 4 detections we report are shown \referee{using red points}. }
\label{fig:L_XRT_ratio}
\end{figure}

In Fig. \ref{fig:L_XRT_ratio}, we show a histogram of the limits ratio of X-ray to UV-optical emission at the same times (transparent bars), and the 4 detections we report (vertical red lines with shaded error bars). Our limits range between $10^{-1}-10^{-4}$ of the optical emission, and the highest detection is $\sim10^{-2}$. In $\S$~\ref{subsec:SNfits}, we derived constraints on the velocity profiles of the SN ejecta through UV-optical light curve fitting. The photon arrival weighted time of our detections (as well as those in the literature) typically correspond to a few days after explosion - probing the forward shock emission in the extended CSM around the progenitor star at $(0.5-2)\times10^{15}\, \rm cm$. We can use these to constrain the CSM density at $\sim10^{15}$ cm and subsequently constrain the mass-loss of the progenitor star prior a few years prior to explosion. 

At a time $t$, a constant velocity shock moving through an optically thin CSM with $v_{s,csm}$ will sweep up a mass:
\begin{equation}
    \frac{M_{CSM}}{M_{\odot}}=2.7\times10^{-4}\,v_{s,csm,9}t_{5d}\rho_{o,-16}
\end{equation}
where   $v_{s,csm,9}=\frac{v_{s,csm}}{10^{9}\,\text{cm\,}\text{s}^{-1}}$,$t_{5d}=\frac{t_{xray}}{5\,\text{d}}$ and $\rho_{o,-16}=\frac{\rho_{o}(r=10^{15}\, \rm cm)}{10^{-16}\,\text{g\,}\text{cm}^{-3}}$. To find the velocity $v_{s,csm}$ we assume it is well approximated by the velocity of the piston (the ejected envelope) at equal mass to the swept of CSM. This is given through the profiles of \cite{Rabinak2011}. Following their notation (their equations 3. and 4.) we find: 
\begin{equation}
\delta_{m,piston}=\frac{M_{csm}}{M_{tot}}=2.7\times10^{-4}\frac{f_{\rho}v_{s,csm,9}t_{5d}\rho_{o,-16}}{f_{\rho}M_{\odot}}
\end{equation}
\begin{equation}
    v_{s,csm}=v_{f}\left(\delta_{m}=\frac{M_{csm}}{M_{tot}}\right)
\end{equation}
As long as the fraction $\frac{M_{CSM}}{M_{tot}}$ is larger then the mass fraction in the breakout shell $\delta_{m,bo}$:
\begin{equation}
\frac{v_{s,csm}^{\left(1\right)}}{\text{cm\,}\text{s}^{-1}}=1.5\times10^{9}\left(\frac{f_{v}}{2}v_{s*,8.5}\right)^{0.9}\left(\frac{t_{5d}\rho_{o,-16}}{f_{\rho}M_{\odot}}\right)^{-0.1}
\end{equation}

Here we took $f_{v}=\frac{v_{f}}{v_{s}}=2$, which is typically the case for small $\delta_{m}<0.01$ \citep{Matzner_1999}. This is in agreement with the the velocity evolution of \cite{Chevalier1994} for a steep post-shock ejecta density profile, as expected here \citep[see e.g.,][and references therin]{Waxman2017}. If $\frac{M_{csm}}{M_{tot}}<\delta_{m,bo}$ we can assume $v_{f}=f_{v}v_{s,bo}$ which is the maximum velocity at which breakout occurs.  In this case: 
\begin{equation}
    \frac{v_{s,csm}^{\left(2\right)}}{\text{cm\,}\text{s}^{-1}}=2\times10^{9}\left(\frac{f_{v}}{2}\right)\left(\kappa_{0.34}f_{\rho}M\right)^{0.13}\left(v_{s*,8.5}\right)^{1.13}R_{13}^{-0.26}
\end{equation}
so that $v_{s,csm}=min\left(v_{s,csm}^{\left(1\right)},v_{s,csm}^{\left(2\right)}\right)$. 

The total luminosity generated by the collisionless shock is given by $L\left(t\right)=2\pi\rho_{csm}r^{2}v_{s,csm}^{3}$. 

Using the derived $v_{s,csm}$ we find:\\
\begin{multline}
\label{eq:Xray_lum}
    L_{X} = 10^{42}\,\rm{erg\, s}^{-1} \times \\ 
    \begin{cases} 2.1\, \rho_{o,-16}^{0.7}v_{s*,8.5}^{2.7}t_{5d}^{-0.3}\left(f_{ \rho }M_{0,\odot}\right)^{0.3} & v_{s,csm}=v_{s,csm}^{\left(1\right)}\\
        0.6\, \rho_{o,-16}\left(\kappa_{0.34}f_{\rho}M_{0,\odot}\right)^{0.4}\\ \times \left(v_{s*,8.5}\right)^{3.4}R_{13}^{-0.8} & v_{s,csm}=v_{s,csm}^{\left(2\right)}
        \end{cases}    
\end{multline}  
Using Eq. \ref{eq:Xray_lum}, we convert our constraints on the XRT luminosity to constraints of the CSM density and mass loss. We assume a Bremsstrahlung spectrum with a temperature $T=200\mu \, (\frac{v_{s,csm}}{10^{9}\ {\rm cm s}^{-1}})^{2}$ keV \citep{Fransson1996,Katz2011}, where $\mu$ is the mean particle weight assumed to be $\mu=0.61$ for an ionized medium with a solar composition. We then correct the observed XRT luminosity to a bolometric X-ray luminosity, with correction factors ranging from 2-6 over our sample. We assume no intrinsic X-ray absorption at the SN site. To estimate the error on the values, the calculation is repeated for 100 points randomly drawn from the posterior sample on the shock-cooling light curve fits, and by randomly drawing points from a Gaussian distribution with a mean and standard deviation representing the X-ray measurements. We calculate $\frac{\dot{M}}{M_{\odot}\text{{\rm yr}}^{-1}}=10^{-4}\rho_{o,-16}v_{w,50}$ where $v_{w,50}$ is the CSM velocity in units of $50\, {\rm km\, s}^{-1}$, assumed to be 1. 

\begin{figure}[t]
\centering
\includegraphics[width=\columnwidth]{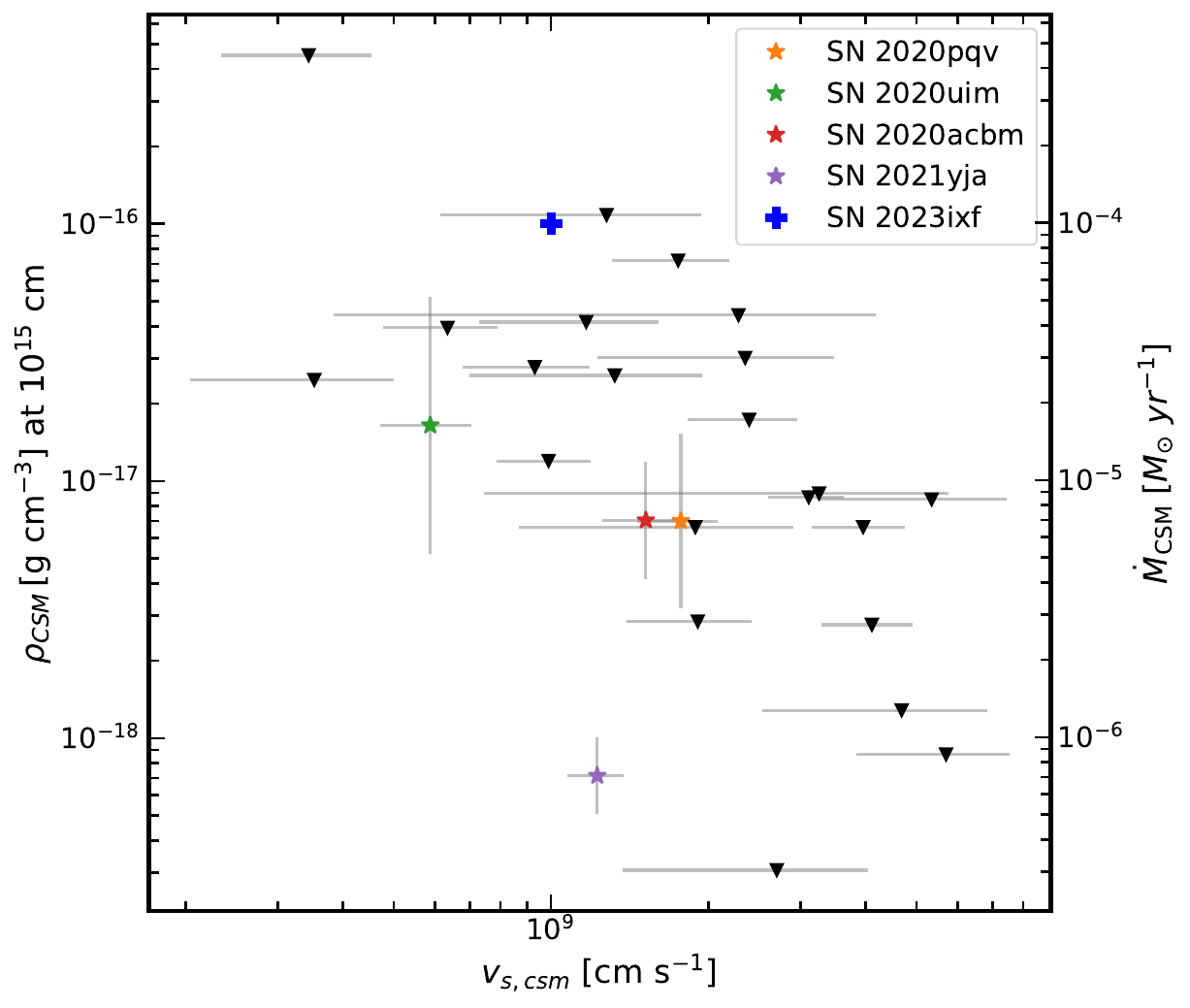} \\
\caption{X-ray limits on the extended ($\sim10^{15}$ cm) CSM density, mass loss and CSM shock velocity. Black triangles represent upper limits, colored points are detections from this work, and the blue plus represents the X-ray constraints of SN\,2023ixf from \cite{Grefenstette2023}. The extended $10^{15}$ cm mass-loss is consistent with field RSG levels.}
\label{fig:rho_csm_dist}
\end{figure}

We show our constraints in Fig. \ref{fig:rho_csm_dist}. Here the colored points represent individual detections, the downward pointing triangles represent upper limits, and the blue plus stands for the estimate of \cite{Grefenstette2023} for the mass-loss of SN\,2023ixf with a shock velocity arbitrarily chosen to be $10^{9}\, {\rm cm\, s}^{-1}$, deduced from the absorbing hydrogen column density between subsequent observations. 

There are 2 main systematics involved in our approach. (1) The emission spectrum of a shock traversing the CSM is highly uncertain, and assuming it will emit with a temperature equal to the plasma temperature is probably inaccurate. For example, \cite{Grefenstette2023} found for SN\,2023ixf a temperature of $35^{+22}_{-12}$ keV, which results in a velocity $v = (0.54^{0.15}_{-0.1})\times10^{9}\, {\rm cm\,s}^{-1}$ which is lower by at least a factor of 2 from the observed photospheric velocity of SN\,2023ixf \citep[][]{Zimmerman2023,JacobsonGalan2023}. Decreasing the temperature of the X-ray spectrum from $>120$ keV to $35$ keV would reduce the bolometric X-ray luminosity by factor $>2$ and subsequently reduce the mass-loss and density. (2) The intrinsic absorption of the CSM could affect the emission. In the case of SN\,2023ixf, \cite{Grefenstette2023} report an absorption column density of $2.6\sim10^{23}$ atoms cm$^{-2}$ at $t=4$ days, and $5\sim10^{22}$ atoms cm$^{-2}$ at $t=11$ days. Using the NASA Portable, Interactive Multi-Mission Simulator\footnote{\href{https://heasarc.gsfc.nasa.gov/cgi-bin/Tools/w3pimms/w3pimms.pl}{https://heasarc.gsfc.nasa.gov/cgi-bin/Tools/w3pimms/w3pimms.pl}}, we estimate our results would change by a factor of $\times 2$ if $N_H=1\times10^{23}$ cm$^{-2}$ in the XRT band. Such a value at the typical photon-weighted XRT observation time would imply a mass loss rate of $\gtrsim 10^{-4}\, M_{\odot}\, {\rm yr}^{-1}$, indicating this will affect only a few of the SNe in our sample. Our limits are consistent with the observed mass-loss of field RSGs \citep{deJager1988,Marshall2004,vanLoon2005}, but lower than inferred through modelling of narrow ``flash-ionization" spectral features, implying mass-loss rates as high as $10^{-2} M_{\odot}\, {\rm yr}^{-1}$ \citep{Dessart2017,Boian2019}, likely since these methods probe different regions of the CSM density profile. This is also the case for SN\,2023ixf: comparisons of the early time spectra performed by \cite{JacobsonGalan2023} and \cite{Bostroem2023b} to the models of \cite{Dessart2017} indicate a mass-loss rate of $10^{-3}-10^{-2}\, M_{\odot}\, {\rm yr}^{-1}$, much higher than those inferred by \cite{Grefenstette2023}, probing the extended CSM. The models of \cite{Dessart2017} introduce a mass-loss rate declining continuously to $10^{-6}\, M_{\odot}\, {\rm yr}^{-1}$ by $r=10^{15}\ \rm cm$, reflecting a dense mass-loss region swept up by the shock in the CSM at early times. Thus they are capable of discriminating between different CSM densities at few $10^{14}\,$cm. 

Since some amount of confined CSM is present in the majority of SNe II \citep{Bruch2021} we consider the effect of such dense CSM on our analysis. We repeat the analysis, but assume that the CSM swept up by the shock at $t<t_{X}$ has a density profile of $10^{-14}\, {\rm g\, cm}^{-3}(\frac{r}{10^{15}\, \rm cm})^{-2}$ ($\dot{M}=10^{-3} M_{\odot}\, {\rm yr}^{-1}$). This weakly decreases $v_{csm}$, and subsequently decreases $L$. For the majority of the sample, our limits do change by more then $50\%$, and at most by a factor of 3. 

Our results independently support the conclusion that by $\sim10^{15}$ cm, the density of the CSM has already declined to typical values observed for RSG stars, and that regions of dense mass loss are confined to the nearby environment of the progenitor star, and probing the final year of its evolution.

\begin{figure}[t]
\centering
\includegraphics[width=\columnwidth]{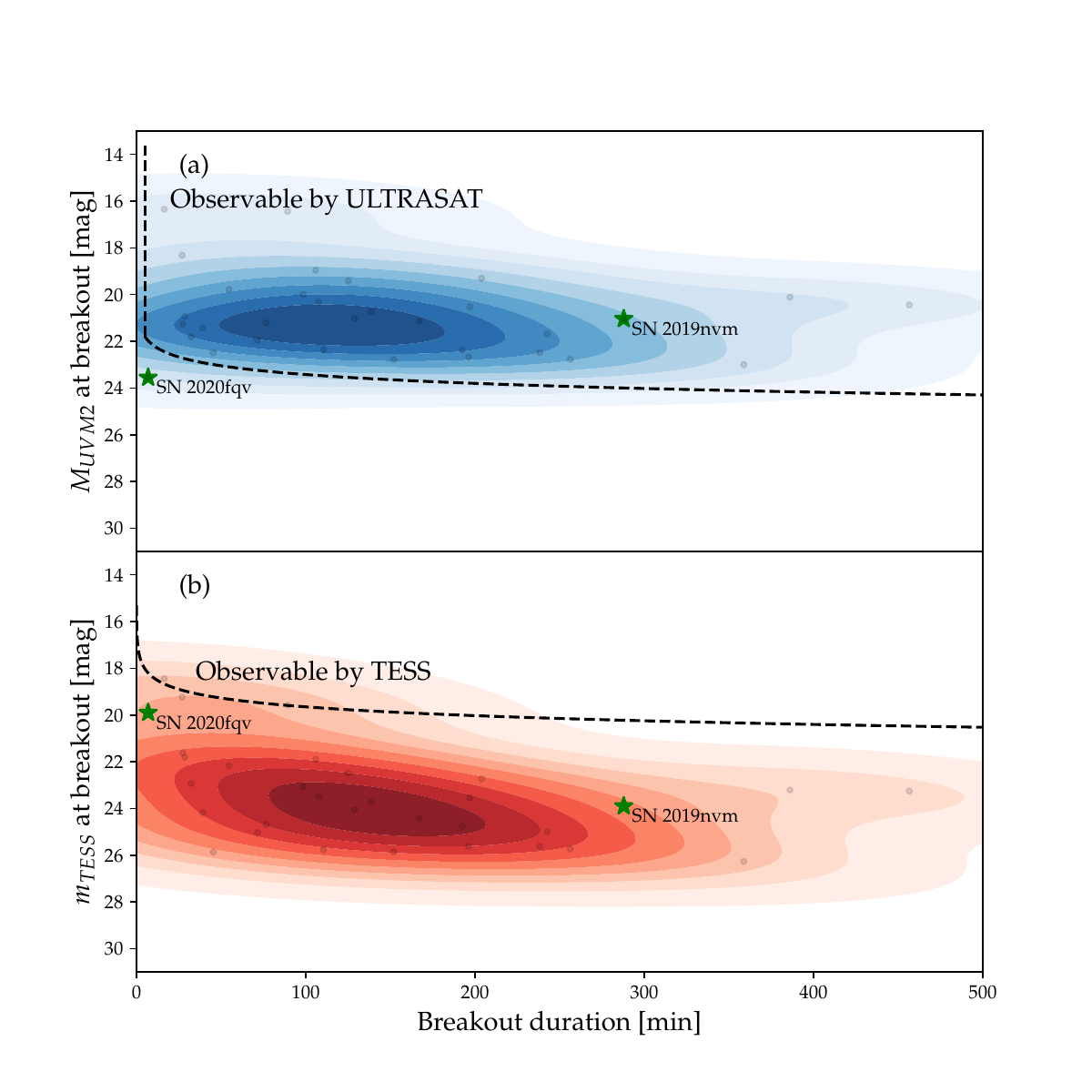} \\
\caption{Prediction for the breakout flare signal from a sample of SNe II in the optical and UV. Panel (a) shows a kernel density estimate (KDE) plot of the \ultrasat\ breakout duration and peak magnitude. Grey points correspond to the prediction of the best fit cooling light curve. The dashed line shows the limiting magnitude of the survey binned to varying degree. Panel (b) shows the same for the  \textit{TESS}\ bandpass, although the prediction for the breakout pulse spectrum are less certain in the optical, and should be treated as lower limits. Our results show it is very difficult to rule out the existence of a breakout pulse in optical wavelengths alone.  }
\label{fig:ultrasat_sbo}
\end{figure}

\begin{figure*}[t]
\centering
\includegraphics[width=\textwidth]{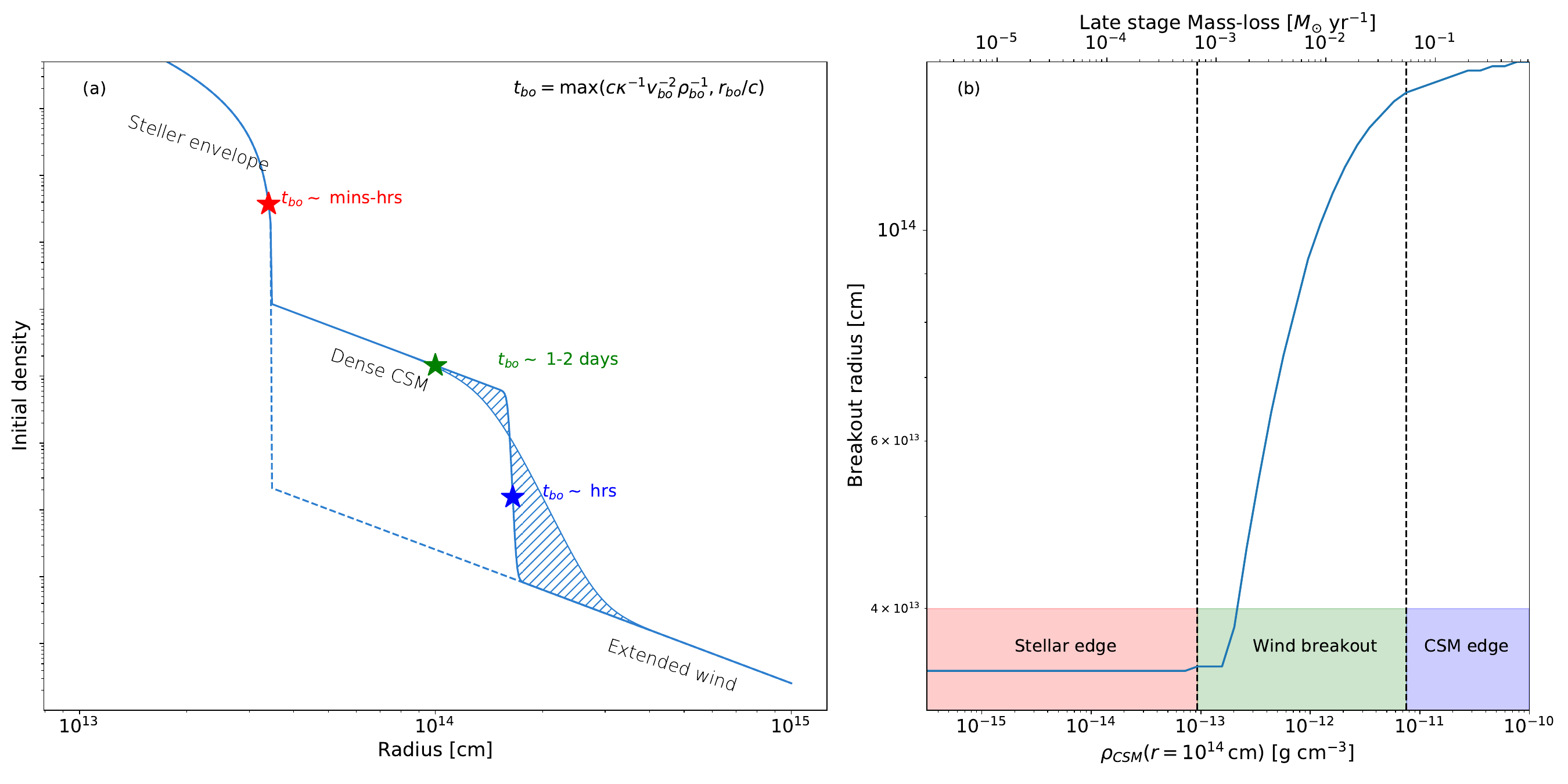} \\
\caption{(a) Schematic illustration of the proposed alternatives for early observations of SNe II. The curves represent possible pre-explosion density profiles of the envelope and CSM, corresponding to the mass-loss history of the progenitor in the months before explosion. Depending on the exact parameters of this profile, the breakout shell can be located in 3 possible locations. The red star represent a breakout radius at the edge of the stellar envelope. Since the density is steeply declining, the shock transition region is narrow, and the duration of breakout will be typically dominated by the light travel time. The blue star corresponds to a breakout at the edge of a dense shell of CSM. The density profile is declining steeply, and the breakout pulse duration can be set either by the light travel time or the shock crossing time, both lasting hours. The green point corresponds to the third option, occurring for a minority of cases. Here the density profile is shallow, and increasing the duration of the breakout pulse to the shock crossing timescale of a few days. The $r\lesssim3\times10^{14}$ cm density will determine the early light curve and spectra, and the $r\gtrsim3\times10^{14}$ cm determines the X-ray emission emerging after the first few days. (b) The breakout radius for a star with a $500R_{\odot}$ and $1 M_{\odot}$ stellar envelope surrounded by varying amounts of CSM confined to $1.5\times10^{14}\rm cm$. The conversion to mass-loss assumed $v_w = 50\, \rm{km\, s}^{-1}$. Increasing the mass of the shell of dense CSM moves the breakout location from the stellar envelope, to the shallow region of the dense CSM, and onward to the edge of the dense CSM, if a significant portion of the envelope was ejected.}
\label{fig:csm_cartoon}
\end{figure*}

\subsection{Observing shock-breakout and shock-cooling with \ultrasat}
\label{subsec:ultrasat_implications}

\ultrasat\ will conduct a high cadence (5 min) UV survey with a 200 $\rm deg^{2}$ field of view (FOV). It will detect tens of shock breakout signatures and hundreds of shock cooling light curves in its first 3 years \citep{Shvartzvald2023}. The high cadence light curves of \ultrasat\ will resolve all phases of the early SN evolution - shock breakout, planar phase and spherical phase shock cooling. While spherical shock cooling alone provides constraints on the progenitor parameters, the planar phase, typically lasting hours, can discriminate between models more finely. Directly observing the breakout pulse can provide independent constraints on the breakout radius, and the velocity of the outermost layers of the ejecta. This can resolve the remaining ambiguity as to the reason for the systematic deviation from the expected planar phase in large radii fits. Observing the early UV-optical color of SNe will discriminate between a light curve rise driven by cooling, following a stellar edge breakout, or by heating of the ejecta, during an extended shock breakout in a shallow density profile (examples of the latter including SN\,2020pni,  \referee{SN\,2018zd} and SN\,2023ixf). For SNe with light curves well matched by a stellar breakout, the velocity and mass of the breakout shell will be constrained by the breakout pulse itself \citep{Sapir2011,sapir2013}.

In combination with X-ray followup and spectral modeling, these can be used to accurately map the CSM density profile, with each tracer probing a different segment of the density profile. While there have been some candidate shock-breakout flares in the optical \citep[]{Garnavich_2016,Bersten2018}, some claims have been disputed \citep{Rubin_2017}, and the sample of \textit{TESS}\ CCSNe of \cite{Vallely2021}, binned to 30-min cadence, show no detection of breakout flares. Breakout flares are expected to peak in the UV or X-ray, but the non-LTE spectral shape makes prediction in the optical highly uncertain \citep{sapir2013,sapir2014}. While initially the number of photons produced is not enough to reach thermal equilibrium, the planar phase temperatures are already close to the equilibrium temperature, and the exact details of this transition can change the optical light curve by orders of magnitude. The UV peak, closer to the peak frequency of the emission, is much better understood.

In order to produce a clear prediction for the \ultrasat\ survey based on the observed sample of SNe II, we calculate the breakout signal in the \textit{TESS}\ and in the UVOT $UVM2$ bandpass ($UVM2$ is chosen since it is closest to the \ultrasat\ bandpass). For every SN we fit in $\S$~\ref{subsec:SC_fitting}, we use breakout properties $\rho_{bo}$, and $v_{s,bo}$ to calculate the luminosity and spectrum at breakout according to the models of \cite{Sapir2011,Katz2012,sapir2013}. We integrate the spectrum and and compute the typical \textit{TESS}\ and \ultrasat\ brightness during breakout, and the duration of the expected breakout. We show the distribution of parameters in Fig. \ref{fig:ultrasat_sbo}. Panel (a) shows a kernel density estimate (KDE) plot of the \ultrasat\ breakout landscape, and panel (b) shows the expected \textit{TESS}\ brightness. We highlight the predictions for SN\,2020fqv and SN\,2020nvm, observed by \textit{TESS}. We stress the optical wavelength predictions are highly uncertain, and should be treated as lower limits. Our results are consistent with the entirety of the breakout flares predicted by our modelling being measured by \ultrasat, and none of the flares being observed in the optical wavelengths.

\section{Conclusions and Summary}
\label{sec:summary}
\begin{itemize}
    \item In this paper we have presented the UV-optical photometry of 34 spectroscopically regular SNe II detected in the ZTF survey and followed up by the \swift\ telescope within 4 days of explosion. In addition to the UV-optical data, we report four XRT detections and 3 $\sigma$ upper limits for the rest of the sample.  
    \item In $\S$~\ref{subsec:color} we analyze the color evolution in of the sample. We show that besides SN\,2020pni, the rest of our sample had UV-optical colors which are becoming redder with time across the entire SED, indicating they are cooling. 
    \item We show that the combination of UV, UV-optical and optical colors can be used as a discriminator between various degree of intrinsic time-dependent deviations from blackbody and host-galaxy extinction with non-MW extinction laws. We show there is no preference in UV-optical color for SNe with flash features, and argue the deviations are consistent with the predictions of shock cooling models. 
    
    \item Using the scatter in early time color, we argue our sample has a host extinction smaller than $E(B-V)=0.2$ mag. Subsequently, we show we can measure the extinction of highly extinguished SNe to better than $0.2$ mag. The average early time colors of the SNe in our sample are provided in Table \ref{tab:colors}.
     
    \item In $\S$~\ \ref{subsec:blackbody} we fit the SEDs of the SNe in our sample at the times of UVOT observations to a blackbody, and recover the evolution of their blackbody radius and temperature. We show that the evolution of these parameters is in excellent agreement with the predictions of spherical phase shock cooling, with a statistically significant difference in the average temperature and radius between object with and without flash features. We also show at least 30\% of the objects in our sample are more luminous than expected from an envelope breakout with $R<10^{14}$ cm- indicating a larger progenitor radius or a higher shock velocity parameter relative to generic expectations. 
    
    \item Motivated by the good agreement with the predictions of spherical phase shock-cooling, we present a method to fit the light curves to latest shock-cooling models in $\S$~\ \ref{subsec:method}, accounting for deviations from blackbody over a large range of parameters, and interpolating between the planar and spherical phase of shock cooling. We demonstrate this method is unbiased when fitting the MG simulations of \refereetwo{M24}, although these have correlated residuals. We demonstrate that we can recover the breakout radius $R_{*}$, the shock velocity parameter $v_{*,8.5}$ describing the velocity profile in the outer regions of the ejecta, and the extinction. We show that we cannot recover the envelope mass $M_{env}$, total mass $M$, or numerical density scaling parameter $f_{\rho}$ using our method. \referee{We conclude that by fitting we can confirm or reject the underlying assumption of shock-cooling following envelope breakout.} 
    
    \item Overall we find the early UV-optical light curves of our sample divides into 3 groups. (1) A majority (33/34) of SNe which cool at early times, which we denote as ``II-C". This group is comprised of (a) SNe that are well fit throughout their evolution, with radii characteristic of the observed RSG radius distribution
    and (b) SNe which are fit by larger radius, more luminous models and which systematically miss the early ($<$1d) rise. We denote these as ``II-C+" and ``II-C-" respectively. (2) The third group is represented by a single object in our sample (SN\,2020pni), which is heating in the first few days. A similar evolution has been observed for the nearby SN\,2023ixf \referee{and for SN\,2018zd}. We denote these as ``II-H". 
    
    \item As we have demonstrated that there is no bias in our fitting method, we argue \referee{this deviation from the predicted rise} reflects a physical difference from an idealized breakout from a polytropic envelope. We speculate this difference could be related to the presence of CSM or an asymmetric shock breakout. We assume the inference of large radii is real, and show that while most of the sample is characterized by a large radius, this is due to a luminosity bias affecting our sample at distance $>70$ Mpc. We show the volume corrected probability peaks at radii similar to those of field RSG. We conclude that while $71^{+7}_{-4}\%$ of observed SNe II are over luminous, with a large radius, the majority ($66^{+11}_{-22}\%$) of exploding RSG have a typical radius at explosion. Since some objects in our sample are also consistent with a smaller radius, this should be treated as a lower limit. 
    \item Using the X-ray limits and the constraints on the velocity profile of the ejecta from the light curve fitting, we derive limits on the CSM density at $0.5-2\times 10^{15}$ cm from the progenitor star, which constrains the mass loss of the progenitors $\sim 3-15$ yrs before the explosion assuming a $50\, \rm{km\, s}^{-1}$ winds. We show the limits and detection are systematically lower than the required mass-loss to explain flash ionization features, supporting the conclusion that these stars undergo increased mass-loss in the final months before explosion. Uncertainties in the spectral shape of the X-ray emission, the amount of CSM below $10^{14}$ cm, and absorption in the CSM will change this result by less than an order of magnitude.

    \item In $\S$~\ \ref{subsec:ultrasat_implications}  we study the predictions of the fit parameter distribution to the landscape of shock breakout flares for the \ultrasat\ mission and high-cadence optical missions such as TESS. We argue the non-detections of breakout flares in the optical surveys is to be expected, and that observations with \ultrasat\ should indeed easily discover the breakout flares from an analogues sample to ours. 
    \item By combining our constraints on the breakout radius and the extended CSM density, we propose a scenario that explains all three groups in our sample in a single framework. By varying the amount of CSM lost in the last year, the breakout radius, duration and temperature change. If a small amount of mass ($\lesssim10^{-3}\, M_{\odot}$) is lost, breakout will occur at the stellar envelope. Its characteristic duration will be minutes to an hour and will peak in the extreme UV. This scenario can explain most SNe II-C+. If the star loses most of its envelope ($\gtrsim0.1\, M_{\odot}$), breakout will occur at the edge of the dense CSM. The characteristic breakout duration will be hours long, and can contaminate the early light curves as it will peak in the far UV. This scenario will explain most II-C-. If the SN loses $\sim0.01\, M_{\odot}$ during the last year, breakout will occur in the dense CSM. Such a breakout will be occur over a timescale of a few days, during which a heating of the breakout region and an increase in luminosity will be observed as the breakout pulse is released, with an SED peaking in the near UV. This scenario will account for SNe II-H. This framework is schematically summarized in Fig. \ref{fig:csm_cartoon}.

\end{itemize}

\section{Data Availability}
All data used in this paper will be made available via WISeREP\footnote{\href{https://www.wiserep.org}{https://www.wiserep.org}} \citep{yaron2012}. We make all figures of all light curve fits and light curve plots available as online figures through the journal website upon publication. \refereetwo{The code used for producing light curve and blackbody fits is released to \href{https://github.com/idoirani}{https://github.com/idoirani} \citep{Irani24code}}.

\section{Acknowledgements }

\textit{Software}: \package{Astropy} \citep{Astropy2013,Astropy2018}, \package{IPython} \citep{Perez2007}, \package{Matpotlib} \citep{Hunter2007}, \package{Numpy} \citep{Oliphant2006}, \package{Scipy} \citep{Virtanen2020}, \package{exctinction} \citep{Barbary2016},
 \package{dynesty} \citep{Skilling2004,Skilling2006,Feroz2009,Higson2019,Speagle2020a} GROWTH marshal \citep{Kasliwal2019}, Fritz/\package{SkyPortal} \citep{vanderWalt2019,Coughlin_2023}, \package{SWarp} \citep{Bertin2010}

\textit{Facilities}: 
P48, \swift\ (UVOT, XRT), P60 (RC), Liverpool telescope (IO:O)

We thank Doron Kushnir, Barak Zackay and Boaz Katz for their insights on the analysis. We are grateful to the staff at the various observatories where data were obtained. This work made use of data supplied by the UK Swift Science Data Centre at the University of Leicester.

Based on observations obtained with the Samuel Oschin Telescope 48-inch and the 60-inch Telescope at the Palomar Observatory as part of the Zwicky Transient Facility project. ZTF is supported by the National Science Foundation under Grants No. AST-1440341 and AST-2034437 and a collaboration including current partners Caltech, IPAC, the Weizmann Institute of Science, the Oskar Klein Center at Stockholm University, the University of Maryland, Deutsches Elektronen-Synchrotron and Humboldt University, the TANGO Consortium of Taiwan, the University of Wisconsin at Milwaukee, Trinity College Dublin, Lawrence Livermore National Laboratories, IN2P3, University of Warwick, Ruhr University Bochum, Northwestern University and former partners the University of Washington, Los Alamos National Laboratories, and Lawrence Berkeley National Laboratories. Operations are conducted by COO, IPAC, and UW.

The ZTF forced-photometry service was funded under the Heising-Simons Foundation grant \#12540303 (PI M. J. Graham). 
The SED Machine at Palomar Observatory is based upon work supported by the NSF under grant 1106171. 
The Gordon and Betty Moore Foundation, through both the Data-Driven Investigator Program and a dedicated grant, provided critical funding for SkyPortal.

The Liverpool Telescope is operated on the island of La Palma by Liverpool John Moores University in the Spanish Observatorio del Roque de los Muchachos of the Instituto de Astrof\'isica de Canarias with financial support from the UK Science and Technology Facilities Council. Partly based on observations made with the Nordic Optical Telescope, operated at the Observatorio del Roque de los Muchachos.

This research has made use of the Spanish Virtual Observatory (https://svo.cab.inta-csic.es) project funded by MCIN/AEI/10.13039/501100011033/ through grant PID2020-112949GB-I00
 A.G-Y.’s research is supported by the EU via ERC grant 725161, the ISF GW excellence center, an IMOS space infrastructure grant and BSF/Transformative and GIF grants, as well as the André Deloro Institute for Advanced Research in Space and Optics, The Helen Kimmel Center for Planetary Science, the Schwartz/Reisman Collaborative Science Program and the Norman E Alexander Family Foundation ULTRASAT Data Center Fund, Minerva and Yeda-Sela;  A.G.-Y. is the incumbent of the Arlyn Imberman Professorial Chair.
E. Waxman’s research is partially supported by grants from the ISF, Norman E Alexander Family M Foundation ULTRASAT Data Center, Nella and Leon Benoziyo Center for Astrophysics, Schwartz Reisman Institute for Theoretical Physics, and by the Max Planck Professorial Chair of Quantum Physics.E.O.O. is grateful for the support of grants from the Benozio center, Willner Family Leadership Institute, Ilan Gluzman (Secaucus NJ), Madame Olga Klein - Astrachan,
Minerva foundation, Israel Science Foundation, BSF-NSF, Israel Ministry of Science, Yeda-Sela, and Weizmann-MIT, and the Rosa and Emilio Segre Research Award.

\bibliographystyle{apj},
\bibliography{bibliograph.bib}

\appendix

\section{}

\subsection{Shock-cooling model}
\label{ap:SC_mod}
\subsubsection{Blackbody evolution}
We fit SN light-curves to the shock-cooling model of \cite{Morag2023}. This model describes the blackbody evolution of a cooling envelope until recombination or sufficient transparency of the envelope, using a set of four free parameters: (1) $R_{13}$, the radius of the progenitor star in units of $10^{13}$ cm. (2) $f_{\rho}M_{0}$ the product of the numeric factor $f_{\rho}$, which describes the structure of the density near the edge of the stellar envelope, and $M_{0}$ which is the progenitor mass prior to the SN in units of \msun. (3) $v_{s*,8.5}$ the shock velocity parameter in units of $10^{8.5} {\rm cm\ s^{-1}}$, which corresponds to $v_{s*,8.5} = 1.05f_{\rho}^{-0.19}\sqrt(E/M)$, roughly equal to $\sim v_{ej}/5$ at early times. (4) $M_{\rm env}$, the envelope mass. $\kappa_{0.34}$ is the opacity in units of $0.34\, \rm cm^2$ g$^{-1}$ and is set to 1 for all cases. $t_{d}/t_{hr}$ is the time since the explosion in units of days or hours respectively. Following their notation, $L,T$ evolve according to:
\begin{equation}
\label{eq:AL_sum}
    L_{SC}=L_{\rm planar}+0.9\exp\left[-\left(\frac{2t}{t_{\rm tr}}\right)^{0.5}\right]\,L_{\rm RW},
\end{equation}
\begin{equation}
\label{eq:AT_col_MSW_2}
    T_{\rm col}=1.1\min\left[T_{\rm ph,planar}\,,\,T_{\rm ph,RW}\right]
\end{equation}
which are valid during 
\begin{equation}
\label{eq:Ats}
   3 R / c = 17\,R_{13} \, {\rm min} < t < \min[t_{\rm 0.7 eV}, t_{\rm tr}/2].
\end{equation}

Where the terms in Eqs.~(\ref{eq:AL_sum} -- \ref{eq:Ats}) are
\begin{align}
\label{eq:AL_planar_powerlaw}
    \frac{L_{{\rm planar}}}{10^{42}\,{\rm erg\,s^{-1}}}=3.01\,R_{13}^{2.46}v_{{\rm s*,8.5}}^{0.60}(f_{\rho}M_{0})^{-0.06}t_{{\rm hr}}^{-4/3}\kappa_{0.34}^{-1.06},
\end{align}
\begin{align}
\label{eq:AT_planar_powerlaw}
    \frac{T_{{\rm ph,planar}}}{\,{\rm eV\,}}=6.94\,R_{13}^{0.12}v_{{\rm s*,8.5}}^{0.15}(f_{\rho}M_{0})^{-0.02}\kappa_{0.34}^{-0.27}t_{{\rm hr}}^{-1/3},
\end{align}
\begin{align}
\label{eq:ALRW}
   \frac{L_{{\rm RW}}}{2.08\times10^{42}\,{\rm erg\,s}^{-1}}=\,R_{13}v_{{\rm s*,8.5}}^{1.91}(f_{\rho}M_{0})^{0.09}\kappa_{0.34}^{-0.91}t_{{\rm d}}^{-0.17},
\end{align}
\begin{align}
\label{eq:ATphRW}
\frac{T_{{\rm ph,RW}}}{\,{\rm \,eV}}=1.66\,R_{13}^{1/4}v_{{\rm s*,8.5}}^{0.07}(f_{\rho}M_{0})^{-0.03}\kappa_{0.34}^{-0.28}t_{{\rm d}}^{-0.45},
\end{align}
\begin{equation} \label{eq:At_0_7eV}
    t_{0.7 \rm eV} = 6.86 \, R_{13}^{0.56} v_{\rm s*,8.5}^{0.16} \kappa_{0.34}^{-0.61} (f_{\rho}M_0)^{-0.06} \rm  \, days
\end{equation}
\begin{equation}
\label{eq:At_transp}
    \begin{split}
        \begin{aligned}
            t_{\rm tr} &= 19.5 \, \sqrt{\frac{\kappa_{0.34}M_{\rm env,0}} {v_{\rm s*,8.5}}}\, \text{days}.
        \end{aligned}
    \end{split}
\end{equation}

\subsubsection{Deviations from blackbody}
\refereetwo{M24} fully relax the assumption of LTE. A temperature, density and wavelength dependent opacity is used to estimate the flux in every wavelength, \referee{accounting for the effects of line emission and absorption}. a semi-analytical model \referee{of the SED is calibrated to} a set of radiation hydrodynamical simulations with multiple photon groups. \refereetwo{M24} show that the SED can be described using:
\begin{equation}
\label{eq:Lnu epsilon final_Appendix}
L_{\nu} = \begin{cases}
\left[L_{\rm BB} (0.85 \, T_{\rm col})^{-m} + L_{\nu,\epsilon}^{-m}\right]^{-1/m} & h\nu<3.5 T_{\rm col} \\
1.2 \times L_{\rm BB}(0.85 R_{13}^{0.13} t_d^{-0.13} \times T_{\rm col}) & h\nu>3.5 T_{\rm col},
\end{cases}
\end{equation}
with $m=5$ and
\begin{equation}
    L_{\rm BB}=L\times\pi B_{\nu}(T_{\rm col})/\sigma T_{\rm col}^{4}
    \label{eq:L_nu_BB_formula_Appendix},
\end{equation}
\begin{equation}
 L_{\nu,\epsilon}=\frac{\left(4\pi\right)^{2}}{\sqrt{3}}r_{col,\nu}^{2}\frac{\sqrt{\epsilon_{\nu}}}{1+\sqrt{\epsilon_{\nu}}}B_{\nu}(T_{col,\nu}),\quad 
  \epsilon_\nu=\frac{\kappa_{\rm ff,\nu}}{\kappa_{\rm ff,\nu}+\kappa_{\rm es}}  \label{eq: epsilon prescription _Appendix}
\end{equation}
and 
\begin{equation}
    r_{\rm col,\nu} = R + 2.18 \times 10^{13} L_{\rm br,42.5}^{0.48} T_{\rm br,5}^{-1.97}
    \kappa_{0.34}^{-0.07} \tilde{t}^{0.80} \nu_{\rm eV}^{-0.08} \, \rm cm,
\end{equation}
\begin{equation}
    T_{\rm col,\nu} = 5.47 \, L_{\rm br,42.5}^{0.05} T_{\rm br,5}^{0.92}
    \kappa_{0.34}^{0.22}
    \tilde{t}^{-0.42} \nu_{\rm eV}^{0.25} \, \rm eV,
\end{equation}
\begin{equation}
    \kappa_{\rm ff} = 0.03 \, L_{\rm br,42.5}^{-0.37} T_{\rm br,5}^{0.56} 
    \kappa_{0.34}^{-0.47}
    \tilde{t}^{-0.19}
    \nu_{\rm eV}^{-1.66} \, \rm cm^2 \, g^{-1}.
\end{equation}
Here $L_{\rm br}=L_{\rm br,42.5} 10^{42.5} \rm \, erg \, s^{-1}$, $T_{col}=5 T_{\rm col,5}$ eV, and $\nu=\nu_{\rm eV}$ eV, and $R$ in terms of the break parameters is:
\begin{equation}
    R = 2.41\times10^{13} \, t_{\rm br,3}^{-0.1} \,L_{br,42.5}^{0.55} \,T_{br,5}^{-2.21} \,\text{cm}.
\end{equation}

\subsubsection{Fitting procedure}

Since the validity of this model is dependent on the model parameters, a $\chi^2$ minimization is not applicable. Instead, we fit this model with a likelihood function adapted for a variable validity domain, as discussed in detail in \cite{Soumagnac2020}:
\begin{equation}
   \mathcal{L}={\rm PDF}\left(\chi^{2},dof\right)=\frac{\left(\chi^{2}\right)^{\frac{N}{2}-1}\exp\left(\frac{\chi^{2}}{2}\right)}{2^{\frac{N}{2}}\Gamma\left(\frac{N}{2}\right)}
\end{equation}
where $\chi^2$ is the $\chi^2$ statistic, PDF is the $\chi^2$ distribution given the number of degrees of freedom, and $\Gamma$ is the gamma function. We calculate the $\chi^2$ statistic using the observational errors and an empirical covariance matrix:
\begin{equation}
    \chi^{2}=\left(f_{i}-m_{i}\right)\left(COV_{ij}\right)^{-1}\left(f_{j}-m_{j}\right)
\end{equation}
 $f_i$ are the observed fluxes, $m_i$ are the integrated synthetic fluxes for the model. The covariance matrix $COV_{ij}$ is calculated using the observational errors $\sigma_i$ and an empirical covariance calculated over the hydrodynamical simulation sample.
 \begin{equation}
    COV=1.5COV_{sys}+\frac{1}{\boldsymbol{\sigma}_{obs}^{2}}
\end{equation}
\begin{equation}
    COV_{sys,ij}=\left\langle r_{i}r_{j}\right\rangle -\left\langle r_{i}\right\rangle \left\langle r_{j}\right\rangle ;\,\,r_{i}=f_{i}-m_{i}
\end{equation}
 Here $\sigma_i$ include a 10 \% systematic error to account for cross-instrument calibration, and we scale $COV_{sys}$ by a factor 1.5 to account for the theoretical uncertainty, assumed to have the same covariance structure as the theoretical calibration uncertainty. This covariance matrix is constructed for every SN data set separately. For every SN, a synthetic dataset in created using the sampling and bands available in its individual dataset. The average \referee{is calculated over the parameters space of valid models (i.e., where $t<{t_validity}$) for each data point.}

\subsection{Early-time Colors}
\begin{figure*}[t]
\centering
\includegraphics[width=\textwidth]{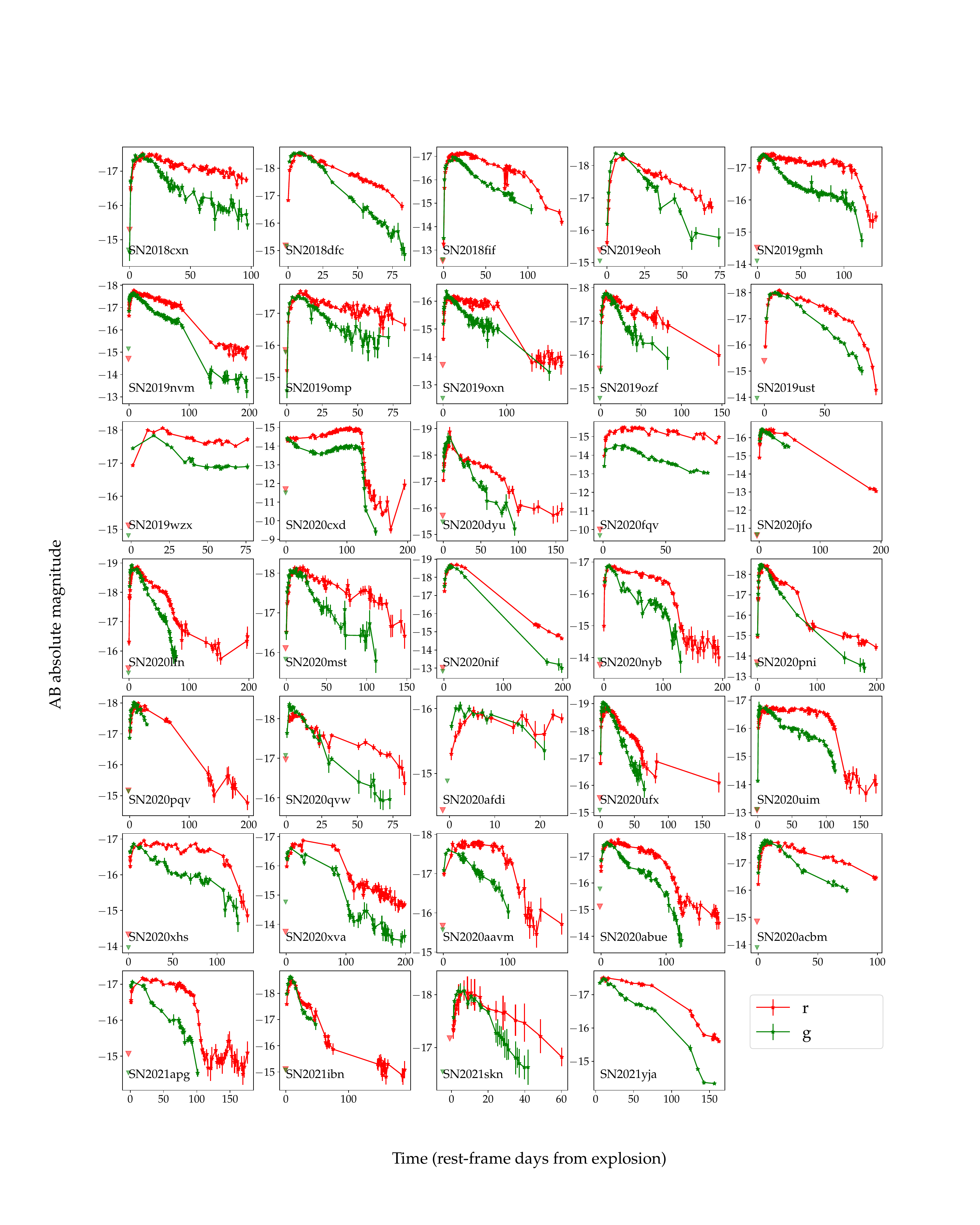} \\
\caption{The full ZTF $g$-band and $r$-band light curves of our sample of 34 SNe II.}
\label{fig:full_lc}
\end{figure*}

\referee{The early time colors of the SNe in our samples excluding the extinguished SN\,2020fqv are shown in Table \ref{tab:colors}. The full ZTF $g,r$ light curve are shown in Fig. \ref{fig:full_lc}}.
\begin{deluxetable*}{clccccccc}
\centering
\label{tab:colors}
\tablecaption{SNe II early-time colors}
\tablewidth{34pt} 
\tablehead{\colhead{t [rest-frame days]} &   \colhead{Filter} &   \colhead{$\lambda_{piv}$ [\AA]} &       \colhead{$W_{eff}$ [\AA]} &  \colhead{Mean $m_{\lambda}-m_{r}$ [mag]} &  \colhead{STD [mag]} &  \colhead{Bluest color [mag]} &  \colhead{Reddest color [mag]} &  \colhead{$N_{SN}$}} 
\tabletypesize{\scriptsize} 
\startdata
       1 &   UVW2 &        2055 &    305 &               -0.70 &       0.27 &      -1.13 &      -0.49 &     5 \\
       1 &   UVM2 &        2246 &    259 &               -0.70 &       0.30 &      -1.12 &      -0.32 &     5 \\
       1 &   UVW1 &        2580 &    397 &               -0.57 &       0.25 &      -0.97 &      -0.35 &     5 \\
       1 &      u &        3467 &    352 &               -0.53 &       0.20 &      -0.82 &      -0.34 &     5 \\
       1 &      g &        4702 &    641 &               -0.24 &       0.09 &      -0.43 &      -0.05 &    18 \\
       1 &      i &        7489 &    767 &                0.27 &       0.15 &       0.14 &       0.60 &     6 \\
       \hline
       2 &   UVW2 &        2055 &    305 &               -0.67 &       0.27 &      -1.30 &      -0.18 &    23 \\
       2 &   UVM2 &        2246 &    259 &               -0.64 &       0.28 &      -1.23 &      -0.02 &    22 \\
       2 &   UVW1 &        2580 &    397 &               -0.59 &       0.22 &      -1.17 &      -0.20 &    23 \\
       2 &      u &        3467 &    352 &               -0.59 &       0.21 &      -0.95 &      -0.12 &    23 \\
       2 &      g &        4702 &    641 &               -0.27 &       0.11 &      -0.44 &       0.06 &    30 \\
       2 &      i &        7489 &    767 &                0.27 &       0.23 &       0.08 &       0.98 &    22 \\
       \hline
       3 &   UVW2 &        2055 &    305 &               -0.33 &       0.30 &      -0.89 &       0.24 &    30 \\
       3 &   UVM2 &        2246 &    259 &               -0.37 &       0.30 &      -0.94 &       0.14 &    30 \\
       3 &   UVW1 &        2580 &    397 &               -0.36 &       0.24 &      -0.93 &       0.12 &    30 \\
       3 &      u &        3467 &    352 &               -0.45 &       0.24 &      -1.05 &       0.16 &    30 \\
       3 &      g &        4702 &    641 &               -0.18 &       0.12 &      -0.45 &       0.10 &    31 \\
       3 &      i &        7489 &    767 &                0.23 &       0.21 &       0.00 &       0.93 &    25 \\
       \hline
       4 &   UVW2 &        2055 &    305 &               -0.01 &       0.35 &      -0.57 &       0.68 &    29 \\
       4 &   UVM2 &        2246 &    259 &               -0.13 &       0.31 &      -0.71 &       0.39 &    29 \\
       4 &   UVW1 &        2580 &    397 &               -0.17 &       0.28 &      -0.73 &       0.29 &    29 \\
       4 &      u &        3467 &    352 &               -0.34 &       0.25 &      -1.01 &       0.28 &    30 \\
       4 &      g &        4702 &    641 &               -0.14 &       0.10 &      -0.39 &       0.02 &    31 \\
       4 &      i &        7489 &    767 &                0.23 &       0.15 &      -0.02 &       0.80 &    24 \\
       \hline
       5 &   UVW2 &        2055 &    305 &                0.31 &       0.35 &      -0.29 &       0.98 &    27 \\
       5 &   UVM2 &        2246 &    259 &                0.13 &       0.29 &      -0.45 &       0.62 &    26 \\
       5 &   UVW1 &        2580 &    397 &                0.01 &       0.22 &      -0.45 &       0.47 &    28 \\
       5 &      u &        3467 &    352 &               -0.25 &       0.21 &      -0.51 &       0.41 &    28 \\
       5 &      g &        4702 &    641 &               -0.12 &       0.09 &      -0.28 &       0.10 &    29 \\
       5 &      i &        7489 &    767 &                0.21 &       0.10 &      -0.00 &       0.42 &    23 \\
\hline
\enddata
\tablenotetext{a}{}
\end{deluxetable*} 

\end{document}